   \documentclass[final,5p,times,twocolumn,authoryear]{elsarticle}

\usepackage{amssymb}
\usepackage{lipsum}
\usepackage{macros} 
\usepackage{amsmath}
\usepackage{booktabs}
\usepackage{array}
\usepackage{tabularx}
\usepackage{multirow}
\usepackage{verbatim}
\usepackage[percent]{overpic}
\usepackage[dvipsnames]{xcolor}

\newcommand{\kms}{km\,s$^{-1}$}

\journal{High Energy Astrophysics}

\begin{document}

\begin{frontmatter}

\title{Can accreting isolated neutron stars be detected?}

\author[first,second]{Marina Afonina\corref{cor1}}
\cortext[cor1]{Corresponding author}
\ead{afoninamd@gmail.com}
\author[third,first]{Anton Biryukov}
\author[first]{Sergei Popov}
\affiliation[first]{organization={Sternberg Astronomical Institute,  Lomonosov Moscow State University},
            addressline={13 Universitetskij pr.}, 
            city={Moscow},
            postcode={119234}, 
            country={Russia}}
\affiliation[second]{organization={Department of Physics, Lomonosov Moscow State University},
            addressline={1/2 Leninskie Gory}, 
            city={Moscow},
            postcode={119991},
            country={Russia}}
\affiliation[third]{organization={The Raymond and Beverly Sackler School of Physics and Astronomy, Tel Aviv University}, 
            addressline={55 Chaim Levanon St.}, 
            city={Tel Aviv},
            postcode={6997801},
            country={Israel}}
\begin{abstract}

We perform population synthesis modeling of isolated neutron stars in the Milky Way over its lifetime. Compared with previous studies, we use more detailed models of the interstellar medium and the magneto-rotational evolution of neutron stars. We demonstrate that presently, the spin-down rate at the propeller stage is the main uncertain factor that influences the number of accreting isolated neutron stars.
If the propeller stage duration allows neutron stars to begin accreting matter from the interstellar medium and if the efficiency of accretion is high, then the number of accreting isolated neutron stars in {\it eROSITA} data can reach $\sim$a few thousand. Still, uncertainties in spin-down at the propeller stage and in the accretion process can drastically decrease this number. We suggest that future observations of neutron stars in wide low-mass binaries recently discovered by {\it Gaia} can clarify these issues.
\end{abstract}




\end{frontmatter}




\section{Introduction}
\label{introduction}

 With the core-collapse rate about one in a few tens of years \citep{2021NewA...8301498R} and the Galactic age slightly above $10^{10}$~yrs, we immediately obtain that there are up to $10^9$ neutron stars (NSs) and black holes (BHs) in our Galaxy. NSs are expected to be more numerous than BHs by a factor of a few, as NSs originate from  less massive supernova (SN) progenitors, which are more numerous (some of BHs can also form without a SN explosion). Mostly, NSs are expected to be isolated objects as binary (or multiple) stellar systems are typically destroyed due to rapid mass loss and natal kick during a SN explosion \citep{2006LRR.....9....6P}. However, the number of observed isolated NSs is $<10^4$. Mostly, they are young (age $\lesssim 10^7$~yrs) objects, and radio pulsars dominate in this population \citep{2005AJ....129.1993M}. Some active radio pulsars are not detected as their beams do not cross Earth, some are too dim. Still, the total Galactic population of radio pulsars is estimated to be $\lesssim 10^5$ objects \citep{2006ApJ...643..332F}. What is the fate of other (older) NSs?

Isolated NSs can demonstrate various types of astrophysical appearance, see e.g. \cite{2023Univ....9..273P} and references therein. However, it is natural to expect that only relatively young objects might appear as bright sources. The reason is that most types of energy resources (thermal, magnetic, rotational) are exhausted after some cosmologically short period of time. Thus, old NSs might be very elusive objects.
Still, more than half a century ago 
\citep{1970ApL.....6..179O, 1970AZh....47..824S} it was hypothesized that old NSs can start to accrete the interstellar medium and thus appear as detectable e.g., X-ray sources. 

Accretion from the interstellar medium is expected to proceed in the Bondi-Hoyle-Littleton regime 
\citep{2004NewAR..48..843E}. 
However, some important deviations can happen, see below.
In this study, we discuss only accretion onto magnetized NSs. 
Isolated BHs can also appear as accreting sources, but this topic is beyond the scope of this paper. An interested reader can address \cite{2021MNRAS.505.4036S} and references therein. 

After the launch of the {\it ROSAT} X-ray surveyor, it was realized that this instrument can discover isolated accreting NSs 
\citep{1991A&A...241..107T}. 
This idea triggered several detailed studies aimed at modeling properties of such sources and estimate their number; see
\cite{1993ApJ...403..690B}, 
\cite{1991ApJ...381..210B}, 
\cite{1994ApJ...423..748M}, 
\cite{1993A&A...269..319T}, 
\cite{1996MNRAS.278..577M}, 
\cite{1996A&A...309..469Z}, 
\cite{1998A&A...331..535P}. 

Optimistic estimates of 1990s were based on three main assumptions:
\begin{itemize}
\item there are many low-velocity isolated NSs;
\item most of NSs start to accrete;
\item accretion proceeds with high efficiency.
\end{itemize}

Despite all the enthusiasm, all observational efforts failed to discover any accreting isolated NS, see e.g.,
\cite{1997A&A...319..525B}, 
\cite{1998A&AS..128..349D}. 
This stage of isolated NSs studies are well summarized by
\cite{2000PASP..112..297T}. 

In the mid-1990s, it was realized that on average NSs are rapidly moving, with typical spatial velocities about 300-400~km~s$^{-1}$ \citep{1994Natur.369..127L}. 
This challenged at least the first two items mentioned above. 
\cite{2000ApJ...530..896P} 
performed a population synthesis study of isolated NSs in the Milky Way aiming to determine the fraction of accreting sources accounting for large average spatial velocities. The authors concluded that for a realistic kick velocity distribution just $\lesssim 10$\% of NSs can reach the stage of accretion from the interstellar medium. 
Still, assuming high efficiency of accretion, this group of authors predicted that 
a small number of accreting isolated NSs can be detected at low fluxes by {\it Chandra} and {\it XMM-Newton} X-ray observatories
\citep{2000ApJ...544L..53P}. 

The question of accretion efficiency was addressed by \cite{2003ApJ...594..936P} 
and in several numerical studies: 
\cite{1999ApJ...517..906T},
\cite{2001ApJ...561..964T},
\cite{2003ApJ...593..472T},
\cite{2012MNRAS.420..810T}.
All of these authors concluded that the luminosity of an accreting isolated NS could be much lower than that predicted by the Bondi formula. 

Typically, NSs with larger magnetic fields evolve faster (see e.g., \citealt{2000ApJ...530..896P}). A fraction of NSs are born with magnetic field significantly larger than standard radio pulsars (see e.g., \citealt{2010MNRAS.401.2675P}). Thus, such compact objects can reach the accretion stage faster increasing the total number of isolated accretors. This motivated a new population synthesis study.
\cite{2010MNRAS.407.1090B} 
demonstrated that indeed the number of accreting isolated NSs significantly increases due to accounting for initially highly magnetized objects. The authors suggest that in the Solar vicinity about one third of isolated NSs are at the accretor stage. However (see below), in this study, some overoptimistic assumptions about the propeller stage have been made.

As it was already noticed above, accretion regime at low mass inflow rates is not certain. In 2012, a new low-rate accretion regime was proposed \citep{2012MNRAS.420..216S}. This {\it settling accretion regime} later was applied to isolated NSs by \cite{2015MNRAS.447.2817P}. One of the proposals made in this study is that an isolated NS in the settling accretion regime can appear as a transient source due to accumulation of matter in an envelope, which can rapidly fall onto the NS when the matter cools down.

Progress in the observational technique provides new opportunities to search for accreting isolated NSs. High hopes were related to the launch of the {\it Spectrum-RG} satellite with the {\it eROSITA} X-ray telescope aboard \citep{2021A&A...647A...1P}. {\it eROSITA} was supposed to conduct eight all-sky surveys in four years. Unfortunately, the scientific program was terminated in February 2022 after just four surveys were completed. 
Observability of isolated NSs of different kind was studied in several papers, see e.g., \cite{2018IAUS..337..112P, 2021ARep...65..615K} and references therein. 
Data reduction of the four {\it eROSITA} surveys is in progress. There is a hope that accreting isolated NSs can be identified in these data. 

In this paper, we present a detailed population synthesis of isolated NSs in the Galaxy. Each NS evolves on a timescale of $13.6$~Gyr, moving through the Milky Way and interacting with the ISM. We consider several evolutionary models: the models of the spin-down at the propeller stage and magnetic field behavior. Additionally, we propose two models of the ISM, a model for a homogeneous ISM and a more sophisticated two-phase model. Most importantly, we calculate the X-ray flux of NSs that have reached the accretor stage, taking interstellar absorption into account. Finally, we provide an upper limit on the number of accreting isolated NSs that can be observed by \textit{eROSITA}.

The plan of the paper is as follows. The next section describes the evolution of the spin period and magnetic field of the NS adopted in this work, two ISM models, the method for calculating the observability of accreting NSs, and the population synthesis algorithm. Sect.~3 presents the population synthesis results, including the estimate of the number of observable accretors within different evolutionary models. In Sect.~4, we discuss the results obtained. Finally, in Sect.~5 summarizes our findings and provides a brief conclusion.




\section{Model}

\subsection{Spin evolution of neutron stars}

\label{sec_spin_evo}


The NSs change their observational appearance over time. The observed properties of an NS depend on its physical parameters (spin period, velocity, magnetic field, etc.), where the spin period is considered as the main parameter. 
In describing the spin evolution of NSs, we follow the general approach presented, for example, by~\cite{1992ans..book.....L, 2024Galax..12....7A}. In this consideration, the NS is characterized by a mass $M=1.4~M_\odot$, a moment of inertia $I = 10^{45}$~g~cm$^2$, a rotation frequency $\omega$ (or a spin period $P=2\pi/\omega$), and a dipolar magnetic field value at the equator $B$ (or a magnetic moment $\mu=BR_\text{NS}^3$, where $R_\text{NS}=10$~km).

Along with the physical parameters of the object, it is important to consider the parameters describing the interaction of the object with the external material. The first parameter is the characteristic velocity of the NS relative to the ISM: $v = \sqrt{v_\infty^2 + c_\text{s}^2}$, where $c_\text{s}$ is the sound speed in the medium, $v_\infty$ takes into account the spatial velocity of the object $\vec{v}_\text{NS}$ and the local velocity of the interstellar material $\vec{v}_\text{ISM}$: $v_\infty = |\vec{v}_\text{ISM} - \vec{v}_\text{NS}|$. The second parameter $\dot{M}$ is the accretion rate. We adopt the Bondi formula \citep{1952MNRAS.112..195B,2025ApJ...980..226T}:
\begin{multline}
\label{eq_M_dot_bondi}
    \dot{M} = \pi R_\text{G}^2\rho v = 4\pi(GM)^2\rho v^{-3}= \\ = 7.3\times10^{8}\left(\frac{n}{1\text{~cm}^{-3}}\right)\left(\frac{v}{100\text{~\kms}}\right)^{-3}\text{g/s},
\end{multline}
where $G$ is the Newton constant, $\rho = n m_\text{p}$ is the mass density of the medium, $n$ is the number density of the medium, $m_\text{p}$ is the proton mass, $R_\text{G}=2GM/v^2=3.7\times10^{12} \text{~cm~}(v/100\text{~\kms})^{-2}$ is the gravitational capture radius, which represents the radius of gravitational influence of the NS. The parameter $\dot{M}$ is used even when accretion is not possible. In this case, $\dot{M}$ characterizes the properties of the external medium.

\begin{figure*}
\centering 
\begin{overpic}[width=\textwidth]{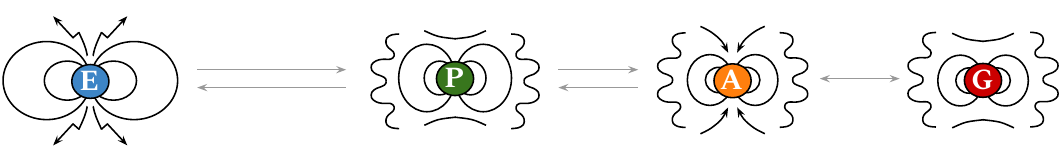}	
    \put(18.5,11.0){$max(R_\text{G}, R_\text{l}) = R_\text{Sh}$}
    \put(18.5,8.5){$min(R_\text{G}, R_\text{l}) = R_\text{Sh}^\text{env}$}
    \put(18.5,3){$max(R_\text{G}, R_\text{l}) = R_\text{Sh}^\text{env}$}
    \put(18.5,0.5){$R_\text{l} = R_\text{m}$}
    \put(53,8.5){$R_\text{c} = R_\text{m}$}
    \put(53,3){$R_\text{c} = R_\text{A} $}
    \put(77.5,8){$R_\text{A} = R_\text{G}$}
\end{overpic}
\caption{Evolutionary stages~-- ejector (E), propeller (P), accretor (A), and georotator (G)~-- and transition conditions between them. For direct transitions (E-P, P-A, and A-G), the left side of each equation must be greater than or equal to the right side. For reverse transitions (P-E, A-P, and G-A), it must be less than the right side. The E-P and P-E transitions require two conditions each. The wavy lines are supposed to illustrate the interaction of the magnetosphere and the external matter. 
}
\label{fig_epag}
\end{figure*}

The way the NS interacts with the external material is determined by the evolutionary stage. We consider four main stages -- ejector, propeller, accretor, and georotator. The transition between stages correspond to the equality of some critical radii, as it is shown in Fig.~\ref{fig_epag}. Let us briefly introduce the spin evolution at the evolutionary stages and the transition conditions along with necessary critical radii.

An NS is usually born with a short spin period, then spins down and switches between evolutionary stages. The spin-down process can be described by the Euler equation: $I\dot{\omega} = -K$, where $K$ is the spin-down torque. For the spin period, the same equation is
\begin{equation}
\label{dot_P}
    \frac{\text{d}P}{\text{d}t}=\frac{P^2}{2\pi I} K.
\end{equation}
\textit{Ejector. }
The first evolutionary stage is usually the ejector stage, where the NS produces the pulsar wind with the luminosity $L = 2\mu^2\omega^4/c^3$ 
{\citep{2014MNRAS.441.1879P}, where factor 2 accounts for the isotropic distribution of the angle between the spin and magnetic axis of a pulsar \citep{2024PASA...41...14A}.}
Considering the pulsar wind generation equal to the rotation energy losses $(I\omega\dot{\omega})$, we obtain the spin-down torque at the ejector stage:
\begin{equation}
\label{K_E}
K_\text{E} = 2 \frac{\mu^2}{R_\mathrm{l}^3}.
\end{equation}
Here, $R_\text{l} = c/\omega=4.8\times10^{9}~(P/1\text{~s})$~cm is the radius of the light cylinder, which is also the maximum distance from the NS center at which closed magnetic field lines can exist. 

Two conditions must be met for the ejector-propeller transition to occur. First, the external material must be gravitationally captured. To write it down in terms of critical radii, we define the characteristic radius of interaction between the pulsar wind and the external matter. It is the so-called Shvartsman radius $R_\text{Sh}$, see \cite{1992ans..book.....L}. If the matter is under the gravitational influence of the NS, then $R_\text{Sh}\le R_\text{G}$ (or $R_\text{Sh} \le R_\text{l}$, if $R_\text{l} > R_\text{G}$).

The Shvartsman radius can be determined as follows. The inner pressure is defined by the pulsar wind generation: $p_\text{in} = L / (4\pi R c)$, here (and below in this subsection) $R$ is the distance from the NS. If the material is located outside of the gravitational capture radius, then the external pressure: $p_\text{out} = \rho v^2$. The characteristic radius of interaction between the pulsar wind and external matter is obtained by equating the external and internal pressures:
\begin{multline}
    \label{R_Sh}
    R_{\text{Sh}}=\left( \frac{8 \mu^2 (GM)^2 \omega^4}{\dot{M} v^5 c^4} \right)^{1/2} = R_\text{G} \left(\frac{2 \mu^2 \omega^4}{\dot{M} v c^4}\right)^{1/2} = \\ = 2.7\times10^{13} \left(\frac{P}{10\text{~s}}\right)^{-2} \left(\frac{B}{10^{12}\text{~G}}\right) \left(\frac{n}{1\text{~cm}^{-3}}\right)^{-1/2} \left(\frac{v}{100\text{~\kms}}\right)^{-1} \text{~cm}.
\end{multline}

Once the material is gravitationally captured, it can form a shell around the magnetosphere with inner radius $R_\text{m}$. Following \cite{1979MNRAS.186..779D, 1981MNRAS.196..209D}, we assume that the material creates a rarefied envelope, extending from the magnetosphere to $R_\text{G}$, with a shallow pressure profile.  In the envelope, the balance of the pulsar wind and the thermal pressure of the material is formally possible. Then the radius at which such equality is fulfilled is the Shvartsman radius in the envelope:
\begin{multline}
\label{R_Sh_env}
    R_\text{Sh}^\text{env} = \left(\frac{2\mu^2\omega^4\sqrt{2G M}}{\dot{M}v^2 c^4}\right)^{2} = R_\text{G}\left(\frac{2\mu^2\omega^4}{\dot{M}vc^4}\right)^2 = \\ = 1.1\times10^{16} \left(\frac{P}{10\text{~s}}\right)^{-8} \left(\frac{B}{10^{12}\text{~G}}\right)^4 \left(\frac{n}{1\text{~cm}^{-3}}\right)^{-2} \left(\frac{v}{100\text{~\kms}}\right)^{2}  \text{~cm}.
\end{multline}

For the second ejector-propeller transition condition, the material must be able to interact with the magnetosphere. We consider this to be equivalent to the condition $R_\text{Sh}^\text{env} \le R_\text{l}$ (or $R^\text{env}_\text{Sh} \le R_\text{G}$, if $R_\text{l} > R_\text{G}$). If this condition is met, the pulsar wind can no longer balance the external pressure, and the propeller stage begins.

\textit{Propeller. }
The mechanisms responsible for the spin-down at the propeller stage are uncertain. At the base of this envelope, the magnetosphere interacts with the material, causing the NS to lose rotational energy. We consider several models of the spin-down torque $K_\text{P}$, which are listed in Table~\ref{tab_prop}.

\begin{table}[t]
    \centering
    \renewcommand{\arraystretch}{1.4}
    \begin{tabular}{clc}
    \hline
         Model & Braking torque $K_\text{P}$ & Typical value, dyn cm  \\ \hline
         A & $\dot{M} \omega R_{\text{m}}^2$ &  $8.6\times10^{29}$\\
         B & $\dot{M} \sqrt{2GMR_{\text{m}}}$ & $5.2 \times 10^{27}$ \\ 
         C & $\dot{M}\, \text{max}[v_\infty^2,~v^2_{\text{ff}}(R_\text{m})] / (2\omega)$ & $1.6\times10^{25}$ \\
         D & $\dot{M} v_\infty^2 / (2\omega)$ & $5.8\times10^{23}$ \\ \hline
    \end{tabular}
    \caption{Spin-down moments $K$ at the propeller stage in different models. The models are listed in the descending order of the spin-down efficiency, so model A has the most efficient spin-down mechanism and leads to the shortest propeller stage. Here, $v_\text{ff}(R_\text{m})=\sqrt{2GM/R_\text{m}}$. The typical value is calculated for the parameters corresponding to the propeller stage: $P=100$~s, $B=10^{12}$~G, $n = 1$~cm$^{-3}$, $v=100$~\kms. This parameters also yield $v_\text{ff}(R_\text{m})>v$. The authors of the models: A~--- \cite{1975SvAL....1..223S}, B~--- \cite{1973ApJ...179..585D}, C~--- \cite{1975AA....39..185I}, D~--- \cite{1981MNRAS.196..209D}.
    }
    \label{tab_prop}
\end{table}

The critical radius of interaction between the magnetosphere and the base of the envelope includes the Alfv{\'e}n radius. This is the distance in the equatorial plane of the NS at which the dynamic pressure of a material falling toward the central object at a velocity of $\sqrt{2GM/R}$ is equal to the magnetic field pressure.
\begin{multline}
\label{R_A}
R_A = \left( \frac{ \mu^2}{2 \dot{M} \sqrt{2GM}} \right)^{2/7} = \\ = 5.4\times10^{10} \left(\frac{B}{10^{12}\text{~G}}\right)^{4/7} \left(\frac{n}{1\text{~cm}^{-3}}\right)^{-2/7} \left(\frac{v}{100\text{~\kms}}\right)^{6/7} \text{~cm}.
\end{multline}
Since the material in the shell is not falling freely, the radius of the interaction differs from the Alfv{\'e}n radius:
\begin{multline}
\label{eq_R_m}
    R_\text{m} = R_\text{A}^{7/9}R_\text{G}^{2/9} = \\ = 1.4 \times 10^{11} \left(\frac{B}{10^{12}\text{~G}}\right)^{4/9} \left(\frac{n}{1\text{~cm}^{-3}}\right)^{-2/9} \left(\frac{v}{100\text{~\kms}}\right)^{2/9}  \text{~cm}.
\end{multline}

For the reverse transition, from the propeller stage to the ejector stage, the magnetosphere should be able to be expand to the light cylinder radius $R_\text{m} = R_\text{l}$. Then the pulsar wind must sweep the material out beyond the gravitational capture radius $R_\text{Sh}^\text{env} > R_\text{G}$ (or $R_\text{Sh}^\text{env} > R_\text{l}$, if $R_\text{l}> R_\text{G}$).

At the propeller stage, the material cannot fall onto the NS surface due to the centrifugal barrier. For the propeller-accretor transition, we adopt the relation $R_\text{m} \le R_\text{c}$, where the centrifugal barrier radius:
\begin{equation}
\label{R_c}
R_{\text{c}} = 0.87\left( \frac{GM}{\omega^2} \right)^{1/3}.
\end{equation}
This radius is usually assumed to be equal to the corotation radius, where the velocity $\omega R$ and the Keplerian velocity $\sqrt{GM/R}$ are equal, but we add the coefficient $0.87$ to account for calculations in the three-dimensional case~\citep{2023MNRAS.520.4315L}.

\textit{Accretor. }
During the accretor stage, we consider spherical accretion of freely falling matter. Thus, the critical radius of interaction with the material is $R_\text{A}$. The corresponding spin-down torque is:
\begin{equation}
    \label{K_A}
    K_{\text{A}}=0.4\frac{\mu^2}{R_{\text{c}}^3}.
\end{equation}

Since the magnetosphere radius at the accretor stage is $R_\text{A}$, the reverse transition to the propeller stage correspond to the condition $R_\text{A} > R_\text{c}$.

\textit{Georotator. }
Accretion is possible if the material is gravitationally captured. Otherwise, if $R_\text{A} \ge R_\text{G}$, the gravitational influence of the NS at the magnetosphere radius is insufficient at the accretor stage. This case corresponds to the georotator stage. Assuming low rotational losses, we set $K_G = 0$. The transition from the georotator stage to the accretor stage is symmetrical with the accretor-georotator transition. This transition occurs when $R_\text{A} < R_\text{G}$.

\subsection{Magnetic field evolution}
\label{sec:field_evol}
In addition to spin evolution, we consider the evolution of the magnetic field. The magnetic field can decay due to ohmic dissipation throughout the evolution of the NS \citep{2021Univ....7..351I}. We consider two models of magnetic field decay. {For the magnetars, the magnetic field evolves rapidly at early stages, so the Hall cascade time scale must be taken into account. Thus, the dipolar magnetic field of a magnetar follows }
\begin{equation}
\label{eq:B_short}
    B=B_0\frac{\exp(-t/\tau_{\text{Ohm}})}{1+(\tau_{\text{Ohm}}/\tau_{\text{Hall}})(1-\exp(-t/\tau_{\text{Ohm}}))}.
\end{equation}
Here $B_0$ is the initial magnetic field value. The Ohmic dissipation time $\tau_{\text{Ohm}} = 10^6$~yrs, $\tau_{\text{Hall}} = 10^4/(B_0/10^{15}\text{~G})$~yrs is the characteristic time of the Hall cascade. The rapid magnetic field evolution terminates when the value $B$ is decayed $\sim20$ times, which is equivalent to three $e$-foldings \citep{2014MNRAS.438.1618G}, at which point the so-called Hall attractor stage is reached \citep{2014PhRvL.112q1101G}. In our model, we use this type of evolution to derive the initial magnetic field and spin period of the magnetars, prepared for the main calculations on the larger timescale {(see Sect.~\ref{sec_distr} below).} 


The evolution of the magnetic field on a large timescale can be expressed as follows:
\begin{equation}
\label{eq:B_long}
B = B_0 \exp(-t/\tau'_{\text{Ohm}}),
\end{equation}
$\tau'_{\text{Ohm}} = 1.5 \times 10^{9}$~yrs. To obtain this value of $\tau'_{\text{Ohm}}$, we consider a typical initial magnetic field $10^{12}$~G, and require that it decrease to $10^8$~G within $13.6$~Gyr. Along with the constant field model (CF), in our work, we will consider the evolution of NSs with this model of an exponentially decaying field (ED). { At the same time, in our model, we neglect the influence of accretion on the field decay. The reasoning is as follows.
Even if an NS has a low spatial velocity and accretes during most of its life at a rate $\sim 10^{11}$ g s$^{-1}$,
it can accrete at most $10^{-5} M_\odot$ of matter. This is $\sim 0.1$\% of the mass of the crust of an NS, see \cite{2008LRR....11...10C}. Such a low accretion rate cannot significantly influence the magnetic field, either by dynamical action (submergence, etc.) or by heating of the crust (see sect. 4.5 in \cite{2004MNRAS.351..569P} for details). }

\subsection{Model of the Milky Way}
\label{model_of_the_milky_way}
In our modeling, the NSs are assumed to move within the Galactic potential {$\Phi(R,z)$} and interact with interstellar material. 
{Hereafter we use cylindrical Galactic coordinates: the galactocentric distance $R$ (calculated in the disc plane), the height $z$ above the plane, and the azimuthal angle $\varphi$.}
To calculate the positions and velocities of objects in the Galaxy, we use ready-made solutions presented in the \verb|gala| \textit{Python} package \citep{2017JOSS....2..388P}. We adopt the gravitational potential of the Galaxy \verb|MilkyWayPotential| \citep{2015ApJS..216...29B} with the components listed in Table~2. 

\begin{table*}[ht]
    \label{tab:galaxy_potential}
    \centering
    \caption{The parameters for each component of the Milky Way gravitational potential.
    }
    \begin{tabular}{@{}lllll@{}}
        \toprule
        Component & Authors of the model & Parameters & &  \\ \midrule
        Disk & \cite{1975PASJ...27..533M} &  $m=6.8\times10^{10}\,M_\odot$ & $a=3.0$~kpc & $b=0.28$~kpc\\
        Buldge & \cite{1990ApJ...356..359H} & $m=5\times10^{9}\,M_\odot$ & $c=1.0$~kpc \\
        Nucleus & \cite{1990ApJ...356..359H} & $m=1.71\times10^{9}\,M_\odot$ & $c=0.07$~kpc & \\
        Halo & \cite{1996ApJ...462..563N} & $m=5.4\times10^{11}\,M_\odot$ & $r_\text{s}=15.62$~kpc & \\
        \bottomrule
    \end{tabular}
\end{table*}


The ISM in the Milky Way is extremely turbulent and inhomogeneous. The temperature and number density in the ISM can differ by orders of magnitude, and the phases are closely adjacent to each other. In order to build a realistic model that is not too complicated, we consider the number density distributions of several ISM components: molecular $n_\text{H2}$, neutral $n_\text{HI}$, warm ionized $n_\text{ion}$, hot gas (coronal and in halo) $n_\text{hot}$, and the dust (cold with mass density $\rho_\text{c}$ and warm with $\rho_\text{w}$). 

To account for the different temperatures, we divide the components of the ISM into two groups. The cold phase consists of molecular gas, atomic and the warm ionized medium. The hot gas includes the coronal gas and the gas in the Galactic halo. We adopt different sound speed values: $(c_\text{s})_\text{cold} = 10$~\kms for the cold phase and $(c_\text{s})_\text{hot} = 100$~\kms for the hot phase.

We then consider two models of the ISM: a one-phase model with $\rho_\text{ISM}$ that includes all components except for $n_\text{hot}$ and a two-phase model with the hot phase and the cold phase, including the dust. The first model consists of a single cold phase with a density of:
\begin{equation}
\label{eq_one_phase}
    \rho_\text{ISM} = 1.33 m_\text{p} \left(n_\text{ion}+n_\text{HI}+2n_\text{H2} \right) + \rho_\text{dust}.
\end{equation}
Here, we put the coefficient $1.33$ to take helium into account.
For the two-phase model, there are two mass density distributions, $\rho_\text{cold}$ and $\rho_\text{hot}$. Moving through the Milky Way, the NS can be in only one-phase at a time and can visit both phases along the trajectory with some probability. This probability is proportional to the local volume filled by each phase (cold and hot). To determine the probability, we first consider density distributions of each component of the ISM.



\cite{2006A&A...459..113M} proposed an axis-symmetric model of interstellar density. According to them, the distribution of molecular hydrogen can be described as follows:
\begin{equation}
\label{eq:misiriotis-H2}
    n_\text{H2} = (n_\text{H2})_0 \exp\left(-\frac{R}{h_\text{H2}} - \frac{|z|}{z_\text{H2}}\right).
\end{equation}

For the neutral hydrogen, we introduce an additional factor $\text{sech}^2\left[{(r-B_\text{d})}/{A_\text{d}}\right]$ to the distribution above to cut the density at further distances from the Galactic center \citep{2017ApJ...835...29Y}:
\begin{equation}
\label{eq:misiriotis-HI}
n_\text{HI} =
    \left\{
    \begin {array} {ll}
    \displaystyle
    0, & r < R_t  \\
    \displaystyle
    (n_\text{HI})_0 \exp\left(-\frac{R}{h_\text{HI}} - \frac{|z|}{z_\text{HI}}\right), & R_\text{t} \le r < B_\text{d} \\ \displaystyle
    (n_\text{HI})_0 \exp\left(-\frac{R}{h_\text{HI}} - \frac{|z|}{z_\text{HI}}\right)\text{sech}^2\left(\frac{r-B_\text{d}}{A_\text{d}}\right), & r \ge B_\text{d} \\
    \end{array}
    \right.
\end{equation}
Here $r=\sqrt{R^2+z^2}$. The parameters in eqs.~\ref{eq:misiriotis-H2} and~\ref{eq:misiriotis-HI} are listed in the Table~\ref{tab:misiriotis}.

The gas between the disk and the halo is the warm ionized medium. The density of ionized hydrogen varies with Galactic height \citep{2008PASA...25..184G}:
\begin{equation}
\label{eq:ion}
    n_\text{ion} = (n_\text{ion})_0 \exp\left(-\frac{|z|}{H_n}\right).
\end{equation}

\begin{table}[ht]
    \centering
    \caption{The parameters for the density distributions of ionized, neutral and molecular hydrogen, cold and warm dust, which are adopted from \cite{2008PASA...25..184G,2006A&A...459..113M,2017ApJ...835...29Y}} 
    \begin{tabular}{@{}lll@{}}
        \toprule
        Parameter & Unit & Value \\ \midrule
        $(n_\text{ion})_0$ & cm$^{-3}$ & 0.014 \\
        $H_n$ & kpc & 1.83 \\ \midrule
        $(n_\text{H2})_0$ & cm$^{-3}$ & 4.06 \\
        $h_\text{H2}$ & kpc & 2.57 \\
        $z_\text{H2}$ & kpc & 0.08 \\ \midrule
        $(n_\text{HI})_0$ & cm$^{-3}$ & 0.32 \\
        $h_\text{HI}$ & kpc & 18.24 \\
        $z_\text{HI}$ & kpc & 0.52 \\ 
        $R_\text{t}$ & kpc & 2.75 \\
        $B_\text{d}$ & kpc & 15 \\
        $A_\text{d}$ & kpc & 2.5\\  \midrule
        $(\rho_\text{c})_0$ & g~cm$^{-3}$ & $1.51\times10^{-25}$\\
        $h_\text{c}$ & kpc & 5.0\\
        $z_\text{c}$ & kpc & 0.1\\
        $(\rho_\text{w})_0$ & g~cm$^{-3}$ & $1.22\times10^{-27}$\\
        $h_\text{w}$ & kpc & 3.3\\
        $z_\text{w}$ & kpc & 0.09\\
        \bottomrule
    \end{tabular}
    \label{tab:misiriotis}
\end{table}

We adopt the model for coronal gas and hot gas in halo from \cite{2024A&A...681A..78L}. The hot gas has a spherical and a cylindrical component:
\begin{equation}
\label{eq_hot}
    n_\text{halo} = (n_\text{halo})_0\left[\left(1+\frac{r^2}{r_0^2}\right)^{-\frac{3\beta}{2}} + \exp\left(-\frac{R}{R_\text{h}} - \frac{|z|}{z_\text{h}}\right)\right].
\end{equation}
The parameters are listed in Table~4.
\begin{table}[ht]
    \label{tab:halo}
    \centering
    \caption{Parameters for the density distribution of the hot gas \citep{2024A&A...681A..78L}. The value of $r_0$ calculated as $r_0 = \left( {C}/{n_0} \right)^{{1}/{(3\beta)}}$.}
    \begin{tabular}{@{}lll@{}}
        \toprule
        Parameter & Unit & Value \\ \midrule
        $\beta$ & - & $0.5$ \\ 
        $C$ & cm$^{-3}$ kpc$^{3\beta}$ & $(4.6 \pm 0.1) \times 10^{-2}$ \\ 
        $r_0$ & kpc & 1.27 \\
        $(n_\text{halo})_0$ & cm$^{-3}$ & $(3.2 \pm 0.4) \times 10^{-2}$ \\ 
        $R_\text{h}$  &kpc & $6.2 \pm 0.4$ \\ 
        $z_\text{h}$ & kpc & $1.1 \pm 0.1$ \\ 
        \bottomrule
    \end{tabular}
\end{table}

In addition to the gas phases, we consider the dust in the ISM, which is distributed as follows:
\begin{equation}
    \rho_\text{c,w} =(\rho_\text{c,w})_0 \, \exp\left(-\frac{R}{h_\text{c,w}} - \frac{|z|}{z_\text{c,w}}\right),
\end{equation}
where the parameters of the cold (c) and warm (w) dust components are shown in Table~\ref{tab:misiriotis}. In calculation,s we sum the density from both components, therefore $\rho_\text{dust} = \rho_\text{c} + \rho_\text{w}$.

The ISM, consisting of the considered components, yields values of the dispersion measure \citep{2023ApJ...946...58C} and the column density \citep{2016A&A...594A.116H} that align with observations.

For the two-phase ISM model, we can now derive the probability for the NS to be in the cold and in the hot phase.  The NS can enter either phase with a probability that is proportional to the local volume occupied by each of two-phases, or the filling factor. The filling factor of the cold phase ranges from $0$ to $1$ and depends on the location in the Galaxy. To derive $f_\text{ISM}(R,~z)$, we assume that the cold phase can form dense structures. Then, we consider pressure equilibrium between the cold and hot phases within the local volume at coordinates $(R,~z)$: $\rho_\text{cold} (c_\text{s})^2_\text{cold} = \rho_\text{hot} (c_\text{s})^2_\text{hot}$, where the local density of the hot gas is $\rho_\text{hot} = 1.33m_\text{p} n_\text{halo}$. Using this equation, we can calculate the local density of the cold gas $\rho_\text{cold}$. Since the speed of sound in the hot medium is ten times higher than in the cold medium, the following density of the cold phase $\rho_\text{cold}\approx 100 \rho_\text{hot}$. The volume occupied by each phase can be found from the fact that the total local density is fixed: $f_\text{ISM}\rho_\text{cold} + (1-f_\text{ISM})\rho_\text{hot} = \rho_\text{tot} = 1.33 m_\text{p} \left(n_\text{ion}+n_\text{HI}+2n_\text{H2} + n_\text{halo} \right)$. The resulting filling factor:
\begin{equation}
\label{eq_fill}
    f_\text{ISM}(R,~z) = \frac{\rho_\text{tot}(R,~z) - \rho_\text{hot}(R,~z)}{\rho_\text{cold}(R,~z) - \rho_\text{hot}(R,~z)}.
\end{equation}
To take the dust density into account, we add $\rho_\text{dust}$ to the cold phase after the filling factor is calculated. If $f_\text{ISM}\ge1$, then the local volume is assumed to be filled with the cold phase only, and its density is considered equal to that of the medium in the one-phase model $\rho_\text{cold} = \rho_\text{ISM}$.

Finally, we consider the spatial motion of the interstellar matter. In general, the gas in the disk rotates around the center of the Galaxy with a velocity close to the circular velocity in the gravitational potential of the Milky Way:
$(v_\text{ISM})_{z=0} \approx v_\text{circ} {=\sqrt{R\cdot\nabla\Phi(R, 0)}}$.
Gas above the plane of the Galaxy rotates more slowly with increasing altitude:
\begin{equation}
\label{eq:vISM}
v_\text{ISM} = v_\text{circ} - z\frac{\partial v}{ \partial z},
\end{equation}
where $\partial v/ \partial z = 15$~\kms~kpc$^{-1}$ \citep{Marasco2011-vd}. 
Far in the Galactic halo, the gas rotates with a velocity $180$~\kms~\citep{2016ApJ...822...21H}. We take it into account, keeping the velocity map continuous. 

\subsection{Distributions of NSs initial parameters}
\label{sec_distr}
Before the calculations, it is necessary to specify the initial coordinates and velocities of NSs with respect to the inertial frame of reference in the Galaxy, as well as the initial spin periods, and the magnetic fields. Below, we describe the distributions adopted
for the generation of all these quantities.

{\it Initial spatial positions.} We start with generating the initial vertical coordinate $z$ of a neutron star following \cite{2006ApJ...643..332F}. In particular, the probability density function:
\begin{equation}
    f(|z|) \propto \exp \left(\frac{|z|}{z_0}\right),
    \label{eq:f_z}
\end{equation}
where $z_0 = 50$~pc. To obtain the distribution over $z$, each value $|z|$ derived from this distribution is given a random sign, so $z = -|z|$ in half the cases.

We consider the positions of OB-stars to be the initial coordinates of NSs in the Galactic plane \citep{2004A&A...422..545Y}, which are distributed as follows:
\begin{equation}
\label{eq:f_r}
    f (R)\propto \left(\frac{R}{R_\odot}\right)^a \exp \left[-b\left(\frac{R}{R_\odot}\right)\right],
\end{equation}
where $R=\sqrt{x^2+y^2}$, while $a = 4$, $b = 6.8$, and $R_\odot=8$~kpc is the galactocentric distance of the Sun.

The initial azimuthal angle $\varphi_0$ is assumed to be distributed uniformly within the interval $0\le\varphi_0<2\pi$.

{\it Initial velocities.} We treat initial velocity $\vec{v}_0$ of an NS as the sum of three components: the circular velocity at a given $R$ in the Galactic plane, the residual (peculiar) velocity of the progenitor star, and the kick velocity gained after the supernova explosion (see a review in \citealt{2025NewAR.10101734P}):
\begin{equation}
    (\vec{v}_\text{NS})_0 = \vec{v}_\text{circ} + \vec{v}_\text{res} + \vec{v}_\text{kick} 
    \label{eq:initial_velocity}
\end{equation}

The vector $\vec{v}_\text{circ}$ lies in the Galactic plane and $|\vec{v}_\text{circ}|$ is derived from the 
gravitational potential. Direction of $\vec{v}_\text{circ}$ coincide with the
local vector of the galactic rotation, so that $\vec{v}_\text{circ} = v_\text{circ} \{-{y}/R,~{x}/{R},~0\}$, if
it expressed in a right-handed Cartesian coordinates $(x,y,z)$, with the origin at the Galactic center, the x-axis pointed in the direction of galactic rotation, and the y-axis pointed toward the Sun.

In turn, the cylindrical components $(v_R,~v_\varphi,~v_z)$ of the residual velocity $\vec{v}_\text{res}$ are independently 
taken from the normal distributions with the standard deviations $(10,10,8)$~\kms respectively \citep{2022AstL...48..243B}. 


Finally, we assumed the isotropic kick velocity $\vec{v}_\text{kick}$. Its absolute value was taken from the sum of two Maxwellian distributions with different standard deviations \citep{2021MNRAS.508.3345I}: $f(v) = wf_{\sigma1}(v) + (1-w)f_{\sigma2}(v)$, where $w=0.2$, $\sigma_1=45$~km/s, $\sigma_2=336$~km/s.

{\it Initial periods and magnetic fields.} 
While generating synthetic neutron stars, we assume that 90\% of them are born as classical pulsars and 10\% as magnetars.
{ A rough estimate of the fraction of magnetars among newborn NSs can be made using their numbers and ages. 
This way, \cite{1999PNAS...96.5351K} obtained an estimate exactly $\sim10$\%.
This number was later supported by a detailed population synthesis modeling by
\cite{2010MNRAS.401.2675P}. Afterwards,
\cite{2019MNRAS.487.1426B} suggested that the fraction of magnetars can be even higher: $40\%^{+60\%}_{-28\%}$. Still, more recent estimates are closer to the classical value.
E.g., \cite{2025ApJ...986...88S} report that according to their modeling the fraction of magnetars is $10.7\%^{+18.8\%}_{-4.4\%}$.
}

According to \cite{2022MNRAS.514.4606I}, the probability density function for the pulsar initial spin period:
\begin{equation}
\label{p_distr}
f (\log_{10} P_0) = \frac{1}{\sigma_{P_0} \sqrt{2 \pi}} \exp \left( {- \frac{(\log_{10} P_0 - \overline{\log_{10} P_0})^2}{2\sigma_{P_0}^2}} \right),
\end{equation}
where $P_0$ is taken in seconds, the mean is $\overline{\log_{10} P_0}= -1.04$, and the standard deviation is $\sigma_{P_0} = 0.53$.

For the initial equatorial (dipole) magnetic fields of pulsars, we use the similar distribution
\begin{equation}
\label{B_distr}
f (\log_{10}B_0) = \frac{1}{\sigma_{B_0} \sqrt{2 \pi}} \exp \left( {- \frac{(\log_{10} B_0 - \overline{\log_{10} B_0})^2}{2\sigma_{B_0}^2}} \right),
\end{equation}
where $B_0$ is taken in Gauss,  $\overline{\log_{10} B}= 12.44$, and standard deviation $\sigma_B = 0.44$ \citep{2022MNRAS.514.4606I}.

In turn, initial parameters of magnetars are calculated in two steps. Indeed, it is believed that magnetars evolve fast on short timescales $\lesssim10^7$~yr after their birth. Using the data for magnetars\footnote{http://www.physics.mcgill.ca/~pulsar/magnetar/main.html} \citep{2014ApJS..212....6O} and considering only the objects with the magnetic field $>10^{13}$~G, we derive the mean and the standard deviation for the initial magnetic field distribution $\log_{10}B_0'$: $\overline{\log_{10}B_0'}=14.33$, $\sigma_{B'_0} = 0.46$. We assume a distribution similar to eq.~(\ref{B_distr}) for the pre-initial magnetic field $B_0'$. We use such $B_0'$ along with the pre-initial spin period $P_0'$ derived from eq.~(\ref{p_distr}) to model the first stage of the spin evolution of a magnetar. At this stage, its magnetic field decays accordingly to eq.~\ref{eq:B_short}, and the spin period evolves as at the ejector stage. As soon as the Hall attractor is reached at time $t_H$, the synthetic magnetar is assigned with the initial spin period $P_0 = P'(t_H)$ and magnetic field $B_0 = B_0'(t_H)$ respectively.

\subsection{Observability of accreting isolated neutron stars}
\label{sect_model_observability}

We assume that the luminosity of the NS at the accretor stage comes from the total potential energy of the matter released by accretion: $L_\text{X} \approx {GM\dot{M}}/{R_\text{NS}}$. The material falls onto the polar caps of NSs, the radius of which is $R_\text{cap} = R_\text{NS} \sqrt{{R_\text{NS}}/{R_\text{A}}}$ \citep{1998astro.ph..4047S}. 

{Due to the time dilation effects, the accretion rate as seen by a local observer is $\dot M_\mathrm{loc} = \dot M/\sqrt{1 - 2r_\mathrm{g}/R_\mathrm{NS}}$ so that the local accretion luminosity is $L_\mathrm{loc} = L_\mathrm{X}/\sqrt{1 - 2r_\mathrm{g}/R_\mathrm{NS}}$. Assuming the blackbody spectrum of the emission, one can then define the local effective temperature $T_\mathrm{loc}$  via $L_\mathrm{loc} = S_\mathrm{cap}\sigma_\mathrm{B} T_\mathrm{loc}^4$, where $\sigma_\mathrm{B}$ is the Stefan-Boltzmann constant. At the same time, the distant observer will detect the temperature of the emitting area to be $T = T_\mathrm{loc}\sqrt{1 - 2r_\mathrm{g}/R_\mathrm{NS}}$. We use the latter to model the observed spectrum of the NS emission.}


{Without absorption, the mean X-ray flux from the source located at a distance $d$:}
\begin{equation}\label{F_x}
F_{\text{X}}(d) = \frac{L_\text{X}\sqrt{1-2r_\text{g}/R_\text{NS}}}{4\pi d^2}.
\end{equation}
{Here, the factor of $\sqrt{1-2r_\text{g}/R_\text{NS}}\approx 0.8$ {is applied again due to} the transformation of the time interval between the reference frame near the surface of the NS and that of a distant observer. 
The distance $d = |\vec{r}-\vec{r}_\odot|$, where $\vec{r}_\odot = \{0, R_\odot,0\}$ is the position of the Sun, $R_\odot=8$~kpc, and $\vec{r}=\{x,~y,~z\}$ is the position of the accreting NS. We do not take the modulations of the measured flux due to the NS orientation, rotation, and light-bending into account, since this phenomenon would introduce only a small factor of the order of unity to the mean flux \citep{2002ApJ...566L..85B, 2020A&A...640A..24P}. We also neglect the possible emission propagation effects in the star's magnetosphere. }

The corresponding photon flux:
\begin{equation}
    N_\text{ph} =  \int_0^{\infty} N_\text{ph}(\nu) \text{d}\nu =  F_\text{X}\frac{\int_0^{\infty} \frac{B_\nu(T)}{h\nu} \text{d}\nu}{\int_0^{\infty} {B_\nu(T)} \text{d}\nu},
\end{equation}
where $B_\nu(T) = {2h\nu^3}/[{c^2} ({\exp\{h\nu / k_\text{B}T\}-1})]$, $h$ is the Planck constant, $k_\text{B}$ is the Boltzmann constant.

Radiation undergoes interstellar absorption, so the number of photons reaching the observer decreases with increasing column density $N_\text{H}$ along the line of sight between the source and the observer:
$
N_\text{ph}'(\nu) = N_\text{ph}(\nu) \exp({-\sigma(\nu)N_\text{H}}).
$
Here, $\sigma(\nu)$ is the absorption cross section of an X-ray photon, calculated per hydrogen atom \citep{1983ApJ...270..119M}. The column density is calculated with the number density $(n_\text{HI}+2n_\text{H2})$. The $N_\text{H}$ map derived is consistent with the observational data of \cite{2016A&A...594A.116H}.

Finally, taking the effective area of \textit{eROSITA} $S_\text{eff}$, we can calculate the count rate of the telescope:
\begin{equation}
    \text{CR}  =  F_\text{X}\frac{\int_0^{\infty} S_\text{eff}(\nu)\frac{B_\nu(T)}{h\nu} \exp({-\sigma(\nu) N_\text{H}}) \text{d}\nu}{\int_0^{\infty} {B_\nu(T)} \text{d}\nu}.
\end{equation}

We consider the NS to be potentially observable if their count rate is higher than $0.01$~cts~s$^{-1}$. According to \cite{2021ARep...65..615K}, it is the minimum count rate required for an object to be observed by \textit{eROSITA} after all the data has been processed. 

\subsection{The population synthesis algorithm}
\label{sec_algorithm}

{The calculations are performed using the original \textit{Python} code\footnote{https://github.com/afoninamd/NS}.} They start by specifying the distribution of initial parameters, which are the initial spin period $P_0$, the initial equatorial dipolar magnetic field on the surface of the NS $B_0$, the initial coordinates $R_0$ and $z_0$, and the velocities $(\vec{v}_\text{NS})_0$ in the Galactic reference frame.

After the arrays of the initial parameters are set, we perform calculations of the coordinates and velocities of the NS tracks versus time. All the tracks are calculated over the lifetime of the Galaxy, which is $13.6$~Gyr. 
{We calculate every trajectory over $10$ thousand points, adopting an integration step of $\sim1$~Myr, which is much less than the typical orbital period in the Galaxy $\sim0.2$~Gyr. For NSs born at $R=8$~kpc, this yields a relative error in the total energy of an object in orbit of $<10^{-3}$ for kick velocities $\lesssim200$~\kms and $\lesssim10^{-2}$ for higher values of $v_\text{kick}$.}
Some of the trajectories reach $100$~kpc from the center of the Galaxy. In this case, we consider the NSs with these parameters to have been thrown out of the Galaxy and do not use this trajectory further in our modeling.

For the tracks of the NSs staying in the Galaxy, we use the derived dependencies of coordinates $\vec{r}$ and velocities $\vec{v}_\text{ISM}$ to calculate the spin evolution in different ISM models and with different magnetic field behaviors (the field is either constant or exponentially decaying, see {Sect.~\ref{sec:field_evol}}). Along the trajectory, we define the properties of the external material, which are the mass density $\rho$ and the parameter $v=\sqrt{c_\text{s}^2+|\vec{v}_\text{NS}-\vec{v}_\text{ISM}|^2}$. We consider two models of the ISM which differ in sound speed $c_\text{s}$ and mass density $\rho$, while similar in $\vec{v}_\text{ISM}$, which is defined in section~\ref{model_of_the_milky_way}. In the single-phase model, the mass density is $\rho_\text{ISM}$ from eq.~\ref{eq_one_phase} and the speed of sound $c_\text{s}=10$~\kms. For the two-phase model, the filling factor $f_\text{ISM}$ is derived for the whole trajectory. Then each point with the probability $f_\text{ISM}$ is labeled as the cold phase ($c_\text{s}=10$~\kms, $\rho_\text{cold}$) or with the probability $(1-f_\text{ISM})$ as the hot phase ($c_\text{s}=100$~\kms, $\rho_\text{hot}$). 

The idea behind the evolutionary track calculations is to use each spatial trajectory to calculate the evolution of many NSs with the same initial parameters, but with different ages. We illustrate this idea using an example. An evolutionary track is an array of parameters, e.g., spin period $P_i$ or evolutionary stage, corresponding to an array of time points along the trajectory. Typically, a trajectory, and consequently an evolutionary track, contains several thousand points. All the points of the evolutionary track together represent a population of several thousand NSs. Each point on the track can be interpreted as an NS with an age $t_i$ born with the same initial parameters $P_0$ and $B_0$, and arranged along a spatial trajectory that corresponds to a single set of initial coordinates and a velocity vector. These initial parameters are used in various evolutionary models: propeller spin-down, ISM, and magnetic field behavior models.


To calculate an evolutionary track, we use a non-uniform time grid to improve the precision of the numerical method at no additional cost.  To determine the value of the time step $\Delta t$ several iterations are performed if needed. For the first iteration, the time step at the beginning of the track is chosen so that the difference in the spin period between neighboring time points $t_i$ and $t_{i+1}$ is equal $\Delta P_i = \Delta P_{i+1}$ as long as the NS is at the ejector stage. For the remainder of the track, the time step $\Delta t$ is uniform. After the first iteration, sharp jumps in the calculated spin period $P(t)$ can occur. If the two neighboring $P_{i+1}$ and $P_i$ differ by more than two times, then additional time steps are added between $t_{i}$ and $t_\text{i+1}$. With the new time grid, the spin evolution is fully recalculated. It is important to note that every change in the time grid demands recalculation of the $\vec{r}(t)$ and $\vec{v}_\text{NS}(t)$ along the trajectory, but does not demand the recalculation of the trajectory itself. We can interpolate $\vec{r}(t)$ and $\vec{v}_\text{NS}(t)$ and use them without modification as functions in every iteration. Three iterations are the maximum, and usually two iterations are enough. This approach helps to take into account the rapid spin-down at the propeller stage as well. Thus, we obtain the evolutionary track, which is the dependence of the spin period, the evolutionary stage, the magnetic field, and the parameters of the external material on the age of the NS. The observability of the sources is calculated using these dependencies. To reduce the computational cost, we do not calculate the observability of the points at the track that give the unabsorbed flux $F_\text{X}<10^{-15}$~erg~cm$^{-2}$~s$^{-1}$, which are too faint to be observed, especially accounting for the interstellar absorption.

Using a variety of evolutionary tracks calculated at a supercomputer cluster, we can determine the properties of the NS population, such as the percentage of the isolated NSs at different evolutionary stages in the Galaxy and the observability of the isolated accreting NSs as soft X-ray sources. We use every point of the track, so every track resembles a population of the NS of different ages born with one set of initial parameters. 

Here, it is important to take the star formation history into account. We adopt the star formation history from the work~\cite{2016A&A...589A..66H} (the 5th curve in panel \textit{a} of their Fig.~4). As the star formation rate was much higher $10-13$~Gyr ago, the NSs born in this period contribute more to the total statistics. 


When calculating the properties of the population, each point of an evolutionary track is assigned a weight $w_\text{i}$, proportional to the time step and the star formation rate. To make all tracks equal, the sum of the weights for one track is always 1. For example, if we want to examine the distribution of the NS in the Galaxy over some parameter $\Pi$, we record the values of this parameter along the tracks and get $\{\Pi_i\}$ that corresponds to the time points $\{t_i\}$. Then, we plot a histogram of the values $\{\Pi_i\}$ with weights $\{w_i\}$, using the points from many evolutionary tracks.

\section{Results}

The main goals of this work are to estimate the total number of isolated accretors, the number of potentially observable old isolated accreting NSs, and to determine their basic properties. Before we can calculate NSs' observability, we first need to determine the number of NSs at the accretor stage in the Galaxy and examine the factors that influence this number. This section, therefore, consists of three parts. We first perform a realistic population synthesis using the known distribution of initial parameters and the laws of NS evolution to determine the general distribution of NSs at different evolutionary stages in the Galaxy. Then, we examine how the time spent at the accretor stage $\tau_\text{A}$ depends on the choice of the model. Using a uniform distribution of initial parameters, we explore typical initial values of the parameters that are most likely to result in the onset of accretion. Finally, we calculate the observability of accreting sources with realistic parameter distributions. 

\subsection{Number of neutron stars in the Galaxy at different evolutionary stages}

\label{res_models}

Let us discuss how differences in evolutionary models influence NS evolution in general, and the percentage of NSs at the accretor stage in particular.

We calculate the percentage of NSs in the Galaxy at each evolutionary stage: ejector, propeller, accretor, and georotator, for different models: four models of spin-down at the propeller stage, two models of the magnetic field evolution (constant field and exponential decay), and two models of the ISM (one-phase and two-phase). The results are shown in Table~\ref{tab:results}. To obtain these results and estimate the error, we calculated $100$ samples of $100$ thousand tracks, making ten million trajectories in total. The error is obtained as a standard deviation of the $100$ values of each evolutionary stage percentage. Approximately $47.3$\% of objects leave the Galaxy, i.e., their final distance from the Galactic center exceeds $100$~kpc. They are not used in further calculations. For the remaining trajectories, we calculate evolutionary tracks in different models using the algorithm for realistic population synthesis described in section~\ref{sec_algorithm}. Thus, we assume that the results in Table~\ref{tab:results} should represent the current distribution of the NSs over evolutionary stages in the Milky Way.

\begin{table*}[ht]
\centering
\caption{The percentage of NSs at every evolutionary stage in different models. Some values are shown only approximately since they are comparable to the margin of error. Zero values indicate that the stage does not occur in the simulation. }
\label{tab:results}
\renewcommand{\arraystretch}{1.1}
\begin{tabularx}{\textwidth}{@{\extracolsep{\fill}}cp{3cm}p{3cm}p{3cm}p{3cm}} 
\toprule
\multicolumn{1}{l}{} & \multicolumn{4}{l}{One-phase ISM, constant field} \\ 
\hline
Propeller model & Ejector & Propeller & Accretor & Georotator \\
A & $34.8 \pm 0.7$ & $4.36 \pm 0.13$ & $44.7 \pm 0.4$ & $16.2 \pm 0.7$ \\
B & $50.4 \pm 0.7$ & $16.3 \pm 0.5$ & $32.5 \pm 0.5$ & $0.83 \pm 0.10$ \\
C & $54.0 \pm 0.8$ & $44.4 \pm 0.8$ & $1.61 \pm 0.15$ & $10^{-7}-10^{-6}$ \\
D & $54.0 \pm 0.9$ & $46.0 \pm 0.9$ & $\sim10^{-5}-10^{-4}$ & $0$ \\
\bottomrule
\addlinespace[4pt]
\multicolumn{1}{l}{} & \multicolumn{4}{l}{One-phase ISM, exponentially decaying field} \\ 
\hline
Propeller model & Ejector & Propeller & Accretor & Georotator \\
A & $15.51 \pm 0.24$ & $31.3 \pm 0.5$ & $52.8 \pm 0.7$ & $0.34 \pm 0.05$ \\
B & $17.00 \pm 0.22$ & $55.4 \pm 0.6$ & $27.6 \pm 0.6$ & $(3.7 \pm 1.7)\times10^{-3}$ \\
C & $17.11 \pm 0.23$ & $77.49 \pm 0.25$ & $5.4 \pm 0.4$ & $0$ \\
D & $17.11 \pm 0.23$ & $82.89 \pm 0.23$ & $\sim10^{-5}-10^{-4}$ & $0$ \\
\bottomrule
\addlinespace[4pt]
\multicolumn{1}{l}{} & \multicolumn{4}{l}{Two-phase ISM, constant field} \\ 
\hline
Propeller model & Ejector & Propeller & Accretor & Georotator \\
A & $42.4 \pm 0.7$ & $6.53 \pm 0.14$ & $36.2 \pm 0.4$ & $14.9 \pm 0.6$ \\
B & $60.8 \pm 0.9$ & $14.6 \pm 0.5$ & $23.6 \pm 0.6$ & $0.96 \pm 0.13$ \\
C & $72.3 \pm 1.1$ & $26.2 \pm 1.0$ & $1.50 \pm 0.14$ & $(5.0 \pm 4.6)\times10^{-3}$ \\
D & $73.0 \pm 1.0$ & $27.0 \pm 1.0$ & $\sim10^{-4}-10^{-3}$ & $10^{-5}-10^{-4}$ \\
\bottomrule
\addlinespace[4pt]
\multicolumn{1}{l}{} & \multicolumn{4}{l}{Two-phase ISM, exponentially decaying field} \\ 
\hline
Propeller model & Ejector & Propeller & Accretor & Georotator \\
A & $18.42 \pm 0.24$ & $40.0 \pm 0.4$ & $41.3 \pm 0.6$ & $0.27 \pm 0.04$ \\
B & $20.98 \pm 0.26$ & $61.7 \pm 0.6$ & $17.3 \pm 0.7$ & $(5.1 \pm 1.8)\times10^{-3}$ \\
C & $21.96 \pm 0.26$ & $75.11 \pm 0.19$ & $2.93 \pm 0.22$ & $10^{-6}-10^{-5}$ \\
D & $21.98 \pm 0.25$ & $78.02 \pm 0.25$ & $\sim10^{-4}-10^{-3}$ & $0$ \\
\bottomrule
\end{tabularx}
\end{table*}

The results of earlier population synthesis studies are in considerable agreement with the stage percentages obtained. Specifically, for the propeller model B, the ISM in one-phase, and the constant field models, we got $35$~\% of the NS in the Milky Way are at the ejector stage, $4$~\% are at the propeller stage, $16$~\% are at the georotator stage, and $45$~\% are accretors. Within similar evolutionary models, results obtained by \cite{2010MNRAS.407.1090B} are different slightly: $55$~\% are ejectors, $30$~\% are accretors, and the percentages of the other stages nearly coincide with those obtained in this work. 

In our model, the most important factor influencing the percentage of accretors among isolated NSs is the spin-down at the propeller stage. In all ISM and magnetic field models used in this study, the highest number of accretors is always in the propeller model A, $36-53$\% of all NSs staying in the Milky Way. This number is $17-33$\% for model B, only a few percent in model C, and several orders of magnitude lower in model D. To explain such a big difference between the models, we examine the duration of the propeller stage for typical parameters of the NS and the ISM.

\begin{figure}
    \centering
    \includegraphics[width=\linewidth]{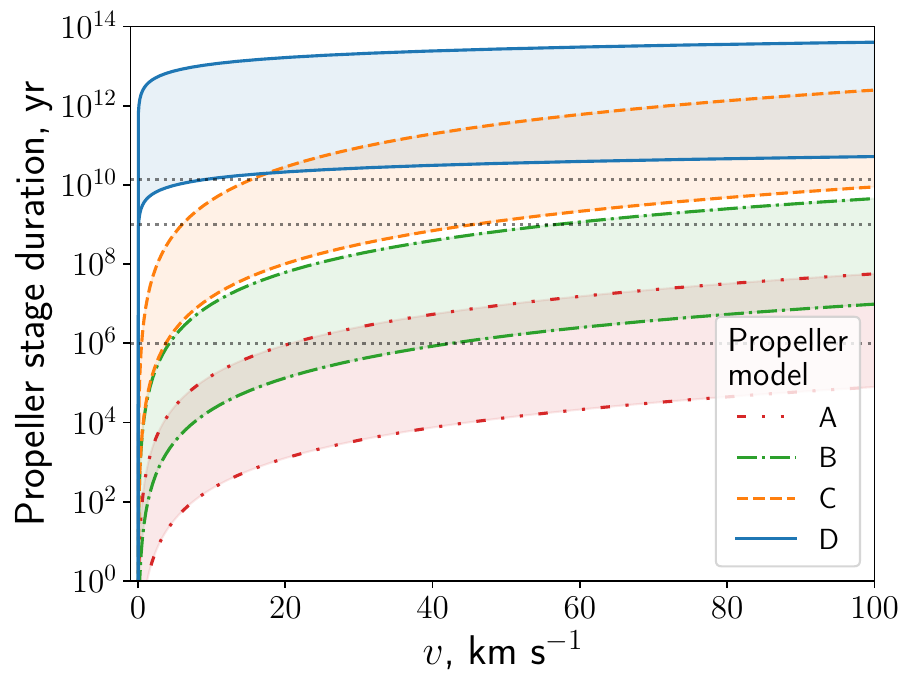}
    \caption{Duration of the propeller stage over the characteristic velocity $v$ of an NS in different propeller models. The magnetic field is constant. Every propeller model is shown as two curves with filled color in between. In each model, the lower curve is for the parameters that promote a rapid evolution: the constant field $B=10^{14}$~G, the number density of the ISM $n=10$~cm$^{-3}$. The upper curve is for average parameters: $B=10^{12}$~G, $n=0.1$~cm$^{-3}$. Three grey dotted horizontal lines indicate $1$~Myr, $1$~Gyr and $13.6$~Gyr.}
    \label{fig_propeller}
\end{figure}

Fig.~\ref{fig_propeller} shows the dependence of the propeller stage duration on the velocity $v$, since the parameter $v$ influences the stage duration the most. The duration is calculated as the time required for the NS with the constant field $B$ to spin down from the ejector–propeller to the propeller–accretor spin period in the medium with constant number density $n$. Two sets of parameters are chosen: the average values $B=10^{12}$~G, $n=0.1$~cm$^{-3}$; and the values $B=10^{14}$~G, $n=10$~cm$^{-3}$, promoting faster evolution.

Referring to Fig.~\ref{fig_propeller}, we can compare the effectiveness of the spin-down mechanism in different propeller models. In all models, the lower the velocity $v$, the higher the magnetic field $B$ and the number density $n$, the faster the NS reaches the accretor stage. 
However, the relationship between the braking torque and these parameters varies across all models, so the propeller stage duration for the models differs drastically. While the propeller stage in model A can realistically end within a few million years, in model D it can easily take more than the age of the Galaxy, preventing the onset of accretion onto the NS. E.g., for a low-velocity NS $v\sim30$~\kms, with parameters $B=10^{12}$~G and $n=0.1$~cm$^{-3}$, the propeller stage will take several million years in model A, $\sim200$~million years in model B, more than $100$~billion years in model C, and over $10^{13}$~years in model D. Therefore, the duration of the propeller stage differs by approximately two orders of magnitude between the models. 

The difference in propeller stage duration by orders of magnitude explains why the number of accretors in models B, C and D differs this much -- the propeller stage is longer, and fewer NSs reach the accretor stage in models with less effective spin-down mechanisms. The accretor percentages in models A and B differ by only a factor of a few. This is because both models have a spin-down mechanism that is efficient enough so that the duration of the propeller stage is already relatively short. 


The efficiency of the spin-down mechanism at the propeller stage determines not only the number of accretors, but also influences the propeller and ejector stages. Firstly, the shorter the propeller stage, the fewer NSs there will be in the Galaxy at this stage. Secondly, with an effective propeller spin-down, the NS is less likely to undergo a reverse transition to the ejector stage. For example, in model A, with a single-phase ISM and a constant field, less than $5$\% of the NSs are at the propeller stage, since the propeller stage is so short.
In model B, there are more NSs at the ejector stage, because some NSs can now switch back to this stage after the first ejector-propeller transition. In models C and D, almost all NSs are either at the ejector or propeller stage. In both models, NSs evolve at the ejector stage until the first ejector-propeller transition, then switch between the ejector and the propeller stages as the external parameters fluctuate. In these models, only a few percent or less of the NS spin down enough to reach accretion.

The studied exponential evolution of the magnetic field has little influence on the number of accreting NSs in the Milky Way. {These results contradict earlier assumptions that the number of accreting NSs could decrease by a factor of a few if the magnetic field decays over long timescales $\sim10^8-10^9$~yr \citep{1998AN....319..109T}.}

Taking the magnetic field decay into account drastically changes the ratio of NSs at the ejector and propeller stages. The difference can be better seen in the case of propeller models C and D, where the number of accretors is small. In a one-phase ISM, in the case of a constant magnetic field, only around $45$\% of NSs are at the propeller stage, compared to over $77$\% if the field is decaying. This is due to the shrinkage of the characteristic radii of interaction with external material, such as the Shvartsman radius and the magnetosphere radius. The smaller the radius, the easier it is to switch from the ejector to the propeller stage. In fact, by the end of the evolution, the ejector-propeller transition period drops to $\lesssim10$~s if the magnetic field has decayed. This spin period is short enough to allow almost all neutron stars to transition to the propeller stage.

The influence of the magnetic field decay on the time of accretion onset is more complicated. On the one hand, the magnetosphere radius is smaller if the field has decayed, making the propeller-accretion transition condition less strict. On the other hand, the NS is still required to spin down significantly from ejector-propeller to the propeller-accretor transition period, and this process can be slower in the case of a decaying magnetic field (except for propeller model D). The interplay of these two effects increases the accretor percentage for propeller models A and C, and decreases it for model B.

The magnetic field decay has the greatest influence on the number of georotators, reducing it by a factor of $20$ or more. Only highly magnetized NSs can switch from the accretor stage to the georotator stage, and this transition is much less probable if the magnetic field has decayed.  

Finally, we consider how the changes in the ISM model affect the NS evolution. In both models, the velocity of the ISM and the average number density are practically identical. However, compared to the one-phase model, in the two-phase model there are more external parameters' fluctuations between the points along the NS trajectory. In the two-phase model, there are more areas with a higher number density, which help the NS to switch to the next evolutionary stage, but there are also more low-density areas where the NS usually evolves much more slowly, especially at the propeller stage. These effects result in a small difference in the accretor percentage: a more realistic two-phase model shows a slightly (less than two times) reduced number of accreting NSs in the Galaxy.






\subsection{The role of initial parameters}

In this section, we present a separate study aimed at determining the main dependencies of the isolated NS evolution on various initial parameters.

We use a uniform grid to derive the dependencies of the time spent at the accretor stage on the initial parameters. These parameters are: the velocity relative to the ISM $(\vec{v}_\text{NS})_0$, the magnetic field $B_0$, the spin period $P_0$, and the position in the Galaxy -- the height above the Galactic plane $z_0$ and the radial distance from the center of the Galaxy in the Galactic plane $R_0$. The initial velocity of NSs $(\vec{v}_\text{NS})_0 = \vec{v}_\text{circ} + \vec{v}_\text{res} + \vec{v}_\text{kick}$ is calculated here with $\vec{v}_\text{res} = 0$, therefore we use $v_\text{kick}$ as the velocity parameter, which is assumed to be isotropic.

\begin{figure*}
\includegraphics[width=\textwidth]{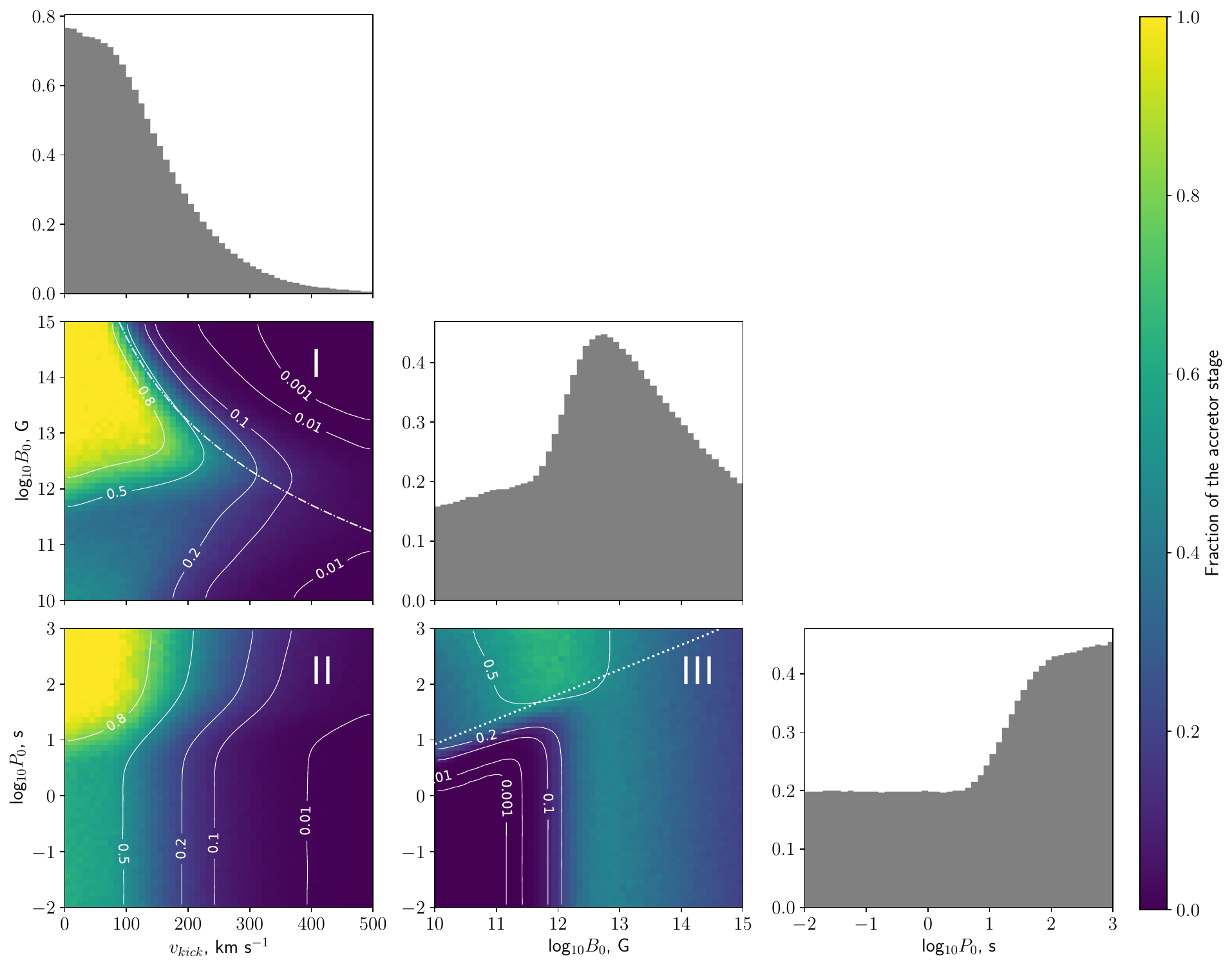}	
\centering 
\caption{Corner diagram illustrating the fraction of time an NS with specific initial values of magnetic field, kick velocity, and spin period spends at the accretor stage $\tau_\text{A}$. Both one- and two-dimensional histograms show $\tau_\text{A}$. 
The dash-dotted line at the panel {I} is the transition condition between the accretor and georotator stages $R_\text{G} = R_\text{A}$ for the number density $n=0.1\text{~cm}^{-3}$. The dotted line in panel {III} separates NSs born at the ejector and propeller stages and corresponds to the ejector-propeller transition with $n=0.1\text{~cm}^{-3}$, $v=100$~\kms. Single-phase ISM, propeller model A, and constant field (CF) models are used. The grey diagrams are normalized so that each bin illustrates an average $\tau_\text{A}$ corresponding to the value of the bin.
}
\label{fig_triangleA}
\end{figure*}

\begin{figure*}
\includegraphics[width=\textwidth]{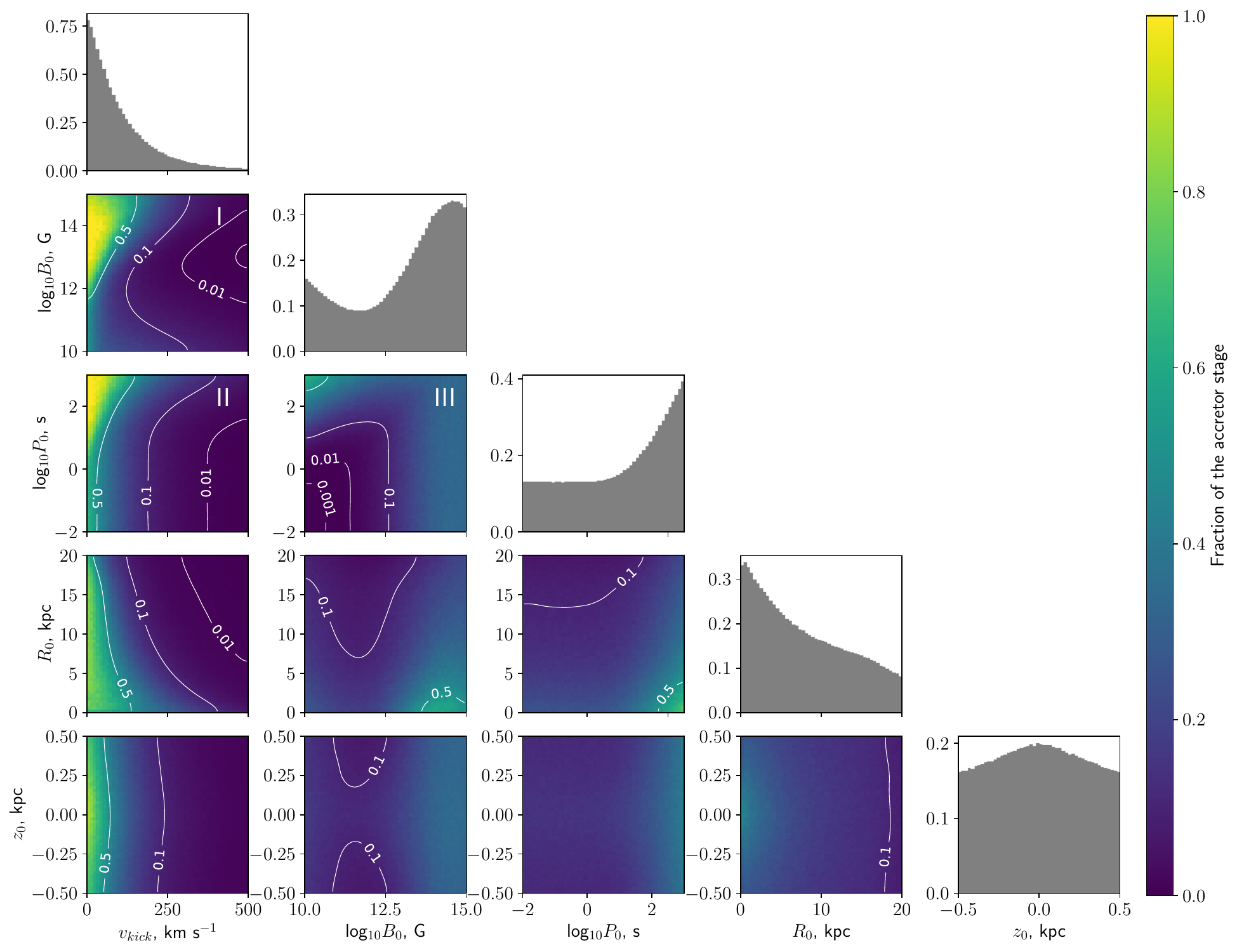}	
\centering 
\caption{Extended corner diagram of the accretion time $\tau_\text{A}$ for the propeller model B, the exponentially decaying magnetic field and the two-phase model of the ISM. The diagram is calculated similarly to Fig.~\ref{fig_triangleA}. The initial parameters are the kick velocity $v_\text{kick}$, the magnetic field $B_0$, the spin period $P_0$, and the coordinates $R_0$ and $z_0$ in the cylindrical coordinate system with the origin at the Galactic center.
}
\label{fig_triangleB}
\end{figure*}

Unlike the realistic population synthesis (sec. 3.1), the sets of the NS parameters are now derived from a uniform distribution: $0<v_\text{kick}<500$~\kms, $10^{10}$~G~$<B_0<10^{15}$~G, $0.01$~s~$<P_0<1000$~s, $0<R_0<20$~kpc, $-0.5$~kpc~$<z_0<0.5$~kpc. For every set of initial parameters, the trajectory is calculated as usual, and after the choice of the evolutionary model (ISM, magnetic field behavior, and the propeller spin-down) the evolutionary track is derived. Then, the fraction of the accretor stage $0\le\tau_\text{A}\le1$ is calculated for each track as the time spent at the accretor stage over the total time of the evolution (the Galactic lifetime). We introduce a grid that divides each interval into $50$ parts, so there are $50^5$ five-dimensional cubes in the parameter space, since the total number of initial parameters is five ($v_\text{kick}$, $B_0,$ $P_0$, $R_0$, $z_0$). Every evolutionary track contributes to some cube, so the value $\tau_\text{A}$ of each cube is the average time spent at the accretor stage of the NSs with the corresponding initial parameters. In this section, we consider one- and two-dimensional projections of this parameter space, Figs.~\ref{fig_triangleA} and~\ref{fig_triangleB}.

There are sixteen combinations of the models that we have considered in the previous subsection: two models of ISM (single and two-phase), four models of the propeller stage (A, B, C, and D), and two models of magnetic field evolution: constant magnetic field (CF) and exponentially decaying magnetic field (ED). In this section, we consider in detail only two sets of models. Figs.~\ref{fig_triangleA} and~\ref{fig_triangleB} illustrate the relationship between $\tau_\text{A}$ and the initial parameters. Each figure is based on $10$ million evolutionary tracks. In Fig.~\ref{fig_triangleA}, we apply propeller model A, a single-phase ISM, and the constant field model. 
The figure is used in the following to analyze and illustrate the dependencies in detail.
In this plot, only the most important parameters are shown: the initial values of the magnetic field, the velocity, and the spin period. In Fig.~\ref{fig_triangleB} we use a more realistic set of models: a two-phase ISM, a less effective propeller model B, and an exponentially decaying field.

First, consider Fig.~\ref{fig_triangleA}.
We use this set of models, since it shows more clearly how the time spent at the accretor stage depends on the initial parameters. These ISM and propeller models result in a greater number of accretors in the Galaxy. In a single-phase model, an NS spends more time in a high-density medium than in a two-phase model. This allows the NS to evolve faster. In propeller model A, an effective spin-down mechanism leads to a short propeller stage and early onset of accretion. Additionally, the constant magnetic field makes it easier to see how differences in initial magnetic field values affect the time an NS spends accreting matter from the ISM.

As can be seen in panels I and II of Figs.~\ref{fig_triangleA},\ref{fig_triangleB}, the time spent at the accretor stage depends on the initial velocity relative to the ISM the most -- the higher the velocity, the fewer NSs reach the accretor stage and/or spend less time accreting. {This result supports the earlier conclusions about the strong dependence of the number of accretors on the characteristic velocity made by \cite{2000A&AT...19..471P}. } There are several important reasons for this dependence. The first one is that objects with a higher velocity have a higher probability of being thrown out of the Galaxy.  
If the kick velocity is parallel to the circular velocity, some objects can reach distances of a hundred kpc from the Galactic center by the end of their evolution, even at relatively low $v_\text{kick} \approx 150$~\kms. With a kick velocity of $450$~\kms, more than half of the NSs leave the Galaxy. 

The second reason is that an important parameter $\dot{M}$ depends heavily on the velocity relative to the surrounding medium: $\dot{M}\propto v^{-3}$. An NS at the ejector stage with a higher $\dot{M}$ has to reach a shorter period to switch to the propeller and, later, to the accretor stage. Thus, the transitions occur earlier. The parameter $\dot{M}$ also plays a role in the spin-down at the propeller stage in every model -- an NS spins down faster with a higher $\dot{M}$. In other words, an NS with a higher $\dot{M}$ (lower $v_\text{kick}$) reaches the accretor stage earlier and spends more time accreting matter from the ISM in general. As can be seen in both Figs.~\ref{fig_triangleA} and~\ref{fig_triangleB}, the majority of accreting neutron stars have $v_0\lesssim 100$~\kms.

Additionally, a higher $\dot{M}$, which corresponds to a lower $v_\text{kick}$, decreases the chance of the accretor-georotator transition. The dash-dotted curve in panel I of Fig.~\ref{fig_triangleA} ($\log_{10} B_0 - v_\text{kick}$) shows the condition separating accretors and georotators. It is the equality of the magnetosphere radius, which is $R_\text{A}$ for the accretor stage and the gravitational capture radius $R_\text{A} = R_\text{G}$. The equation of the curve is derived for typical parameters $n=0.1$~cm$^{-3}$, $v=100$~\kms:
\begin{equation}
    B(R_\text{A}=R_\text{G}) \approx 5\times 10^{14} \left(\frac{v}{100\text{~\kms}}\right)^{-5} \left(\frac{n}{0.1}\right)^{-1/2}\text{~G}.
\end{equation}
While the NS with lower magnetic field (and smaller magnetosphere $R_\text{A}$) can spend $\approx80$\% of their time as accretors, the percentage of accretors for $R_\text{A}>R_\text{G}$ drops to $10$\%, because with a larger magnetosphere the NSs switch to the georotator stage.  However, this effect is absent in a more realistic model in Fig.~\ref{fig_triangleB} due to the decay of the magnetic field, which makes the magnetosphere to shrink with time.

Now, we will consider how the time spent at the accretor stage depends on the initial spin period and magnetic field.  In panel III of Fig.~\ref{fig_triangleA} ($\log_{10} P_0 - \log_{10} B_0$), there is an area with a low percentage of accretors. These NSs have short initial spin periods $P_0$ and low magnetic fields $B_0$. The area is bounded by a diagonal line from the top and a vertical line from the right. 

The diagonal line in the III diagram separates the NSs born at the ejector and propeller stages. The NSs above the line are born with spin periods already higher than required for the ejector-propeller transition. In the propeller model A, almost all of the NSs born at the propeller stage reach accretion. Therefore, the objects above the line start to accrete, while the ones below generally do not.


The vertical boundary in Fig.~\ref{fig_triangleA} is related not just to the initial condition (as the diagonal one), but to the spin-down evolution. It corresponds to the time of the accretion onset. Since the propeller stage duration in model A is, for most realistic parameters, negligibly small compared to the Galactic age, we consider only the time required to reach the propeller stage instead of the time of accretion onset. The time spent in the ejector stage $t_\text{E}$ until the NS reaches a spin period $P$ is proportional to the period $t_\text{E}\propto P^2$ if the initial spin period is relatively short $P_0\ll P$:
\begin{equation}
    t_\text{E}(P) = \frac{Ic^3}{8\pi^2\xi \mu^2 } \left(P^2 - P_0^2\right)\approx \frac{Ic^3}{8\pi^2\xi \mu^2 } P^2.
\end{equation}
In panel III, for initial spin periods $P_0<10$~s, the accretors are separated with a vertical line, which corresponds to the value of $B_0\approx2\times10^{12}$~G for the $n=0.1\text{~cm}^{-3}$, $v=100$~\kms.

In Fig.~\ref{fig_triangleB}, both diagonal and vertical boundaries also exist, but they are less noticeable. Compared to Fig.~\ref{fig_triangleA} with propeller model A, the propeller stage in model B lasts longer, so not all NSs born at the propeller stage would start to accrete. The delay between the birth of the object and the start of the accretor stage shifts the diagonal line upward. The vertical line now corresponds to the higher values of $B_0$ due to the magnetic field decay. Generally, in a more realistic propeller model, the higher the initial spin period and the magnetic field values are, the earlier the accretor stage starts. This dependence of accretion onset time on $P_0$ and $B_0$ results in gradients on the diagram rather than clear lines.

In addition to the initial values of velocity, magnetic field and spin period, Fig~\ref{fig_triangleB} shows how $\tau_\text{A}$ depends on the initial position in the Galaxy. It can be concluded that the NS in general spends more time at the accretor stage if it is born closer to the center of the Galaxy and closer to the Galactic plane. Firstly, with the same initial velocity, the farther the NS is born from the Galactic center or the Galactic plane, the more elongated the orbit will be, and the easier it will be to leave the Galaxy. Additionally, the average number density of the ISM is lower at high altitudes and farther from the center, which leads to a lower $\dot{M}$ and, consequently, to a slower evolution of the NS. This reduces the time spent at the accretor stage, because the onset of accretion is delayed. Thus, Fig~\ref{fig_triangleB} shows that the NSs born within the first kpc from the center spend twice as much time at the accretor stage as objects born on the outskirts of the Galaxy at $R_0=20$~kpc. Additionally, the NSs born at the Galactic plane spend $\approx1.5$ times more time at the accretor stage.

\subsection{Modeling observations with \textit{eROSITA}}

\begin{figure*}
\includegraphics[width=\textwidth]{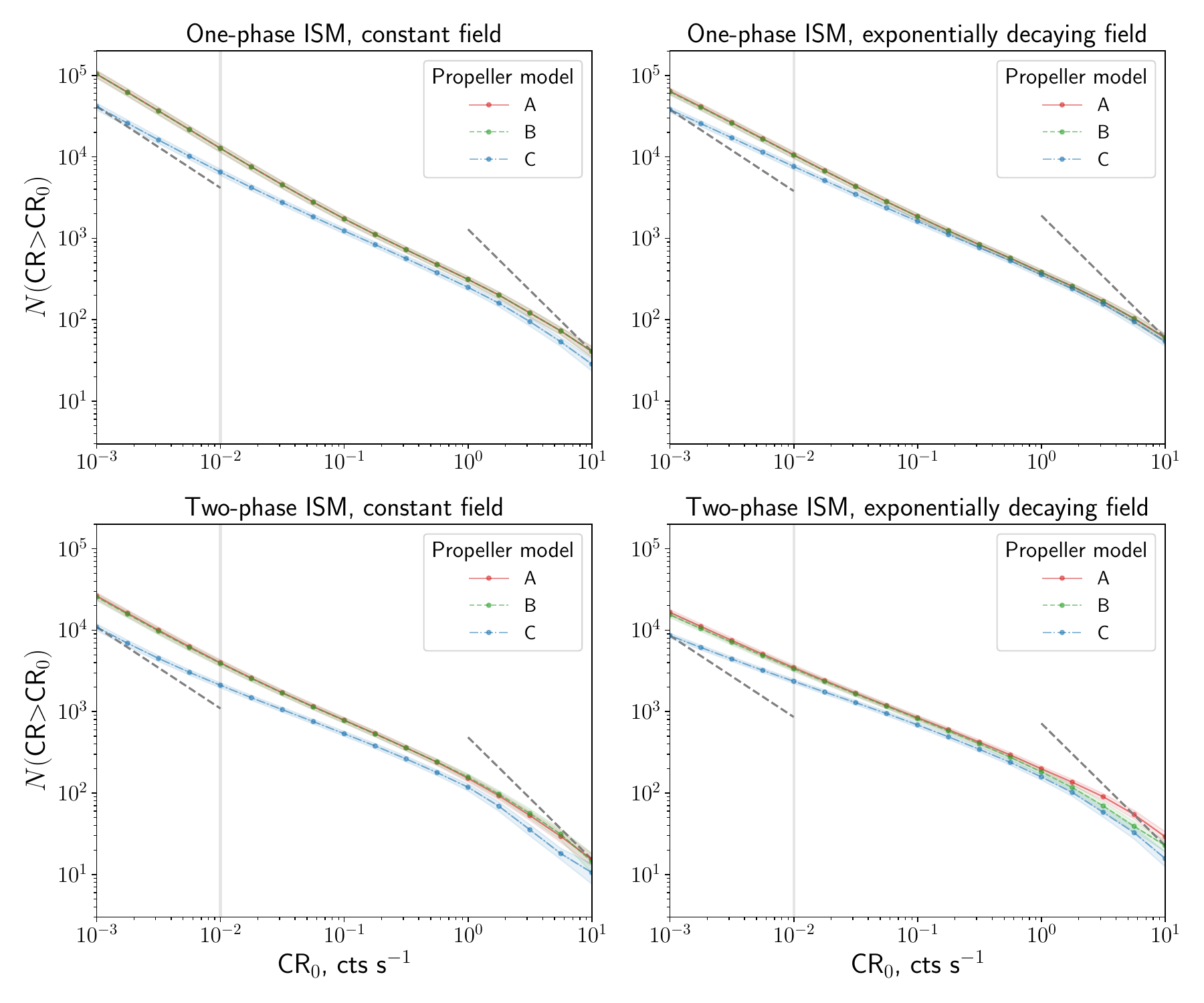}	
\centering 
\caption{The number of the accreting neutron stars $N$ producing the count rate CR$_0$ or brighter versus CR$_0$. In each panel, only three out of four propeller models are shown, since models A, B, and C produce enough accreting NSs that might be observable, while model D does not. Propeller models A and B yield similar results, so their curves are nearly identical. Objects potentially visible to the \textit{eROSITA} telescope are located to the right of the faint vertical line at $10^{-2}$~cts~s$^{-1}$. The dashed line on the left in each panel shows the dependence $ N(\text{CR}>\text{CR}_0)\propto \text{CR}_0^{-1}$, the one on the right is $\propto \text{CR}_0^{-3/2}$. The narrow colored area shows the range of values within one standard deviation of the mean.
}
\label{fig_counts}
\end{figure*}

In this subsection, we estimate the number of the old accreting isolated NSs $N = N(CR >$~CR$_0)$, with a certain \textit{eROSITA} count rate CR$_0$ or higher and discuss the number of the differences in the number of potentially observable sources, $N_\text{obs} = N(CR >$~$0.01$~cts~s$^{-1})$, between the models used. The number of potentially observed sources is listed in Table~\ref{tab:observ}, the $\log N-\log\text{CR}_0$ diagram is shown in Fig.~\ref{fig_counts}.

The results are obtained the following way. As in section~\ref{res_models}, the realistic initial parameter distribution is used. It includes the pulsar and magnetar populations.  We create $100$ samples of $100$~thousand sets of initial parameters and calculate the trajectories and evolutionary tracks, taking into account the star formation history. 
There are $10$~million trajectories in total, but evolutionary tracks are calculated only for the objects that stay in the Galaxy (i.e., they do not reach $100$~kpc from the Galactic center), which is $52.7$\%. Then, following section~\ref{sect_model_observability}, we calculate the count rate for the points of the evolutionary tracks, corresponding to the accretor stage. Each sample results in an array $N$ over the same set of the values of CR$_0$. The numbers of the sources are normalized, assuming that there are $300$~million NSs staying in the Galaxy at the moment. The error for each value of $N$ is calculated as the standard deviation of the $100$ values from the $100$ samples.

\begin{table}[ht]
\centering
\caption{The number of old isolated accreting NSs potentially observable by \textit{eROSITA}. Only three out of four propeller models are shown, since in propeller model D, the spin-down mechanism is so inefficient that it prevent accretion in most cases and results in zero NSs with the count rate $\text{CR}>0.01$~cts~s$^{-1}$.}
\label{tab:observ}
\renewcommand{\arraystretch}{1.1}
\begin{tabular}{ccc}
\toprule
\multicolumn{1}{l}{} & \multicolumn{2}{l}{One-phase ISM} \\ 
\hline
Propeller model & Constant field & Exponential decay \\
A & $13000 \pm 1400$ & $11000 \pm 1000$ \\
B & $13000 \pm 1400$ & $10000 \pm 1000$ \\
C & $6500 \pm 600$ & $7600 \pm 700$ \\
\bottomrule
\multicolumn{1}{l}{} & \multicolumn{2}{l}{Two-phase ISM} \\ 
\hline
Propeller model & Constant field & Exponential decay \\
A & $4000 \pm 400$ & $3470 \pm 260$ \\
B & $4000 \pm 400$ & $3350 \pm 250$ \\
C & $2100 \pm 150$ & $2360 \pm 150$ \\
\bottomrule 
\end{tabular}
\end{table}

Let us first discuss the possible number of potentially observable X-ray sources with \textit{eROSITA}, which is shown in Table~\ref{tab:observ}. In different evolutionary models, this number varies significantly. The uncertainty in the spin-down at the propeller stage is the most important factor introducing the range of the observable old NSs, influencing the number of accreting NSs. In the most optimistic case proposing rapid evolution (one-phase ISM, constant field, propeller model A) there are $\approx13300$ sources, and it can be a few thousand in more realistic cases. In all magnetic field and ISM models, this number can be even reduced to zero by using propeller model D, which has the least effective spin-down mechanism.

As discussed earlier, velocity is the most important parameter influencing the number of accreting isolated neutron stars in the Galaxy. The number of potentially observable NSs derived depends on velocity even more strongly. Lower velocity values favor the onset of accretion and consequently lead to a higher number of observable accretors. Additionally, velocity influences the X-ray flux, which is proportional to the X-ray luminosity $L_\text{X}$. The flux is much higher for low-velocity objects: $L_\text{X}\propto\dot{M}\propto v^{-3}$. NSs with a higher flux are visible from a greater distance, meaning that the objects with lower velocities are more prevalent in the population of observable accreting NSs. In all ISM and magnetic field models, $~99$\% of the potentially visible accreting NSs have characteristic velocities relative to the ISM $v\lesssim60-70$~\kms in propeller models A and B, and $\lesssim40$~\kms in model C. In every evolutionary model, for half of the population $v\lesssim20$~\kms. The characteristic velocity cannot be lower than the speed of sound, which is $100$~\kms for the hot phase of the ISM, which means that in the two-phase ISM model accretors are only visible if they are located in the cold phase of the ISM.

The decrease in flux with increasing characteristic speed $v$ can be a stronger factor in limiting the number of potentially observable sources in comparison to the dependence of the total number of accreting NSs in the Galaxy on $v$. This is evidenced by the fact that the number of potentially observable accretors in propeller models A, B, and C does not differ as significantly as the number of accretors in the Galaxy in general.

Across all magnetic field and ISM models, propeller models A and B lead to almost the same $(N-\text{CR}_0)$ dependence and produce very similar numbers of observable sources, even though the total number of accretors in the Galaxy differs by a factor of $1.5-2$ between the two. Models A and C also produce observable accretor numbers that vary only by a factor of a few, while their total Galactic accretor populations differ far more significantly -- by $\sim20$ times.  According to Fig.~\ref{fig_propeller}, the propeller stage duration for the constant velocity $60$~\kms, magnetic field $10^{12}$~G in the medium with $n=0.1$~cm$^{-3}$ is less than one billion years for propeller models A and B. It means that, in both propeller models, the accretor stage starts on average in the first billion years, while the overall timescale of the calculations is $13.6$~Gyr. Taking the star formation history into account, which decreases the number of NSs with ages $<10$~Gyr, makes the difference between the two propeller models even less significant, so the number of potentially observable accreting NSs in both cases is similar. In the case of propeller models A and C, the difference between the number of observable accretors is much more noticeable. In model C, at velocities $60$~\kms the duration of the propeller stage can still exceed the age of the Galaxy, so the dependence of the overall number of accretors in the Galaxy on the velocity still plays an important role in the number of observable sources. Thus, in propeller model C, some of the potentially observable NSs from models A and B do not start to accrete. In model D, the duration of the propeller stage is high even for low-velocity objects. Consequently, the number of accreting isolated neutron stars is so small that there are no observable sources.


In our model, all observable sources produce thermal X-ray emission as a black body with the effective temperature $T_\text{eff}$. Results show that in the case of the constant magnetic field $T_\text{eff}\sim10^6-10^7$~K, the maximum of the distribution of observable accretors over $T_\text{eff}$ is $5\times10^{6}$~K regardless of the propeller and ISM model. If the field exponentially decays, the effective temperature is generally lower: $T_\text{eff}\sim6\times10^{5} - 9\times10^{6}$ with the maximum corresponding to $T_\text{eff}=2\times10^{6}$~K. This is the effect of different average magnetic field values in two models. The effective temperature depends on the polar cap radius $R_\text{cap}$ and the X-ray luminosity $L_\text{X}$, which is proportional to the accretion rate, $\dot{M}$. Since all observable accretors are located in the medium with similar parameters $c_\text{s}$ and $n$ (in the two-phase model, only objects from the cold phase are visible) and all have low velocities, the values of $\dot{M}$ are roughly similar. The polar cap radii in the two models of the magnetic field behavior are generally different. During $13.6$~Gyr, the magnetic field value $B$ decreases four orders of magnitude in the case of the decaying field. The Alfv{\'e}n radius depends on the magnetic field as $R_\text{A}\propto B^{4/7}$, so the magnetosphere of the accreting NSs shrinks with time, which affects the polar cap radius. The complete dependence of the effective temperature on the magnetic field value is $T_\text{eff}\propto S_\text{cap}^{-1/4}\propto R_\text{cap}^{-1/2}\propto R_\text{A}^{1/4}\propto B^{1/7}$. Therefore, if the magnetic field decays $10^4$ times, the effective temperature of the accreting NS decreases by about half an order of magnitude. This explains why the values of $T_\text{eff}$ are slightly lower in the case of a decaying field than in the case of a constant field.

There is also a slight difference in the slope of the curves in the two magnetic field evolution models, which is also connected to the different effective temperatures. This is because interstellar absorption is more intense for lower temperatures: the cross section $\sigma(\nu)$ in the ISM increases as the energy of the radiation decreases. In the case of a decaying magnetic field, X-ray emission is generally less energetic and absorbed more readily, resulting in a shallower curve on the $\log N-\log\text{CR}_0$ diagram.

Observed accreting isolated neutron stars are located close to the Solar System. The spatial distribution of potentially observable sources is similar in all evolutionary models. In propeller models A and B, half of the objects are located within $410$~pc of the Sun, and in model C, this distance is $550$~pc. Almost all of the objects, $99$\%, are within $1.5$~kpc in all models except model D, which results in zero observable sources. 

According to the distribution of the sources over the galactic height $z$, $99$\% of them are close to the Galactic plane, with $z<470$~pc. Half of these sources are distributed in a thin disk with $z\approx60$~pc. 

The slope of the curves in Fig.~\ref{fig_counts} is influenced by the spatial distribution of the potentially observable NSs. The slope becomes steeper with increasing count rate. NSs with a lower count rate are located farther from the Sun and are distributed in a disc. For a count rate CR$_0=10^{-2}$~cts~s$^{-1}$, the hight of the disc is $h\approx60$~pc, while the distance from the Sun is $\approx400-500$~pc. Without interstellar absorption, the count rate is proportional to the flux, which depends on the distance from the Sun, $d=|\vec{r}-\vec{r}_\odot|$, as $F_\text{x} \propto d^{-2}$. The number of the observed sources with increasing $d$ increases proportionally to the area of the disc $\propto d^2$. This means that, when the disc is thin, $d\gg h$, the number of sources brighter some flux $(F_\text{X})_0$ would depend on this flux as $N_0\propto (F_\text{X})_0^{-1}$, since $N_0\propto d^2$ and $(F_\text{X})_0 \propto d^{-2}$.  Therefore, the slope coefficient of the curve on the $\log N-\log\text{CR}_0$ diagram is close to $-1$ if the count rate is low, i.e., most of the observable NSs are located relatively far. As the distance decreases, which corresponds to an increase in CR$_0$, the slope becomes steeper. When $d\sim h$, the spatial distribution of the accreting NSs is closer to a spherical, so the number of the sources $\propto d^3$, while the flux $(F_\text{X})_0 \propto d^{-2}$. Thus, the higher the flux (or the count rate), the closer the slope coefficient is to $-3/2$.

\section{Discussion}

{ In this section, we briefly discuss several issues related to the evolution of isolated NSs and our modeling. We pay special attention to some uncertainties and simplifications of our model, as well as to some applications of our results.}

\subsection{Prospects for detection of isolated accreting neutron stars}


{ The discovery of accreting isolated NSs is a long-sought goal. Let us briefly summarize some properties of these sources, which might help to identify them among weak X-ray sources.

Basically, the strategy is similar to that used in searches for isolated radioquiet thermally emitting (i.e., cooling) NSs such as the Magnificent Seven (see e.g., \citealt{2026A&A...705A.148K}). 
Accreting isolated NSs are expected to be weak soft thermal X-ray sources without bright counterparts at any wavelengths. Similarly to thermally emitting NSs, they can have only very weak optical/UV counterparts, if any.

Two main differences between the expected properties of accreting and thermally emitting NSs are related to their variability. E.g., all but one of the Magnificent Seven objects demonstrate clear spin periods $\sim 10$~s. As thermally emitting NSs might be young sources with ages $\lesssim 10^6$~yrs, their spin periods are expected to be relatively short. In contrast, accreting isolated NSs are long-period sources. The lower boundary is defined by the critical period of the transition from the propeller to the accretor stage.
In our model, it is defined as: }
\begin{equation}
    P_\text{PA} 
     \approx 1.3\times10^{4} \left(\frac{v}{10\text{~\kms}}\right)^{1/3} \left(\frac{B}{10^{12}\text{~G}}\right)^{2/3}\left(\frac{n}{1\text{~cm}^{-3}}\right)^{-1/3}\text{~s}.
\end{equation}
{This parameter is obtained from the condition $R_\text{m} = R_\text{c}$, which corresponds to the direct transition from the propeller stage. Here, $R_\text{m}$ is calculated using eq.~(\ref{eq_R_m}).}

{ After the stage of accretion is reached, the period continues to grow. Finally, it reaches very large values and is regulated by properties of turbulence in the ISM. 
It would be extremely difficult to detect the spin variability of an accreting isolated NS.
Oppositely, these sources might demonstrate irregular variability on a time scale $\gtrsim R_\mathrm{G}/v$ due to fluctuations in the ISM properties. This time scale is of the order months-years for most of accretors. This is much different from expectations for thermally emitting NSs, which are expected to have stable luminosities on this time scale.

Finally, measurements of proper motions will help to distinguish between cooling and accreting isolated NSs. The luminosity of a thermally emitting isolated NS is not related to its velocity,
which is not the case for an accreting source. Large proper motions will undoubtedly point to the cooling nature of the source.
}

\subsection{Kick velocity distribution}

 The evolution and observational appearance of isolated NSs strongly depend on their spatial velocities. 
 These velocities are mainly determined by the kick which a compact object receives at birth (additional contributions can be due to the progenitor's velocity and orbital motion if the supernova exploded in a binary, which we do not include in the present study). 
 Here, we are mainly interested in the fraction of low-velocity NSs. 
 This parameter is poorly constrained by observations of radio pulsars and other isolated sources, as it is difficult to measure small velocities. On the other hand, it is not possible to apply estimates of the number of small kicks obtained from observations of binary systems. In this case, low velocities can be related to the so-called $e^-$-capture SNe \cite{1984ApJ...277..791N}.  Such SNe mainly occur in binary systems that remain gravitationally bound after explosions \cite{2002ApJ...574..364P}, i.e., NSs born in the $e^-$-capture SNe do not contribute much to the statistics of isolated objects. 

 Lacking strong observational constraints, the shape of the velocity distribution in the low-velocity part is typically determined just by the assumption about the analytical approximation for the general distribution. This introduces some uncertainty to our results. However, at very low values of the effective relative velocity of a NS and the ISM ($\lesssim 20-30$~km~s$^{-1}$), contributions from the progenitor velocity, see eq~(\ref{eq:initial_velocity}), and the sound velocity in the medium are dominant, not the kick. 

 In our opinion, uncertainties in the kick velocity distribution do not currently dominate the error budget of population synthesis studies of isolated NSs. From the point of view of the luminosity of accreting compact objects, a poor understanding of accretion at low rates is more crucial. And from the evolutionary point of view, uncertainties in the propeller stage behavior are much more important.

\subsection{Propeller regime}

The results of the present work demonstrate that the choice of the propeller model is the most important factor determining the number of old isolated NSs observable as soft X-ray sources. {As was also shown in earlier studies (e.g. \citealt{1993ApJ...403..690B}), the duration of the propeller stage can make the onset of the accretion onto an isolated NS impossible.}
Unfortunately, due to the lack of observational constraints, the situation with these models is still rather uncertain.
In addition to the propeller stage, there are several analogous regimes that can reduce the number of observable accretors: the subsonic propeller stage and the settling accretion regime. We briefly comment on them below. 

The subsonic propeller stage is an evolutionary stage that can follow the propeller regime considered in the present work. It is examined, e.g., by \cite{1981MNRAS.196..209D}. According to these authors, accretion does not start immediately when the condition $R_\text{m}=R_\text{c}$ is reached. Instead, the NS switches from the propeller to the subsonic propeller stage. The stage proceeds, delaying the onset of accretion until another critical spin period is reached. This period is determined by the relation between energy input and energy losses in the envelope around the magnetosphere \citep{2001A&A...368L...5I}. If this stage is present, the NS begins accreting later, thereby reducing the number of observable X-ray sources.

Instead of the onset of Bondi accretion with the rate $\dot{M}$, 
the subsonic settling accretion, proposed by \cite{2012MNRAS.420..216S}, 
can be established. 
This regime leads to significantly 
lower X-ray luminosity. 
There are studies proposing the discovery of NSs at this stage, e.g, 
\cite{2017MNRAS.465L.119P}. 
However, later studies have pointed out inconsistencies in the interpretation of these observations \citep{2017MNRAS.469.1502S}. 

More recently, a candidate for the NS at the propeller stage in a symbiotic X-ray binary was proposed by \cite{2024MNRAS.528L..38D}. 
The modeling of the evolution of the system and further X-ray observations suggest that this NS more likely accretes the material from the wind of the secondary component rather than experiencing the propeller regime \citep{2024Univ...10..205A,2025A&A...693A.260Z}. 
Thus, observations still do not allow us to constrain the properties of the propeller stage or to support the presence of a similar regime of interaction between the magnetosphere and external matter.

Numerical studies of the propeller stage favor the spin-down torque similar to the propeller model A \citep{1985A&A...151..361W,2003ApJ...588..400R}. 
However, these calculations consider NSs with relatively high accretion rates compared to the ISM values of $\dot{M}$. To study the propeller regime under conditions similar to the ISM, we can hope for observations of NSs in wide low-mass binaries, where the compact objects interact with material from the stellar wind of a Sun-like companion and evolve similarly to isolated NSs. 

\subsection{Magnetic field evolution}

The evolution of the magnetic field is one of the important components in our modeling.
The possibility that the neutron star's field decays was proposed soon after the radio pulsars' discovery \citep{1969ApJ...157.1395O}. These authors advocated field decay in normal radio pulsars with a timescale $\sim 4\times 10^6$~yrs. Later studies demonstrated that such a rapid decrease in the standard pulsars' field cannot be sustained for a long time, but still a brief period of field decay is possible
(see, e.g., \cite{2014MNRAS.444.1066I, 2015AN....336..831I} and references therein). Moreover, several studies suggest that the radio pulsar population with typical ages $\lesssim 10^7$~yrs can be explained with a constant field \citep{2006ApJ...643..332F}.
 Still, up to now, the physics of NS field evolution has not been well understood, even for relatively young objects
(see reviews in \citealt{2021Univ....7..351I, 2025arXiv250906699P}). For the purposes of our population synthesis, we need to specify the field evolution on a scale of billions of years, which is even less certain. 

 In this study, we used two simple approaches to account for the field evolution. In the first one, the field is assumed to be constant. Obviously, this is an oversimplification, so this scenario can be used mostly for illustrative purposes. In the second case, we used a two-stage field evolution. At first, if the NS belongs to the magnetar population, there is an episode of relatively rapid field decay due to the Hall cascade. The timescale is inversely proportional to the initial field. This phase saturates after $\sim 3~e$-foldings. Then, the field decays exponentially with a constant time scale $1.5 \times 10^9$~yrs to be consistent with the field measurements for millisecond radio pulsars. I.e., in this approach, all old NSs have relatively low fields in comparison to normal radio pulsars. This also can be an oversimplification as the long-term behavior of the magnetic field in the absence of strong accretion (like in the systems that produce millisecond pulsars) is very uncertain since we do not know of any object to probe this feature. Thus, field measurements for old (e.g., accreting) isolated NSs (or NSs in wide non-interacting binaries, see below) are very much welcomed. 

The behavior of the magnetic field on a long timescale can influence the isolated NS evolution in two ways. For a smaller field, the rate of spin-down is smaller in both -- ejector and propeller -- stages preceding accretion. Therefore, field decay can postpone accretion. In contrast, for smaller fields, the critical periods for a stage transition become smaller. E.g., imagine such a strong decay that the magnetospheric radius $R_\mathrm{m}$ becomes smaller than the NS radius. Then, the NS immediately starts to accrete. The exact influence of the field evolution depends on the interplay of these two main effects. 

If the initial spin period is much smaller than the critical period $p_\mathrm{E}$, then the duration of the ejector stage only increases for a decaying magnetic field. Thus, the field decay can speed up the transition to the propeller stage only if the initial spin period is sufficiently large ($\gtrsim 10$~s for realistic conditions), or if the field decays to very low values. As is visible in Fig.~3 (panel III), our model of decay does not overproduce accretors for standard initial spin periods. 

Different models of the propeller spin-down demonstrate different dependencies on the magnetic field: for models A and B, the spin-down is more rapid for a larger field, while for models C and D the rate does not depend on the field. 
The propeller-accretor transition period grows for a larger magnetic field. 
In general, field decay might shorten
the propeller stage duration across all the models we used. However, the effect is not sufficiently strong to influence the number of accretors in the considered cases. One can hardly introduce a more effective monotonic field decay, as we use the exponential one and apply the lowest known fields as the bottom value. Slower field decay might not significantly modify our conclusions but can result in a different spin period distribution of old NSs and influence the characteristics of accretion.

In our study, we do not intend to apply multiple models of late field evolution due to the discussed uncertainties. In contrast, we hope that in the near future, observations can provide clues to select a proper model of the magnetic field behavior. 


\subsection{Efficiency of accretion}
{The actual accretion rate of the material that reaches the surface of the neutron star can be expressed as $\dot M = \zeta  \dot M_\text{B}$, where $\dot M_B$ is the maximum (Bondi) accretion rate (see eq.~\ref{eq_M_dot_bondi}). Thus, $0< \zeta \le 1$. The particular value of $\zeta$ determines the X-ray luminosity $L_\text{X}=GM \dot{M}/R_\text{NS}$. In our study, we assume $\zeta \equiv 1$, which not might be the case.}
{In this subsection} we consider how the equation for $\dot{M}$ can vary from the Bondi value. 

The choice of the equation for $\dot{M}$ does not introduce much uncertainty to the calculations, despite the fact that eq.~\ref{eq_M_dot_bondi} for calculating the accretion rate, $\dot{M} = \beta \pi R_\text{G}^2 \rho v$, where $\beta=1$, can be considered somewhat simplified. The value of $\dot{M}$ may vary along different Mach numbers $\mathcal{M} = |\vec{v}_\text{NS}-\vec{v}_\text{ISM}|/c_\text{s}=v_\infty/c_\text{s}$ and different gas adiabatic indices $\gamma$. For typical parameters considered in our model, $v_\infty\gtrsim c_\text{s}$ , the coefficient $\beta$ can vary from $2\sqrt{2}$ to $4$ \citep{1979MNRAS.188...83H, 1985MNRAS.217..367S}. More recent numerical estimates show that the dependence of the coefficient $\alpha_\text{acc}$ on the Mach number is non-monotonic: for values $\mathcal{M} < 1$ and $\gamma = 5/3$, the coefficient $\beta = 1/4$. At Mach numbers~$\sim 1-2$ the accretion rate reaches a maximum and $\beta$ is almost equal to $1$; then it decreases to $1/2$ at higher $\mathcal{M}$ \citep{1997A&A...320..342F}. For the subsonic case, the results of numerical studies agree that if $v_\infty<c_\text{s}$, the best approximation for $\gamma=5/3$ remains $\dot{M}\approx\pi(GM)^2\rho c_\text{s}^{-3}$ \citep{1971MNRAS.154..141H, 1994A&AS..106..505R, 1997A&A...320..342F, 2024arXiv240210341P}. 
Thus, $\dot{M}$ in the cases considered by us may differ from the adopted equation only by a factor of a few.

The {case of $\zeta = 1$} corresponds to the standard Bondi accretion of the free-falling material onto the NS surface. However, this accretion regime occurs only if the surrounding matter cools effectively. For low values of $\dot{M}$, a more accurate description of the accretion regime can be the subsonic settling accretion with {$\zeta \ll 1$}.


Settling accretion is characterized by the existence of a hot quasi-static shell extending from $R_{\text{A}}$ to $R_{\text{G}}$. The material cools ineffectively and, according to \cite{2015ARep...59..645S}, slowly flows toward the NS's surface with an average radial velocity $u = (t_{\text{ff}} / t_{\text{cool}})^{1/3} v_{\text{ff}}$, where $v_{\text{ff}}$ is the free-fall velocity and $t_\text{cool}$ is the characteristic plasma cooling time. The maximum possible accretion rate onto the surface is $\dot{M}_{\text{SA}} \sim (t_{\text{ff}} / t_{\text{cool}})^{1/3} \dot{M}$. Since $\dot{M}_{\text{SA}} \ll \dot{M}$ ({i.e. $\zeta \ll 1$}), the X-ray luminosity of observable sources may be several orders of magnitude lower than predicted in our calculations. According to the results presented in Fig.~\ref{fig_counts}, in this case, the number of observable sources will also decrease by several orders of magnitude, which may reduce the population of observable accretors to small numbers or completely prevent the possibility of observing isolated accreting NSs.

\subsection{Non-thermal component of the X-ray emission}

{In addition to the thermal component of the spectrum that has been considered, there may also be non-thermal radiation from the hot corona surrounding accreting NSs \citep{2001ApJ...547..355P, 2007A&ARv..15....1D}.}

{
Studies of accretion processes in low-mass X-ray binaries (LMXBs) show that the corona can form, if the accretion disc is present \citep{2026MNRAS.545f2263M}. The contribution of the hard coronal emission and the emission of the boundary layer between the disc and the surface of the NS to the total flux depends on the accretion rate and is smaller when the accretion rate is low. The ISM  provides even lower accretion rates than in LMXBs, so we do not include the luminosity of the hot gas surrounding the disc in our calculations.}

{ 
Another reason for not taking the hot gas into account is that the number of isolated accreting NSs with discs is small. In isolated NSs, the disc could form due to the turbulence momentum from the ISM  \citep{2002A&A...381.1000P}. The accreted material from the ISM carries the angular momentum
\begin{equation}
    j_\text{t} = v_\text{t} R_\text{t}^{-1/3} R_\text{G}^{4/3}.
\end{equation}
If the turbulent momentum $j_\text{t}$ exceeds the momentum $j_\text{K}$ = $\sqrt{GM_\text{NS}R_\text{A}}$, then the disc forms. The lifetime of the disc could be estimated as a time required for the NS to pass through a convective cell of size $R_\text{G}$: $R_\text{G}/v \approx 11(v/10\text{~\kms})^{-3}$~yr. Applying the requirement $j_\text{t} \ge j_\text{K}$ for the disc formation to our calculations, we estimate that the number of NSs with accretion discs in the Galaxy is $\lesssim1$\% of the total number of accreting NSs. 
}

\subsection{Evolution of the Milky Way}

In our model, we assume that the Galactic gravitational potential has cylindrical symmetry and is constant in time. Also, selecting birth positions of NSs, we apply the same distributions for all NSs. 
Thus, we neglect changes in the Galactic structure over time and do not account for the bar contribution, Galactic wrap, and some other features.
Obviously, this is a simplification compared to the real situation.

Our initial spatial distribution of NSs mainly follows the present-day structure of the Galaxy. 
Thus, we do not consider the compact objects formed from Population II progenitors. However, during the early period of the Milky Way formation, the spatial distribution of stars was much different (mostly, more compact), e.g. \cite{2022MNRAS.514..689B}, 
\cite{2024ApJ...972..112C}. 
Still, the fraction of Population II stars is just about 10\%,
see \cite{2008A&ARv..15..145H}. 
In our calculations, only $\lesssim 10$\% of the present-day accreting NSs were formed
in the first billion years of the Galactic evolution. We conclude that neglect of details of the early starformation in the Galaxy might not modify our general results significantly, accounting for all other uncertainties. The number of accreting NSs in the central few kpc might show a stronger dependence on the assumptions about the initial spatial distribution. 
Still, as potentially detectable accretors are situated not that far from the Sun, the results presented in Sec.~3.3 might be specifically insensitive to the details of the early starformation distribution.

In the Galactic gravitational potential model that we used to calculate the trajectories of isolated NSs, we do not account for the bar. This structure originated about 8-10 Gyr ago
and 
its appearance resulted in stellar inward and outward migration \citep{2024A&A...690A.147H}. 
In addition, the presence of the bar results in resonances in the stellar motion \citep{2022A&A...663A..38K}. It also must complicate the relative motion of NSs and the ISM. Accounting for these effects is beyond the scope of our study.

\subsection{Neutron stars in wide low-mass binary systems}

Presently, there are about two dozen wide non-interacting low-mass binary systems with an NS as one of the companions discovered by astrometric methods in the {\it Gaia} data \citep{2024OJAp....7E..58E}.
Recently, \cite{2025arXiv250821805C} 
estimated that in the final {\it Gaia} catalogue there will be $\approx 1500$-5000 binary systems with NSs. This opens interesting possibilities for studying the long-term evolution of NSs under conditions similar to those in the ISM. 

The normal components in these systems are mainly Main Sequence stars with masses $\lesssim 1\, M_\odot$. Thus, the average ages of NSs in these binaries might be about a few Gyr. As systems are wide and components have small masses, the rate of matter outflow due to the stellar wind is moderate, with typical velocities of about a few hundred \kms. Then, the evolution of NSs in such binaries in many respects resembles the evolution of isolated NSs in the ISM \citep{2025NewA..11902401A}. 

Currently, NSs in astrometric binaries remain invisible. Therefore, it is impossible to estimate their parameters (spin, magnetic field, etc.), as well as their evolutionary status. Observations of these compact objects can also be very useful for our understanding of the isolated NSs evolution, as they can shed light on the long-term behavior of the magnetic field, the rate of spin-down on the propeller stage, and other evolutionary properties of NSs. 

As the wide binaries with NS components are well-localized, it is possible to perform deep observations of these systems at various wavelengths \citep{2024A&A...686A.299S, 2026NewA..12302500B} and to identify counterparts for yet-invisible compact objects. 
Thus, there is a significant possibility that weakly accreting NSs or NSs at the propeller stage in low-density environments will be first discovered in non-interacting binary systems. This would help a lot to advance the strategy of searching for isolated evolved NSs.

\subsection{Isolated neutron stars and long-period radio transients}

Recently, a new class of sources has been discovered -- long-period radio transients (LPRTs). Currently, it includes about a dozen sources; see e.g., a brief review in Sec.~3.9 of \cite{2025arXiv251110785M}. They are characterized by pulsed radio emission with periodicity from several minutes up to several hours. The nature of these sources is not known, and there is a possibility that they belong to at least two different groups. In two cases (GLEAM-X J0704-37, ILT J1101 + 5521), the sources are identified with low-mass binary systems containing white dwarfs. Thus, they can be similar to the well-known radio-emitting low-mass binaries AR Sco \citep{2016Natur.537..374M} and/or AE Aqr \cite{2022heas.confE..46M}. In several other cases, LPRTs can be isolated NSs, but the emission mechanism remains unknown. 

In \cite{2023AstL...49..553A, 2024PASA...41...14A} we analyzed the hypothesis that LPRTs can be relatively young long-period NSs. Very long spin periods can be gained, e.g., due to the interaction between the NS's magnetosphere and a fallback disc \citep{2022ApJ...934..184R}. This peculiar feature might influence the NS evolution significantly. In particular, we concluded that such sources have a high probability of becoming isolated accretors. 

In a recent paper, \cite{2025arXiv251115154F} 
proposed that some of LPRTs can actually be accreting isolated NSs even now. 
The author suggested that NSs with an accretion rate from the interstellar medium $\dot M\sim10^{10\pm1}$~g~s$^{-1}$ can be observed as radio emitters due to the electron cyclotron maser emission mechanism. This hypothesis is testable, as in the case of isolated accreting NSs a weak soft X-ray emission might be detectable. Also, if the sources demonstrate significant proper motion, then accretion-powered emission can be rejected, as $\dot M \propto v^{-3}$. 

\subsection{Neutron stars and microlensing}

Several isolated NS candidates have been discovered in microlensing observations, already \citep{2021AcA....71...89M}. In the near future, this number will increase due to ground-based and space observations. 
The expected number of lensing on NSs and BHs was calculated in several studies, see e.g., \cite{2010A&A...523A..33S}, 
\cite{2020ApJ...889...31L}, \cite{2024MNRAS.531.2433S} 
 and references therein. 

  \cite{2024MNRAS.531.2433S} estimated that $\sim15$-20 observable events (photometric and astrometric) with NSs as lenses can be detected in a year across the sky. Then, future missions such as {\it Gaia NIR} can detect many hundreds of isolated NSs. However, e.g., {\it Gaia} is expected to yield only a few NS lensing events even with the DR5 data. In addition, several tens of NS and BH lensing can be detected by {\it Roman} space telescope \citep{2023arXiv230612514L, 2023AJ....165...96S}.
  Despite their unimpressive statistics, these objects will be perfect targets for follow-up observations to detect emission from old NSs or to provide useful upper limits. Thus, it is useful to discuss the expected evolutionary status of such NSs. 

  The results of our study do not provide very optimistic estimates, as for distant isolated NSs, only in rare cases do we expect significant luminosity in any spectral range (see Sec.~3.3). Still, the precise localization of isolated NSs observed as dark lenses and subsequent deep observations are much appreciated. 





\section{Summary and conclusions}

 In this paper, we presented an advanced population synthesis of isolated neutron stars in the Milky Way. The most important novelty of our approach is the use of several models for the propeller spin-down. Uncertainties in the duration of the propeller stage are the main factor that prevents a precise estimation of the number of accreting isolated neutron stars. 
Our calculations show that in the case of persistent accretion with high efficiency, {\it eROSITA} can detect up to several thousand accreting isolated neutron stars as dim sources at count rates $\sim 0.01$~cts~s$^{-1}$ if the propeller spin-down is sufficiently rapid. Oppositely, inefficient loss of angular momentum by a neutron star at the propeller stage, prescribed by \cite{1981MNRAS.196..209D}, results in a very small number of accretors, which is virtually zero if we consider objects detectable by {\it eROSITA}. 

In addition to uncertainties in the propeller stage, modeling the observable population of isolated accretors also suffers from limited knowledge of the properties of the accretion regime at low rates. In the near future, deep observations of neutron stars in wide low-mass non-interacting binary systems might provide new information that is necessary for understanding the evolution of neutron stars in a low-density medium.

\section*{Acknowledgements}
The work of M.A. and S.P. was supported by the RSF grant 25-12-00012.
We thank the referee for useful comments and suggestions.



\bibliographystyle{elsarticle-harv} 
\bibliography{example}

@ARTICLE{2015ApJS..216...29B,
       author = {{Bovy}, Jo},
        title = "{galpy: A python Library for Galactic Dynamics}",
      journal = {\apjs},
         year = 2015,
        month = feb,
       volume = {216},
       number = {2},
          eid = {29},
        pages = {29},
          doi = {10.1088/0067-0049/216/2/29},
archivePrefix = {arXiv},
       eprint = {1412.3451},
 primaryClass = {astro-ph.GA},
       adsurl = {https://ui.adsabs.harvard.edu/abs/2015ApJS..216...29B}
}

@ARTICLE{1984ApJ...277..791N,
       author = {{Nomoto}, K.},
        title = "{Evolution of 8-10 solar mass stars toward electron capture supernovae. I - Formation of electron-degenerate O + NE + MG cores.}",
      journal = {\apj},
         year = 1984,
        month = feb,
       volume = {277},
        pages = {791-805},
          doi = {10.1086/161749},
       adsurl = {https://ui.adsabs.harvard.edu/abs/1984ApJ...277..791N}
}

@ARTICLE{2026NewA..12302500B,
       author = {{Brylyakova}, Elena and {Afonina}, Marina and {Tyul'basheva}, Gayane and {Popov}, Sergei B. and {Tyul'bashev}, Sergei},
        title = "{Low-frequency observations of low-mass binary systems with neutron star candidates}",
      journal = {\na},
         year = 2026,
        month = feb,
       volume = {123},
          eid = {102500},
        pages = {102500},
          doi = {10.1016/j.newast.2025.102500},
archivePrefix = {arXiv},
       eprint = {2509.03001},
 primaryClass = {astro-ph.HE},
       adsurl = {https://ui.adsabs.harvard.edu/abs/2026NewA..12302500B}
}

@ARTICLE{2002ApJ...574..364P,
       author = {{Pfahl}, Eric and {Rappaport}, Saul and {Podsiadlowski}, Philipp and {Spruit}, Hendrik},
        title = "{A New Class of High-Mass X-Ray Binaries: Implications for Core Collapse and Neutron Star Recoil}",
      journal = {\apj},
         year = 2002,
        month = jul,
       volume = {574},
       number = {1},
        pages = {364-376},
          doi = {10.1086/340794},
archivePrefix = {arXiv},
       eprint = {astro-ph/0109521},
 primaryClass = {astro-ph},
       adsurl = {https://ui.adsabs.harvard.edu/abs/2002ApJ...574..364P}
}

@ARTICLE{2025NewAR.10101734P,
       author = {{Popov}, Sergei and {M{\"u}ller}, Bernhard and {Mandel}, Ilya},
        title = "{Natal kicks of compact objects}",
      journal = {\nar},
         year = 2025,
        month = dec,
       volume = {101},
          eid = {101734},
        pages = {101734},
          doi = {10.1016/j.newar.2025.101734},
archivePrefix = {arXiv},
       eprint = {2509.01430},
 primaryClass = {astro-ph.HE},
       adsurl = {https://ui.adsabs.harvard.edu/abs/2025NewAR.10101734P}
}

@ARTICLE{1990ApJ...356..359H,
       author = {{Hernquist}, Lars},
        title = "{An Analytical Model for Spherical Galaxies and Bulges}",
      journal = {\apj},
         year = 1990,
        month = jun,
       volume = {356},
        pages = {359},
          doi = {10.1086/168845},
       adsurl = {https://ui.adsabs.harvard.edu/abs/1990ApJ...356..359H}
}

@ARTICLE{1975PASJ...27..533M,
       author = {{Miyamoto}, M. and {Nagai}, R.},
        title = "{Three-dimensional models for the distribution of mass in galaxies.}",
      journal = {\pasj},
         year = 1975,
        month = jan,
       volume = {27},
        pages = {533-543},
       adsurl = {https://ui.adsabs.harvard.edu/abs/1975PASJ...27..533M}
}

@ARTICLE{1999ApJ...517..906T,
       author = {{Toropin}, Yu. M. and {Toropina}, O.~D. and {Savelyev}, V.~V. and {Romanova}, M.~M. and {Chechetkin}, V.~M. and {Lovelace}, R.~V.~E.},
        title = "{Spherical Bondi Accretion onto a Magnetic Dipole}",
      journal = {\apj},
         year = 1999,
        month = jun,
       volume = {517},
       number = {2},
        pages = {906-918},
          doi = {10.1086/307229},
archivePrefix = {arXiv},
       eprint = {astro-ph/9811272},
 primaryClass = {astro-ph},
       adsurl = {https://ui.adsabs.harvard.edu/abs/1999ApJ...517..906T}
}

@ARTICLE{2001ApJ...561..964T,
       author = {{Toropina}, O.~D. and {Romanova}, M.~M. and {Toropin}, Yu. M. and {Lovelace}, R.~V.~E.},
        title = "{Propagation of Magnetized Neutron Stars through the Interstellar Medium}",
      journal = {\apj},
         year = 2001,
        month = nov,
       volume = {561},
       number = {2},
        pages = {964-979},
          doi = {10.1086/323233},
archivePrefix = {arXiv},
       eprint = {astro-ph/0105422},
 primaryClass = {astro-ph},
       adsurl = {https://ui.adsabs.harvard.edu/abs/2001ApJ...561..964T}
}

@ARTICLE{2003ApJ...593..472T,
       author = {{Toropina}, O.~D. and {Romanova}, M.~M. and {Toropin}, Yu. M. and {Lovelace}, R.~V.~E.},
        title = "{Magnetic Inhibition of Accretion and Observability of Isolated Old Neutron Stars}",
      journal = {\apj},
         year = 2003,
        month = aug,
       volume = {593},
       number = {1},
        pages = {472-480},
          doi = {10.1086/376447},
       adsurl = {https://ui.adsabs.harvard.edu/abs/2003ApJ...593..472T}
}

@ARTICLE{2012MNRAS.420..810T,
       author = {{Toropina}, O.~D. and {Romanova}, M.~M. and {Lovelace}, R.~V.~E.},
        title = "{Bondi-Hoyle accretion on to a magnetized neutron star}",
      journal = {\mnras},
         year = 2012,
        month = feb,
       volume = {420},
       number = {1},
        pages = {810-816},
          doi = {10.1111/j.1365-2966.2011.20093.x},
archivePrefix = {arXiv},
       eprint = {1111.2460},
 primaryClass = {astro-ph.HE},
       adsurl = {https://ui.adsabs.harvard.edu/abs/2012MNRAS.420..810T}
}

@ARTICLE{2003ApJ...594..936P,
       author = {{Perna}, Rosalba and {Narayan}, Ramesh and {Rybicki}, George and {Stella}, Luigi and {Treves}, Aldo},
        title = "{Bondi Accretion and the Problem of the Missing Isolated Neutron Stars}",
      journal = {\apj},
         year = 2003,
        month = sep,
       volume = {594},
       number = {2},
        pages = {936-942},
          doi = {10.1086/377091},
archivePrefix = {arXiv},
       eprint = {astro-ph/0305421},
 primaryClass = {astro-ph},
       adsurl = {https://ui.adsabs.harvard.edu/abs/2003ApJ...594..936P}
}

@ARTICLE{2006LRR.....9....6P,
       author = {{Postnov}, Konstantin A. and {Yungelson}, Lev R.},
        title = "{The Evolution of Compact Binary Star Systems}",
      journal = {Living Reviews in Relativity},
         year = 2006,
        month = dec,
       volume = {9},
       number = {1},
          eid = {6},
        pages = {6},
          doi = {10.12942/lrr-2006-6},
archivePrefix = {arXiv},
       eprint = {astro-ph/0701059},
 primaryClass = {astro-ph},
       adsurl = {https://ui.adsabs.harvard.edu/abs/2006LRR.....9....6P}
}

@ARTICLE{2006ApJ...643..332F,
       author = {{Faucher-Gigu{\`e}re}, Claude-Andr{\'e} and {Kaspi}, Victoria M.},
        title = "{Birth and Evolution of Isolated Radio Pulsars}",
      journal = {\apj},
         year = 2006,
        month = may,
       volume = {643},
       number = {1},
        pages = {332-355},
          doi = {10.1086/501516},
archivePrefix = {arXiv},
       eprint = {astro-ph/0512585},
 primaryClass = {astro-ph},
       adsurl = {https://ui.adsabs.harvard.edu/abs/2006ApJ...643..332F}
}

@ARTICLE{2017ApJ...835...29Y,
       author = {{Yao}, J.~M. and {Manchester}, R.~N. and {Wang}, N.},
        title = "{A New Electron-density Model for Estimation of Pulsar and FRB Distances}",
      journal = {\apj},
         year = 2017,
        month = jan,
       volume = {835},
       number = {1},
          eid = {29},
        pages = {29},
          doi = {10.3847/1538-4357/835/1/29},
archivePrefix = {arXiv},
       eprint = {1610.09448},
 primaryClass = {astro-ph.GA},
       adsurl = {https://ui.adsabs.harvard.edu/abs/2017ApJ...835...29Y}
}

@ARTICLE{2004A&A...422..545Y,
       author = {{Yusifov}, I. and {K{\"u}{\c{c}}{\"u}k}, I.},
        title = "{Revisiting the radial distribution of pulsars in the Galaxy}",
      journal = {\aap},
         year = 2004,
        month = aug,
       volume = {422},
        pages = {545-553},
          doi = {10.1051/0004-6361:20040152},
archivePrefix = {arXiv},
       eprint = {astro-ph/0405559},
 primaryClass = {astro-ph},
       adsurl = {https://ui.adsabs.harvard.edu/abs/2004A&A...422..545Y}
}

@ARTICLE{2021MNRAS.508.3345I,
       author = {{Igoshev}, Andrei P. and {Chruslinska}, Martyna and {Dorozsmai}, Andris and {Toonen}, Silvia},
        title = "{Combined analysis of neutron star natal kicks using proper motions and parallax measurements for radio pulsars and Be X-ray binaries}",
      journal = {\mnras},
         year = 2021,
        month = dec,
       volume = {508},
       number = {3},
        pages = {3345-3364},
          doi = {10.1093/mnras/stab2734},
archivePrefix = {arXiv},
       eprint = {2109.10362},
 primaryClass = {astro-ph.HE},
       adsurl = {https://ui.adsabs.harvard.edu/abs/2021MNRAS.508.3345I}
}

@ARTICLE{2022MNRAS.514.4606I,
       author = {{Igoshev}, Andrei P. and {Frantsuzova}, Anastasia and {Gourgouliatos}, Konstantinos N. and {Tsichli}, Savina and {Konstantinou}, Lydia and {Popov}, Sergei B.},
        title = "{Initial periods and magnetic fields of neutron stars}",
      journal = {\mnras},
         year = 2022,
        month = aug,
       volume = {514},
       number = {3},
        pages = {4606-4619},
          doi = {10.1093/mnras/stac1648},
archivePrefix = {arXiv},
       eprint = {2205.06823},
 primaryClass = {astro-ph.HE},
       adsurl = {https://ui.adsabs.harvard.edu/abs/2022MNRAS.514.4606I}
}

@ARTICLE{2006A&A...459..113M,
       author = {{Misiriotis}, A. and {Xilouris}, E.~M. and {Papamastorakis}, J. and {Boumis}, P. and {Goudis}, C.~D.},
        title = "{The distribution of the ISM in the Milky Way. A three-dimensional large-scale model}",
      journal = {\aap},
         year = 2006,
        month = nov,
       volume = {459},
       number = {1},
        pages = {113-123},
          doi = {10.1051/0004-6361:20054618},
archivePrefix = {arXiv},
       eprint = {astro-ph/0607638},
 primaryClass = {astro-ph},
       adsurl = {https://ui.adsabs.harvard.edu/abs/2006A&A...459..113M}
}

@ARTICLE{2024MNRAS.528L..38D,
       author = {{De}, Kishalay and {Daly}, Fiona A. and {Soria}, Roberto},
        title = "{Infrared spectroscopy of SWIFT J0850.8-4219: identification of the second red supergiant X-ray binary in the Milky Way}",
      journal = {\mnras},
         year = 2024,
        month = feb,
       volume = {528},
       number = {1},
        pages = {L38-L44},
          doi = {10.1093/mnrasl/slad164},
archivePrefix = {arXiv},
       eprint = {2309.07833},
 primaryClass = {astro-ph.SR},
       adsurl = {https://ui.adsabs.harvard.edu/abs/2024MNRAS.528L..38D}
}

@ARTICLE{1952MNRAS.112..195B,
       author = {{Bondi}, H.},
        title = "{On spherically symmetrical accretion}",
      journal = {\mnras},
         year = 1952,
        month = jan,
       volume = {112},
        pages = {195},
          doi = {10.1093/mnras/112.2.195},
       adsurl = {https://ui.adsabs.harvard.edu/abs/1952MNRAS.112..195B}
}

@ARTICLE{2024arXiv240210341P,
       author = {{Prust}, Logan J. and {Glanz}, Hila and {Bildsten}, Lars and {Perets}, Hagai B. and {Roepke}, Friedrich K.},
        title = "{Morphology and Mach Number Dependence of Subsonic Bondi-Hoyle Accretion}",
      journal = {arXiv e-prints},
         year = 2024,
        month = feb,
          eid = {arXiv:2402.10341},
        pages = {arXiv:2402.10341},
          doi = {10.48550/arXiv.2402.10341},
archivePrefix = {arXiv},
       eprint = {2402.10341},
 primaryClass = {astro-ph.HE},
       adsurl = {https://ui.adsabs.harvard.edu/abs/2024arXiv240210341P}
}

@ARTICLE{1997A&A...320..342F,
       author = {{Foglizzo}, T. and {Ruffert}, M.},
        title = "{An analytic study of Bondi-Hoyle-Lyttleton accretion. I. Stationary flows.}",
      journal = {\aap},
         year = 1997,
        month = apr,
       volume = {320},
        pages = {342-361},
          doi = {10.48550/arXiv.astro-ph/9604160},
archivePrefix = {arXiv},
       eprint = {astro-ph/9604160},
 primaryClass = {astro-ph},
       adsurl = {https://ui.adsabs.harvard.edu/abs/1997A&A...320..342F}
}

@ARTICLE{2015MNRAS.447.2817P,
       author = {{Popov}, S.~B. and {Postnov}, K.~A. and {Shakura}, N.~I.},
        title = "{Settling accretion on to isolated neutron stars from interstellar medium}",
      journal = {\mnras},
         year = 2015,
        month = mar,
       volume = {447},
       number = {3},
        pages = {2817-2820},
          doi = {10.1093/mnras/stu2643},
archivePrefix = {arXiv},
       eprint = {1412.4066},
 primaryClass = {astro-ph.HE},
       adsurl = {https://ui.adsabs.harvard.edu/abs/2015MNRAS.447.2817P}
}

@ARTICLE{2012MNRAS.420..216S,
       author = {{Shakura}, N. and {Postnov}, K. and {Kochetkova}, A. and {Hjalmarsdotter}, L.},
        title = "{Theory of quasi-spherical accretion in X-ray pulsars}",
      journal = {\mnras},
         year = 2012,
        month = feb,
       volume = {420},
       number = {1},
        pages = {216-236},
          doi = {10.1111/j.1365-2966.2011.20026.x},
archivePrefix = {arXiv},
       eprint = {1110.3701},
 primaryClass = {astro-ph.HE},
       adsurl = {https://ui.adsabs.harvard.edu/abs/2012MNRAS.420..216S}
}

@INPROCEEDINGS{2018IAUS..337..112P,
       author = {{Pires}, Adriana M.},
        title = "{What will eROSITA reveal among X-ray faint isolated neutron stars?}",
    booktitle = {Pulsar Astrophysics the Next Fifty Years},
         year = 2018,
       editor = {{Weltevrede}, P. and {Perera}, B.~B.~P. and {Preston}, L.~L. and {Sanidas}, S.},
       series = {IAU Symposium},
       volume = {337},
        month = aug,
        pages = {112-115},
          doi = {10.1017/S1743921317009590},
       adsurl = {https://ui.adsabs.harvard.edu/abs/2018IAUS..337..112P}
}

@ARTICLE{2023MNRAS.520.4315L,
       author = {{Lyutikov}, Maxim},
        title = "{Centrifugal barriers in magnetospheric accretion}",
      journal = {\mnras},
         year = 2023,
       volume = {520},
       number = {3},
        pages = {4315-4323},
          doi = {10.1093/mnras/stad284},
archivePrefix = {arXiv},
       eprint = {2210.00300},
 primaryClass = {astro-ph.HE},
       adsurl = {https://ui.adsabs.harvard.edu/abs/2023MNRAS.520.4315L}
}

@ARTICLE{1999PNAS...96.5351K,
       author = {{Kouveliotou}, Chryssa},
        title = "{Magnetars}",
      journal = {Proceedings of the National Academy of Science},
         year = 1999,
        month = may,
       volume = {96},
       number = {10},
        pages = {5351-5352},
          doi = {10.1073/pnas.96.10.5351},
       adsurl = {https://ui.adsabs.harvard.edu/abs/1999PNAS...96.5351K}
}

@ARTICLE{2010MNRAS.401.2675P,
       author = {{Popov}, S.~B. and {Pons}, J.~A. and {Miralles}, J.~A. and {Boldin}, P.~A. and {Posselt}, B.},
        title = "{Population synthesis studies of isolated neutron stars with magnetic field decay}",
      journal = {\mnras},
         year = 2010,
        month = feb,
       volume = {401},
       number = {4},
        pages = {2675-2686},
          doi = {10.1111/j.1365-2966.2009.15850.x},
archivePrefix = {arXiv},
       eprint = {0910.2190},
 primaryClass = {astro-ph.HE},
       adsurl = {https://ui.adsabs.harvard.edu/abs/2010MNRAS.401.2675P}
}

@ARTICLE{1975SvAL....1..223S,
       author = {{Shakura}, N.~I.},
        title = "{The long-period X-ray pulsar 3U 0900-40 as a neutron star with an abnormally strong magnetic field.}",
      journal = {Soviet Astronomy Letters},
         year = 1975,
        month = dec,
       volume = {1},
        pages = {223-225},
       adsurl = {https://ui.adsabs.harvard.edu/abs/1975SvAL....1..223S}
}

@ARTICLE{1979MNRAS.186..779D,
       author = {{Davies}, R.~E. and {Fabian}, A.~C. and {Pringle}, J.~E.},
        title = "{Spindown of neutron stars in close binary systems.}",
      journal = {\mnras},
         year = 1979,
        month = mar,
       volume = {186},
        pages = {779-782},
          doi = {10.1093/mnras/186.4.779},
       adsurl = {https://ui.adsabs.harvard.edu/abs/1979MNRAS.186..779D}
}

@ARTICLE{1981MNRAS.196..209D,
       author = {{Davies}, R.~E. and {Pringle}, J.~E.},
        title = "{Spindown of neutron stars in close binary systems - II.}",
      journal = {\mnras},
         year = 1981,
        month = jul,
       volume = {196},
        pages = {209-224},
          doi = {10.1093/mnras/196.2.209},
       adsurl = {https://ui.adsabs.harvard.edu/abs/1981MNRAS.196..209D}
}

@ARTICLE{1973ApJ...179..585D,
       author = {{Davidson}, Kris and {Ostriker}, Jeremiah P.},
        title = "{Neutron-Star Accretion in a Stellar Wind: Model for a Pulsed X-Ray Source}",
      journal = {\apj},
         year = 1973,
        month = jan,
       volume = {179},
        pages = {585-598},
          doi = {10.1086/151897},
       adsurl = {https://ui.adsabs.harvard.edu/abs/1973ApJ...179..585D}
}

@ARTICLE{1975AA....39..185I,
       author = {{Illarionov}, A.~F. and {Sunyaev}, R.~A.},
        title = "{Why the Number of Galactic X-ray Stars Is so Small?}",
      journal = {\aap},
         year = 1975,
        month = feb,
       volume = {39},
        pages = {185},
       adsurl = {https://ui.adsabs.harvard.edu/abs/1975A&A....39..185I}
}

@ARTICLE{1971MNRAS.154..141H,
       author = {{Hunt}, R.},
        title = "{A fluid dynamical study of the accretion process}",
      journal = {\mnras},
         year = 1971,
        month = jan,
       volume = {154},
        pages = {141},
          doi = {10.1093/mnras/154.2.141},
       adsurl = {https://ui.adsabs.harvard.edu/abs/1971MNRAS.154..141H}
}

@ARTICLE{1979MNRAS.188...83H,
       author = {{Hunt}, R.},
        title = "{Accretion of gas having specific heat ratio 3/3 by a moving gravitating body.}",
      journal = {\mnras},
         year = 1979,
        month = jul,
       volume = {188},
        pages = {83-91},
          doi = {10.1093/mnras/188.1.83},
       adsurl = {https://ui.adsabs.harvard.edu/abs/1979MNRAS.188...83H}
}

@ARTICLE{1985MNRAS.217..367S,
       author = {{Shima}, E. and {Matsuda}, T. and {Takeda}, H. and {Sawada}, K.},
        title = "{Hydrodynamic calculations of axisymmetric accretion flow}",
      journal = {\mnras},
         year = 1985,
        month = nov,
       volume = {217},
        pages = {367-386},
          doi = {10.1093/mnras/217.2.367},
       adsurl = {https://ui.adsabs.harvard.edu/abs/1985MNRAS.217..367S}
}

@ARTICLE{1994A&AS..106..505R,
       author = {{Ruffert}, M.},
        title = "{Three-dimensional hydrodynamic Bondi-Hoyle accretion. III. Mach 0.6, 1.4 and 10; {\ensuremath{\gamma}}=5/3.}",
      journal = {\aaps},
         year = 1994,
        month = sep,
       volume = {106},
        pages = {505-522},
       adsurl = {https://ui.adsabs.harvard.edu/abs/1994A&AS..106..505R}
}

@ARTICLE{2008PASA...25..184G,
       author = {{Gaensler}, B.~M. and {Madsen}, G.~J. and {Chatterjee}, S. and {Mao}, S.~A.},
        title = "{The Vertical Structure of Warm Ionised Gas in the Milky Way}",
      journal = {\pasa},
         year = 2008,
        month = nov,
       volume = {25},
       number = {4},
        pages = {184-200},
          doi = {10.1071/AS08004},
archivePrefix = {arXiv},
       eprint = {0808.2550},
 primaryClass = {astro-ph},
       adsurl = {https://ui.adsabs.harvard.edu/abs/2008PASA...25..184G}
}

@ARTICLE{Marasco2011-vd,
  title     = "Modelling the {H} i halo of the Milky Way",
  author    = "Marasco, A and Fraternali, F",
  journal   = "Astron. Astrophys.",
  publisher = "EDP Sciences",
  volume    =  525,
  pages     = "A134",
  month     =  jan,
  year      =  2011
}

@ARTICLE{2023ApJ...946...58C,
       author = {{Cook}, Amanda M. and {Bhardwaj}, Mohit and {Gaensler}, B.~M. and {Scholz}, Paul and {Eadie}, Gwendolyn M. and {Hill}, Alex S. and {Kaspi}, Victoria M. and {Masui}, Kiyoshi W. and {Curtin}, Alice P. and {Dong}, Fengqiu Adam and {Fonseca}, Emmanuel and {Herrera-Martin}, Antonio and {Kaczmarek}, Jane and {Lanman}, Adam E. and {Lazda}, Mattias and {Leung}, Calvin and {Meyers}, Bradley W. and {Michilli}, Daniele and {Pandhi}, Ayush and {Pearlman}, Aaron B. and {Pleunis}, Ziggy and {Ransom}, Scott and {Rahman}, Mubdi and {Sand}, Ketan R. and {Shin}, Kaitlyn and {Smith}, Kendrick and {Stairs}, Ingrid and {Stenning}, David C.},
        title = "{An FRB Sent Me a DM: Constraining the Electron Column of the Milky Way Halo with Fast Radio Burst Dispersion Measures from CHIME/FRB}",
      journal = {\apj},
         year = 2023,
        month = apr,
       volume = {946},
       number = {2},
          eid = {58},
        pages = {58},
          doi = {10.3847/1538-4357/acbbd0},
archivePrefix = {arXiv},
       eprint = {2301.03502},
 primaryClass = {astro-ph.GA},
       adsurl = {https://ui.adsabs.harvard.edu/abs/2023ApJ...946...58C}
}

@ARTICLE{2016ApJ...822...21H,
       author = {{Hodges-Kluck}, Edmund J. and {Miller}, Matthew J. and {Bregman}, Joel N.},
        title = "{The Rotation of the Hot Gas around the Milky Way}",
      journal = {\apj},
         year = 2016,
        month = may,
       volume = {822},
       number = {1},
          eid = {21},
        pages = {21},
          doi = {10.3847/0004-637X/822/1/21},
archivePrefix = {arXiv},
       eprint = {1603.07734},
 primaryClass = {astro-ph.GA},
       adsurl = {https://ui.adsabs.harvard.edu/abs/2016ApJ...822...21H}
}

@ARTICLE{2024A&A...681A..78L,
       author = {{Locatelli}, N. and {Ponti}, G. and {Zheng}, X. and {Merloni}, A. and {Becker}, W. and {Comparat}, J. and {Dennerl}, K. and {Freyberg}, M.~J. and {Sasaki}, M. and {Yeung}, M.~C.~H.},
        title = "{The warm-hot circumgalactic medium of the Milky Way as seen by eROSITA}",
      journal = {\aap},
         year = 2024,
        month = jan,
       volume = {681},
          eid = {A78},
        pages = {A78},
          doi = {10.1051/0004-6361/202347061},
archivePrefix = {arXiv},
       eprint = {2310.10715},
 primaryClass = {astro-ph.GA},
       adsurl = {https://ui.adsabs.harvard.edu/abs/2024A&A...681A..78L}
}

@ARTICLE{2016A&A...594A.116H,
       author = {{HI4PI Collaboration} and {Ben Bekhti}, N. and {Fl{\"o}er}, L. and {Keller}, R. and {Kerp}, J. and {Lenz}, D. and {Winkel}, B. and {Bailin}, J. and {Calabretta}, M.~R. and {Dedes}, L. and {Ford}, H.~A. and {Gibson}, B.~K. and {Haud}, U. and {Janowiecki}, S. and {Kalberla}, P.~M.~W. and {Lockman}, F.~J. and {McClure-Griffiths}, N.~M. and {Murphy}, T. and {Nakanishi}, H. and {Pisano}, D.~J. and {Staveley-Smith}, L.},
        title = "{HI4PI: A full-sky H I survey based on EBHIS and GASS}",
      journal = {\aap},
         year = 2016,
        month = oct,
       volume = {594},
          eid = {A116},
        pages = {A116},
          doi = {10.1051/0004-6361/201629178},
archivePrefix = {arXiv},
       eprint = {1610.06175},
 primaryClass = {astro-ph.GA},
       adsurl = {https://ui.adsabs.harvard.edu/abs/2016A&A...594A.116H}
}

@ARTICLE{2014MNRAS.441.1879P,
       author = {{Philippov}, Alexander and {Tchekhovskoy}, Alexander and {Li}, Jason G.},
        title = "{Time evolution of pulsar obliquity angle from 3D simulations of magnetospheres}",
      journal = {\mnras},
         year = 2014,
        month = jul,
       volume = {441},
       number = {3},
        pages = {1879-1887},
          doi = {10.1093/mnras/stu591},
archivePrefix = {arXiv},
       eprint = {1311.1513},
 primaryClass = {astro-ph.HE},
       adsurl = {https://ui.adsabs.harvard.edu/abs/2014MNRAS.441.1879P}
}

@BOOK{1992ans..book.....L,
       author = {{Lipunov}, Vladimir M.},
        title = "{Astrophysics of Neutron Stars}",
         year = 1992,
         series={Astronomy and Astrophysics Library},
        publisher={Springer-Verlag},
         address   = {Berlin  Heidelberg},
        isbn  = {978-3-642-76352-6},
       adsurl = {https://ui.adsabs.harvard.edu/abs/1992ans..book.....L}
}

@ARTICLE{2024Galax..12....7A,
       author = {{Abolmasov}, Pavel and {Biryukov}, Anton and {Popov}, Sergei B.},
        title = "{Spin Evolution of Neutron Stars}",
      journal = {Galaxies},
         year = 2024,
       volume = {12},
       number = {1},
          eid = {7},
        pages = {7},
archivePrefix = {arXiv},
       eprint = {2402.04331},
 primaryClass = {astro-ph.HE},
       adsurl = {https://ui.adsabs.harvard.edu/abs/2024Galax..12....7A}
}

@ARTICLE{2023Univ....9..273P,
       author = {{Popov}, Sergei B.},
        title = "{The Zoo of Isolated Neutron Stars}",
      journal = {Universe},
         year = 2023,
       volume = {9},
       number = {6},
          eid = {273},
        pages = {273},
          doi = {10.3390/universe9060273},
archivePrefix = {arXiv},
       eprint = {2306.02084},
 primaryClass = {astro-ph.HE},
       adsurl = {https://ui.adsabs.harvard.edu/abs/2023Univ....9..273P}
}

@ARTICLE{2021Univ....7..351I,
       author = {{Igoshev}, Andrei P. and {Popov}, Sergei B. and {Hollerbach}, Rainer},
        title = "{Evolution of Neutron Star Magnetic Fields}",
      journal = {Universe},
         year = 2021,
        month = sep,
       volume = {7},
       number = {9},
          eid = {351},
        pages = {351},
          doi = {10.3390/universe7090351},
archivePrefix = {arXiv},
       eprint = {2109.05584},
 primaryClass = {astro-ph.HE},
       adsurl = {https://ui.adsabs.harvard.edu/abs/2021Univ....7..351I}
}

@ARTICLE{2014PhRvL.112q1101G,
       author = {{Gourgouliatos}, Konstantinos N. and {Cumming}, Andrew},
        title = "{Hall Attractor in Axially Symmetric Magnetic Fields in Neutron Star Crusts}",
      journal = {\prl},
         year = 2014,
        month = may,
       volume = {112},
       number = {17},
          eid = {171101},
        pages = {171101},
          doi = {10.1103/PhysRevLett.112.171101},
archivePrefix = {arXiv},
       eprint = {1311.7345},
 primaryClass = {astro-ph.SR},
       adsurl = {https://ui.adsabs.harvard.edu/abs/2014PhRvL.112q1101G}
}

@ARTICLE{2014MNRAS.438.1618G,
       author = {{Gourgouliatos}, K.~N. and {Cumming}, A.},
        title = "{Hall effect in neutron star crusts: evolution, endpoint and dependence on initial conditions}",
      journal = {\mnras},
         year = 2014,
        month = feb,
       volume = {438},
       number = {2},
        pages = {1618-1629},
          doi = {10.1093/mnras/stt2300},
archivePrefix = {arXiv},
       eprint = {1311.7004},
 primaryClass = {astro-ph.SR},
       adsurl = {https://ui.adsabs.harvard.edu/abs/2014MNRAS.438.1618G}
}

@ARTICLE{1994Natur.369..127L,
       author = {{Lyne}, A.~G. and {Lorimer}, D.~R.},
        title = "{High birth velocities of radio pulsars}",
      journal = {\nat},
         year = 1994,
        month = may,
       volume = {369},
       number = {6476},
        pages = {127-129},
          doi = {10.1038/369127a0},
       adsurl = {https://ui.adsabs.harvard.edu/abs/1994Natur.369..127L}
}

@ARTICLE{2019MNRAS.487.1426B,
       author = {{Beniamini}, Paz and {Hotokezaka}, Kenta and {van der Horst}, Alexander and {Kouveliotou}, Chryssa},
        title = "{Formation rates and evolution histories of magnetars}",
      journal = {\mnras},
         year = 2019,
        month = jul,
       volume = {487},
       number = {1},
        pages = {1426-1438},
          doi = {10.1093/mnras/stz1391},
archivePrefix = {arXiv},
       eprint = {1903.06718},
 primaryClass = {astro-ph.HE},
       adsurl = {https://ui.adsabs.harvard.edu/abs/2019MNRAS.487.1426B}
}

@ARTICLE{2025ApJ...980..226T,
       author = {{Tejeda}, Emilio and {Toal{\'a}}, Jes{\'u}s A.},
        title = "{Geometric Correction for Wind Accretion in Binary Systems}",
      journal = {\apj},
         year = 2025,
        month = feb,
       volume = {980},
       number = {2},
          eid = {226},
        pages = {226},
          doi = {10.3847/1538-4357/ada953},
archivePrefix = {arXiv},
       eprint = {2411.01755},
 primaryClass = {astro-ph.HE},
       adsurl = {https://ui.adsabs.harvard.edu/abs/2025ApJ...980..226T}
}

@ARTICLE{1983ApJ...270..119M,
       author = {{Morrison}, R. and {McCammon}, D.},
        title = "{Interstellar photoelectric absorption cross sections, 0.03-10 keV.}",
      journal = {\apj},
         year = 1983,
        month = jul,
       volume = {270},
        pages = {119-122},
          doi = {10.1086/161102},
       adsurl = {https://ui.adsabs.harvard.edu/abs/1983ApJ...270..119M}
}

@ARTICLE{2016A&A...589A..66H,
       author = {{Haywood}, M. and {Lehnert}, M.~D. and {Di Matteo}, P. and {Snaith}, O. and {Schultheis}, M. and {Katz}, D. and {G{\'o}mez}, A.},
        title = "{When the Milky Way turned off the lights: APOGEE provides evidence of star formation quenching in our Galaxy}",
      journal = {\aap},
         year = 2016,
        month = may,
       volume = {589},
          eid = {A66},
        pages = {A66},
          doi = {10.1051/0004-6361/201527567},
archivePrefix = {arXiv},
       eprint = {1601.03042},
 primaryClass = {astro-ph.GA},
       adsurl = {https://ui.adsabs.harvard.edu/abs/2016A&A...589A..66H}
}

@ARTICLE{1998astro.ph..4047S,
       author = {{Sahrling}, Mikael},
        title = "{Ohmic Decay of Magnetic Fields due to non-spherical accretion in the Crusts of Neutron Stars}",
      journal = {arXiv e-prints},
         year = 1998,
        month = apr,
          eid = {astro-ph/9804047},
        pages = {astro-ph/9804047},
          doi = {10.48550/arXiv.astro-ph/9804047},
archivePrefix = {arXiv},
       eprint = {astro-ph/9804047},
 primaryClass = {astro-ph},
       adsurl = {https://ui.adsabs.harvard.edu/abs/1998astro.ph..4047S}
}

@ARTICLE{2008A&ARv..15..145H,
       author = {{Helmi}, Amina},
        title = "{The stellar halo of the Galaxy}",
      journal = {\aapr},
         year = 2008,
        month = jun,
       volume = {15},
       number = {3},
        pages = {145-188},
          doi = {10.1007/s00159-008-0009-6},
archivePrefix = {arXiv},
       eprint = {0804.0019},
 primaryClass = {astro-ph},
       adsurl = {https://ui.adsabs.harvard.edu/abs/2008A&ARv..15..145H}
}

@ARTICLE{2022MNRAS.514..689B,
       author = {{Belokurov}, Vasily and {Kravtsov}, Andrey},
        title = "{From dawn till disc: Milky Way's turbulent youth revealed by the APOGEE+Gaia data}",
      journal = {\mnras},
         year = 2022,
        month = jul,
       volume = {514},
       number = {1},
        pages = {689-714},
          doi = {10.1093/mnras/stac1267},
archivePrefix = {arXiv},
       eprint = {2203.04980},
 primaryClass = {astro-ph.GA},
       adsurl = {https://ui.adsabs.harvard.edu/abs/2022MNRAS.514..689B}
}

@ARTICLE{2024ApJ...972..112C,
       author = {{Chandra}, Vedant and {Semenov}, Vadim A. and {Rix}, Hans-Walter and {Conroy}, Charlie and {Bonaca}, Ana and {Naidu}, Rohan P. and {Andrae}, Ren{\'e} and {Li}, Jiadong and {Hernquist}, Lars},
        title = "{The Three-phase Evolution of the Milky Way}",
      journal = {\apj},
         year = 2024,
        month = sep,
       volume = {972},
       number = {1},
          eid = {112},
        pages = {112},
          doi = {10.3847/1538-4357/ad5b60},
archivePrefix = {arXiv},
       eprint = {2310.13050},
 primaryClass = {astro-ph.GA},
       adsurl = {https://ui.adsabs.harvard.edu/abs/2024ApJ...972..112C}
}

@ARTICLE{2025arXiv251110785M,
       author = {{Murphy}, Tara and {Kaplan}, David L.},
        title = "{The Dawes Review 13: A New Look at The Dynamic Radio Sky}",
      journal = {arXiv e-prints},
         year = 2025,
        month = nov,
          eid = {arXiv:2511.10785},
        pages = {arXiv:2511.10785},
          doi = {10.48550/arXiv.2511.10785},
archivePrefix = {arXiv},
       eprint = {2511.10785},
 primaryClass = {astro-ph.SR},
       adsurl = {https://ui.adsabs.harvard.edu/abs/2025arXiv251110785M}
}

@ARTICLE{2016Natur.537..374M,
       author = {{Marsh}, T.~R. and {G{\"a}nsicke}, B.~T. and {H{\"u}mmerich}, S. and {Hambsch}, F. -J. and {Bernhard}, K. and {Lloyd}, C. and {Breedt}, E. and {Stanway}, E.~R. and {Steeghs}, D.~T. and {Parsons}, S.~G. and {Toloza}, O. and {Schreiber}, M.~R. and {Jonker}, P.~G. and {van Roestel}, J. and {Kupfer}, T. and {Pala}, A.~F. and {Dhillon}, V.~S. and {Hardy}, L.~K. and {Littlefair}, S.~P. and {Aungwerojwit}, A. and {Arjyotha}, S. and {Koester}, D. and {Bochinski}, J.~J. and {Haswell}, C.~A. and {Frank}, P. and {Wheatley}, P.~J.},
        title = "{A radio-pulsing white dwarf binary star}",
      journal = {\nat},
         year = 2016,
        month = sep,
       volume = {537},
       number = {7620},
        pages = {374-377},
          doi = {10.1038/nature18620},
archivePrefix = {arXiv},
       eprint = {1607.08265},
 primaryClass = {astro-ph.SR},
       adsurl = {https://ui.adsabs.harvard.edu/abs/2016Natur.537..374M}
}

@ARTICLE{2022A&A...663A..38K,
       author = {{Khoperskov}, Sergey and {Gerhard}, Ortwin},
        title = "{Chemo-kinematics of the Milky Way spiral arms and bar resonances: Connection to ridges and moving groups in the solar vicinity}",
      journal = {\aap},
         year = 2022,
        month = jul,
       volume = {663},
          eid = {A38},
        pages = {A38},
          doi = {10.1051/0004-6361/202141836},
archivePrefix = {arXiv},
       eprint = {2111.15211},
 primaryClass = {astro-ph.GA},
       adsurl = {https://ui.adsabs.harvard.edu/abs/2022A&A...663A..38K}
}

@ARTICLE{2024A&A...690A.147H,
       author = {{Haywood}, Misha and {Khoperskov}, Sergey and {Cerqui}, Valeria and {Di Matteo}, Paola and {Katz}, David and {Snaith}, Owain},
        title = "{Timing the Milky Way bar formation and the accompanying radial migration episode}",
      journal = {\aap},
         year = 2024,
        month = oct,
       volume = {690},
          eid = {A147},
        pages = {A147},
          doi = {10.1051/0004-6361/202348767},
archivePrefix = {arXiv},
       eprint = {2403.08963},
 primaryClass = {astro-ph.GA},
       adsurl = {https://ui.adsabs.harvard.edu/abs/2024A&A...690A.147H}
}

@ARTICLE{2022ApJ...934..184R,
       author = {{Ronchi}, M. and {Rea}, N. and {Graber}, V. and {Hurley-Walker}, N.},
        title = "{Long-period Pulsars as Possible Outcomes of Supernova Fallback Accretion}",
      journal = {\apj},
         year = 2022,
        month = aug,
       volume = {934},
       number = {2},
          eid = {184},
        pages = {184},
          doi = {10.3847/1538-4357/ac7cec},
archivePrefix = {arXiv},
       eprint = {2201.11704},
 primaryClass = {astro-ph.HE},
       adsurl = {https://ui.adsabs.harvard.edu/abs/2022ApJ...934..184R}
}

@INPROCEEDINGS{2022heas.confE..46M,
       author = {{Madzime}, S.~T. and {Meintjes}, P. and {van Heerden}, H. and {Singh}, K.~K. and {Buckley}, D. and {Woudt}, P.~A. and {Fender}, R.},
        title = "{The detection of pulsed emission at the spin period of the white dwarf in AE Aquarii in MeerKAT and Fermi-LAT data}",
    booktitle = {High Energy Astrophysics in Southern Africa 2021},
         year = 2022,
        month = may,
          eid = {46},
        pages = {46},
          doi = {10.22323/1.401.0046},
       adsurl = {https://ui.adsabs.harvard.edu/abs/2022heas.confE..46M}
}

@ARTICLE{2023AstL...49..553A,
       author = {{Afonina}, M.~D. and {Biryukov}, A.~V. and {Popov}, S.~B.},
        title = "{Evolutionary Status of Long-Period Radio Pulsars}",
      journal = {Astronomy Letters},
         year = 2023,
        month = nov,
       volume = {49},
       number = {10},
        pages = {553-559},
          doi = {10.1134/S1063773723090013},
archivePrefix = {arXiv},
       eprint = {2309.12080},
 primaryClass = {astro-ph.HE},
       adsurl = {https://ui.adsabs.harvard.edu/abs/2023AstL...49..553A}
}

@ARTICLE{2025arXiv251115154F,
       author = {{Ferrario}, Lilia},
        title = "{Electron Cyclotron Maser Emission as the Driving Mechanism in Long-Period Radio Transients}",
      journal = {arXiv e-prints},
         year = 2025,
        month = nov,
          eid = {arXiv:2511.15154},
        pages = {arXiv:2511.15154},
          doi = {10.48550/arXiv.2511.15154},
archivePrefix = {arXiv},
       eprint = {2511.15154},
 primaryClass = {astro-ph.HE},
       adsurl = {https://ui.adsabs.harvard.edu/abs/2025arXiv251115154F}
}

@ARTICLE{2000PASP..112..297T,
       author = {{Treves}, Aldo and {Turolla}, Roberto and {Zane}, Silvia and {Colpi}, Monica},
        title = "{Isolated Neutron Stars: Accretors and Coolers}",
      journal = {\pasp},
         year = 2000,
        month = mar,
       volume = {112},
       number = {769},
        pages = {297-314},
          doi = {10.1086/316529},
archivePrefix = {arXiv},
       eprint = {astro-ph/9911430},
 primaryClass = {astro-ph},
       adsurl = {https://ui.adsabs.harvard.edu/abs/2000PASP..112..297T}
}

@ARTICLE{2015AN....336..831I,
       author = {{Igoshev}, A.~P. and {Popov}, S.~B.},
        title = "{Magnetic field decay in normal radio pulsars}",
      journal = {Astronomische Nachrichten},
         year = 2015,
        month = nov,
       volume = {336},
       number = {8-9},
        pages = {831},
          doi = {10.1002/asna.201512232},
archivePrefix = {arXiv},
       eprint = {1507.07962},
 primaryClass = {astro-ph.HE},
       adsurl = {https://ui.adsabs.harvard.edu/abs/2015AN....336..831I}
}

@ARTICLE{2024OJAp....7E..58E,
       author = {{El-Badry}, Kareem and {Rix}, Hans-Walter and {Latham}, David W. and {Shahaf}, Sahar and {Mazeh}, Tsevi and {Bieryla}, Allyson and {Buchhave}, Lars A. and {Andrae}, Ren{\'e} and {Yamaguchi}, Natsuko and {Isaacson}, Howard and {Howard}, Andrew W. and {Savino}, Alessandro and {Ilyin}, Ilya V.},
        title = "{A population of neutron star candidates in wide orbits from Gaia astrometry}",
      journal = {The Open Journal of Astrophysics},
         year = 2024,
       volume = {7},
          eid = {58},
        pages = {58},
       adsurl = {https://ui.adsabs.harvard.edu/abs/2024OJAp....7E..58E}
}

@ARTICLE{2024A&A...686A.299S,
       author = {{Sbarufatti}, B. and {Coti Zelati}, F. and {Marino}, A. and {Mereghetti}, S. and {Rea}, N. and {Treves}, A.},
        title = "{Swift X-ray and UV observations of six Gaia binaries supposedly containing a neutron star}",
      journal = {\aap},
         year = 2024,
       volume = {686},
          eid = {A299},
        pages = {A299},
archivePrefix = {arXiv},
       eprint = {2404.16170},
 primaryClass = {astro-ph.HE},
       adsurl = {https://ui.adsabs.harvard.edu/abs/2024A&A...686A.299S}
}

@ARTICLE{2024Univ...10..205A,
       author = {{Afonina}, Marina D. and {Popov}, Sergei B.},
        title = "{Probing the Propeller Regime with Symbiotic X-ray Binaries}",
      journal = {Universe},
         year = 2024,
        month = may,
       volume = {10},
       number = {5},
          eid = {205},
        pages = {205},
          doi = {10.3390/universe10050205},
archivePrefix = {arXiv},
       eprint = {2404.17549},
 primaryClass = {astro-ph.HE},
       adsurl = {https://ui.adsabs.harvard.edu/abs/2024Univ...10..205A}
}

@ARTICLE{1996A&A...309..469Z,
       author = {{Zane}, S. and {Zampieri}, L. and {Turolla}, R. and {Treves}, A.},
        title = "{The observability of old neutron stars accreting the interstellar medium. III. The solar vicinity.}",
      journal = {\aap},
         year = 1996,
        month = may,
       volume = {309},
        pages = {469-473},
          doi = {10.48550/arXiv.astro-ph/9509094},
archivePrefix = {arXiv},
       eprint = {astro-ph/9509094},
 primaryClass = {astro-ph},
       adsurl = {https://ui.adsabs.harvard.edu/abs/1996A&A...309..469Z}
}

@ARTICLE{2021ARep...65..615K,
       author = {{Khokhryakova}, A.~D. and {Biryukov}, A.~V. and {Popov}, S.~B.},
        title = "{Observability of Single Neutron Stars at SRG/eROSITA}",
      journal = {Astronomy Reports},
         year = 2021,
        month = jul,
       volume = {65},
       number = {7},
        pages = {615-630},
          doi = {10.1134/S1063772921080060},
       adsurl = {https://ui.adsabs.harvard.edu/abs/2021ARep...65..615K}
}

@ARTICLE{2000ApJ...530..896P,
       author = {{Popov}, S.~B. and {Colpi}, M. and {Treves}, A. and {Turolla}, R. and {Lipunov}, V.~M. and {Prokhorov}, M.~E.},
        title = "{The Neutron Star Census}",
      journal = {\apj},
         year = 2000,
        month = feb,
       volume = {530},
       number = {2},
        pages = {896-903},
          doi = {10.1086/308408},
archivePrefix = {arXiv},
       eprint = {astro-ph/9910114},
 primaryClass = {astro-ph},
       adsurl = {https://ui.adsabs.harvard.edu/abs/2000ApJ...530..896P}
}

@ARTICLE{2000ApJ...544L..53P,
       author = {{Popov}, S.~B. and {Colpi}, M. and {Prokhorov}, M.~E. and {Treves}, A. and {Turolla}, R.},
        title = "{The LOG N-LOG S Distributions of Accreting and Cooling Isolated Neutron Stars}",
      journal = {\apjl},
         year = 2000,
        month = nov,
       volume = {544},
       number = {1},
        pages = {L53-L56},
          doi = {10.1086/317295},
archivePrefix = {arXiv},
       eprint = {astro-ph/0009225},
 primaryClass = {astro-ph},
       adsurl = {https://ui.adsabs.harvard.edu/abs/2000ApJ...544L..53P}
}

@ARTICLE{2004NewAR..48..843E,
       author = {{Edgar}, Richard},
        title = "{A review of Bondi-Hoyle-Lyttleton accretion}",
      journal = {\nar},
         year = 2004,
        month = sep,
       volume = {48},
       number = {10},
        pages = {843-859},
          doi = {10.1016/j.newar.2004.06.001},
archivePrefix = {arXiv},
       eprint = {astro-ph/0406166},
 primaryClass = {astro-ph},
       adsurl = {https://ui.adsabs.harvard.edu/abs/2004NewAR..48..843E}
}

@ARTICLE{2021MNRAS.505.4036S,
       author = {{Scarcella}, Francesca and {Gaggero}, Daniele and {Connors}, Riley and {Manshanden}, Julien and {Ricotti}, Massimo and {Bertone}, Gianfranco},
        title = "{Multiwavelength detectability of isolated black holes in the Milky Way}",
      journal = {\mnras},
         year = 2021,
        month = aug,
       volume = {505},
       number = {3},
        pages = {4036-4047},
          doi = {10.1093/mnras/stab1533},
archivePrefix = {arXiv},
       eprint = {2012.10421},
 primaryClass = {astro-ph.HE},
       adsurl = {https://ui.adsabs.harvard.edu/abs/2021MNRAS.505.4036S}
}

@ARTICLE{2025NewA..11902401A,
       author = {{Afonina}, Marina and {Popov}, Sergei},
        title = "{Evolution of neutron stars in wide eccentric low-mass binary systems}",
      journal = {\na},
         year = 2025,
        month = oct,
       volume = {119},
          eid = {102401},
        pages = {102401},
          doi = {10.1016/j.newast.2025.102401},
archivePrefix = {arXiv},
       eprint = {2501.15918},
 primaryClass = {astro-ph.HE},
       adsurl = {https://ui.adsabs.harvard.edu/abs/2025NewA..11902401A}
}

@ARTICLE{2024PASA...41...14A,
       author = {{Afonina}, M.~D. and {Biryukov}, A.~V. and {Popov}, S.~B.},
        title = "{Early accretion onset in long-period isolated pulsars}",
      journal = {\pasa},
         year = 2024,
        month = mar,
       volume = {41},
          eid = {e014},
        pages = {e014},
          doi = {10.1017/pasa.2024.12},
archivePrefix = {arXiv},
       eprint = {2310.14844},
 primaryClass = {astro-ph.HE},
       adsurl = {https://ui.adsabs.harvard.edu/abs/2024PASA...41...14A}
}

@ARTICLE{2002A&A...381.1000P,
       author = {{Prokhorov}, M.~E. and {Popov}, S.~B. and {Khoperskov}, A.~V.},
        title = "{The period distribution of old accreting isolated neutron stars}",
      journal = {\aap},
         year = 2002,
        month = jan,
       volume = {381},
        pages = {1000-1006},
          doi = {10.1051/0004-6361:20011529},
archivePrefix = {arXiv},
       eprint = {astro-ph/0108503},
 primaryClass = {astro-ph},
       adsurl = {https://ui.adsabs.harvard.edu/abs/2002A&A...381.1000P}
}

@ARTICLE{1998A&A...331..535P,
       author = {{Popov}, S.~B. and {Prokhorov}, M.~E.},
        title = "{Spatial distribution of the accretion luminosity of isolated neutron stars and black holes in the Galaxy}",
      journal = {\aap},
         year = 1998,
        month = mar,
       volume = {331},
        pages = {535-540},
          doi = {10.48550/arXiv.astro-ph/9705236},
archivePrefix = {arXiv},
       eprint = {astro-ph/9705236},
 primaryClass = {astro-ph},
       adsurl = {https://ui.adsabs.harvard.edu/abs/1998A&A...331..535P}
}

@ARTICLE{2015ARep...59..645S,
       author = {{Shakura}, N.~I. and {Postnov}, K.~A. and {Kochetkova}, A. Yu. and {Hjalmarsdotter}, L. and {Sidoli}, L. and {Paizis}, A.},
        title = "{Wind accretion: Theory and observations}",
      journal = {Astronomy Reports},
         year = 2015,
        month = jul,
       volume = {59},
       number = {7},
        pages = {645-655},
          doi = {10.1134/S1063772915070112},
archivePrefix = {arXiv},
       eprint = {1407.3163},
 primaryClass = {astro-ph.HE},
       adsurl = {https://ui.adsabs.harvard.edu/abs/2015ARep...59..645S}
}

@ARTICLE{1970ApL.....6..179O,
       author = {{Ostriker}, J.~P. and {Rees}, M.~J. and {Silk}, J.},
        title = "{Some Observable Consequences of Accretion by Defunct Pulsars}",
      journal = {\aplett},
         year = 1970,
        month = jul,
       volume = {6},
        pages = {179},
       adsurl = {https://ui.adsabs.harvard.edu/abs/1970ApL.....6..179O}
}

@ARTICLE{1970AZh....47..824S,
       author = {{Shvartsman}, V.~G.},
        title = "{Ionization Zones around Neutron Stars: H{\ensuremath{\alpha}} Emission, Heating of the Interstellar Medium, and the Influence on Accretion.}",
      journal = {\azh},
         year = 1970,
        month = aug,
       volume = {47},
        pages = {824},
       adsurl = {https://ui.adsabs.harvard.edu/abs/1970AZh....47..824S}
}

@ARTICLE{1991A&A...241..107T,
       author = {{Treves}, A. and {Colpi}, M.},
        title = "{The observability of old isolated neutron stars.}",
      journal = {\aap},
         year = 1991,
        month = jan,
       volume = {241},
        pages = {107},
       adsurl = {https://ui.adsabs.harvard.edu/abs/1991A&A...241..107T}
}

@ARTICLE{2010MNRAS.407.1090B,
       author = {{Boldin}, P.~A. and {Popov}, S.~B.},
        title = "{The evolution of isolated neutron stars until accretion: the role of the initial magnetic field}",
      journal = {\mnras},
         year = 2010,
        month = sep,
       volume = {407},
       number = {2},
        pages = {1090-1097},
          doi = {10.1111/j.1365-2966.2010.16910.x},
archivePrefix = {arXiv},
       eprint = {1004.4805},
 primaryClass = {astro-ph.HE},
       adsurl = {https://ui.adsabs.harvard.edu/abs/2010MNRAS.407.1090B}
}

@ARTICLE{1998A&AS..128..349D,
       author = {{Danner}, R.},
        title = "{Searching for old neutron stars with ROSAT. II. Soft X-ray sources in galactic dark clouds}",
      journal = {\aaps},
         year = 1998,
        month = mar,
       volume = {128},
        pages = {349-357},
          doi = {10.1051/aas:1998384},
       adsurl = {https://ui.adsabs.harvard.edu/abs/1998A&AS..128..349D}
}

@ARTICLE{1997A&A...319..525B,
       author = {{Belloni}, T. and {Zampieri}, L. and {Campana}, S.},
        title = "{Search for old neutron stars in molecular clouds: Cygnus rift and Cygnus OB7.}",
      journal = {\aap},
         year = 1997,
        month = mar,
       volume = {319},
        pages = {525-534},
          doi = {10.48550/arXiv.astro-ph/9611087},
archivePrefix = {arXiv},
       eprint = {astro-ph/9611087},
 primaryClass = {astro-ph},
       adsurl = {https://ui.adsabs.harvard.edu/abs/1997A&A...319..525B}
}

@ARTICLE{1993A&A...269..319T,
       author = {{Treves}, A. and {Colpi}, M. and {Lipunov}, V.~M.},
        title = "{Old isolated neutron stars - Fire burns and cauldron bubbles}",
      journal = {\aap},
         year = 1993,
        month = mar,
       volume = {269},
       number = {1-2},
        pages = {319-324},
       adsurl = {https://ui.adsabs.harvard.edu/abs/1993A&A...269..319T}
}

@ARTICLE{1996MNRAS.278..577M,
       author = {{Manning}, R.~A. and {Jeffries}, R.~D. and {Willmore}, A.~P.},
        title = "{Are there any isolated old neutron stars in the ROSAT Wide Field Camera survey?}",
      journal = {\mnras},
         year = 1996,
        month = jan,
       volume = {278},
       number = {2},
        pages = {577-585},
          doi = {10.1093/mnras/278.2.577},
       adsurl = {https://ui.adsabs.harvard.edu/abs/1996MNRAS.278..577M}
}

@ARTICLE{2021NewA...8301498R,
       author = {{Rozwadowska}, Karolina and {Vissani}, Francesco and {Cappellaro}, Enrico},
        title = "{On the rate of core collapse supernovae in the milky way}",
      journal = {\na},
         year = 2021,
        month = feb,
       volume = {83},
          eid = {101498},
        pages = {101498},
          doi = {10.1016/j.newast.2020.101498},
archivePrefix = {arXiv},
       eprint = {2009.03438},
 primaryClass = {astro-ph.HE},
       adsurl = {https://ui.adsabs.harvard.edu/abs/2021NewA...8301498R}
}

@ARTICLE{2005AJ....129.1993M,
       author = {{Manchester}, R.~N. and {Hobbs}, G.~B. and {Teoh}, A. and {Hobbs}, M.},
        title = "{The Australia Telescope National Facility Pulsar Catalogue}",
      journal = {\aj},
         year = 2005,
        month = apr,
       volume = {129},
       number = {4},
        pages = {1993-2006},
          doi = {10.1086/428488},
archivePrefix = {arXiv},
       eprint = {astro-ph/0412641},
 primaryClass = {astro-ph},
       adsurl = {https://ui.adsabs.harvard.edu/abs/2005AJ....129.1993M}
}

@ARTICLE{1994ApJ...423..748M,
       author = {{Madau}, Piero and {Blaes}, Omer},
        title = "{Constraints on Accreting, Isolated Neutron Stars from the ROSAT and EUVE Surveys}",
      journal = {\apj},
         year = 1994,
        month = mar,
       volume = {423},
        pages = {748},
          doi = {10.1086/173854},
       adsurl = {https://ui.adsabs.harvard.edu/abs/1994ApJ...423..748M}
}

@ARTICLE{1991ApJ...381..210B,
       author = {{Blaes}, O. and {Rajagopal}, M.},
        title = "{The Statistics of Slow Interstellar Accretion onto Neutron Stars}",
      journal = {\apj},
         year = 1991,
        month = nov,
       volume = {381},
        pages = {210},
          doi = {10.1086/170642},
       adsurl = {https://ui.adsabs.harvard.edu/abs/1991ApJ...381..210B}
}

@ARTICLE{1993ApJ...403..690B,
       author = {{Blaes}, Omer and {Madau}, Piero},
        title = "{Can We Observe Accreting, Isolated Neutron Stars?}",
      journal = {\apj},
         year = 1993,
        month = feb,
       volume = {403},
        pages = {690},
          doi = {10.1086/172240},
       adsurl = {https://ui.adsabs.harvard.edu/abs/1993ApJ...403..690B}
}

@ARTICLE{2021A&A...647A...1P,
       author = {{Predehl}, P. and {Andritschke}, R. and {Arefiev}, V. and {Babyshkin}, V. and {Batanov}, O. and {Becker}, W. and {B{\"o}hringer}, H. and {Bogomolov}, A. and {Boller}, T. and {Borm}, K. and {Bornemann}, W. and {Br{\"a}uninger}, H. and {Br{\"u}ggen}, M. and {Brunner}, H. and {Brusa}, M. and {Bulbul}, E. and {Buntov}, M. and {Burwitz}, V. and {Burkert}, W. and {Clerc}, N. and {Churazov}, E. and {Coutinho}, D. and {Dauser}, T. and {Dennerl}, K. and {Doroshenko}, V. and {Eder}, J. and {Emberger}, V. and {Eraerds}, T. and {Finoguenov}, A. and {Freyberg}, M. and {Friedrich}, P. and {Friedrich}, S. and {F{\"u}rmetz}, M. and {Georgakakis}, A. and {Gilfanov}, M. and {Granato}, S. and {Grossberger}, C. and {Gueguen}, A. and {Gureev}, P. and {Haberl}, F. and {H{\"a}lker}, O. and {Hartner}, G. and {Hasinger}, G. and {Huber}, H. and {Ji}, L. and {Kienlin}, A. v. and {Kink}, W. and {Korotkov}, F. and {Kreykenbohm}, I. and {Lamer}, G. and {Lomakin}, I. and {Lapshov}, I. and {Liu}, T. and {Maitra}, C. and {Meidinger}, N. and {Menz}, B. and {Merloni}, A. and {Mernik}, T. and {Mican}, B. and {Mohr}, J. and {M{\"u}ller}, S. and {Nandra}, K. and {Nazarov}, V. and {Pacaud}, F. and {Pavlinsky}, M. and {Perinati}, E. and {Pfeffermann}, E. and {Pietschner}, D. and {Ramos-Ceja}, M.~E. and {Rau}, A. and {Reiffers}, J. and {Reiprich}, T.~H. and {Robrade}, J. and {Salvato}, M. and {Sanders}, J. and {Santangelo}, A. and {Sasaki}, M. and {Scheuerle}, H. and {Schmid}, C. and {Schmitt}, J. and {Schwope}, A. and {Shirshakov}, A. and {Steinmetz}, M. and {Stewart}, I. and {Str{\"u}der}, L. and {Sunyaev}, R. and {Tenzer}, C. and {Tiedemann}, L. and {Tr{\"u}mper}, J. and {Voron}, V. and {Weber}, P. and {Wilms}, J. and {Yaroshenko}, V.},
        title = "{The eROSITA X-ray telescope on SRG}",
      journal = {\aap},
         year = 2021,
        month = mar,
       volume = {647},
          eid = {A1},
        pages = {A1},
          doi = {10.1051/0004-6361/202039313},
archivePrefix = {arXiv},
       eprint = {2010.03477},
 primaryClass = {astro-ph.HE},
       adsurl = {https://ui.adsabs.harvard.edu/abs/2021A&A...647A...1P}
}

@ARTICLE{2025arXiv250906699P,
       author = {{Pons}, Jos{\'e} A. and {Dehman}, Clara and {Vigan{\`o}}, Daniele},
        title = "{Magnetic, thermal and rotational evolution of isolated neutron stars}",
      journal = {arXiv e-prints},
         year = 2025,
        month = sep,
          eid = {arXiv:2509.06699},
        pages = {arXiv:2509.06699},
          doi = {10.48550/arXiv.2509.06699},
archivePrefix = {arXiv},
       eprint = {2509.06699},
 primaryClass = {astro-ph.HE},
       adsurl = {https://ui.adsabs.harvard.edu/abs/2025arXiv250906699P}
}

@ARTICLE{1969ApJ...157.1395O,
       author = {{Ostriker}, J.~P. and {Gunn}, J.~E.},
        title = "{On the Nature of Pulsars. I. Theory}",
      journal = {\apj},
         year = 1969,
        month = sep,
       volume = {157},
        pages = {1395},
          doi = {10.1086/150160},
       adsurl = {https://ui.adsabs.harvard.edu/abs/1969ApJ...157.1395O}
}

@ARTICLE{2014MNRAS.444.1066I,
       author = {{Igoshev}, A.~P. and {Popov}, S.~B.},
        title = "{Modified pulsar current analysis: probing magnetic field evolution}",
      journal = {\mnras},
         year = 2014,
        month = oct,
       volume = {444},
       number = {2},
        pages = {1066-1076},
          doi = {10.1093/mnras/stu1496},
archivePrefix = {arXiv},
       eprint = {1407.6269},
 primaryClass = {astro-ph.HE},
       adsurl = {https://ui.adsabs.harvard.edu/abs/2014MNRAS.444.1066I}
}

@ARTICLE{2025arXiv250821805C,
       author = {{Chawla}, Chirag and {Chatterjee}, Sourav and {Breivik}, Katelyn},
        title = "{Gaia's promise to detect compact-object binaries: where we stand with the third data release}",
      journal = {arXiv e-prints},
         year = 2025,
        month = aug,
          eid = {arXiv:2508.21805},
        pages = {arXiv:2508.21805},
          doi = {10.48550/arXiv.2508.21805},
archivePrefix = {arXiv},
       eprint = {2508.21805},
 primaryClass = {astro-ph.SR},
       adsurl = {https://ui.adsabs.harvard.edu/abs/2025arXiv250821805C}
}

@ARTICLE{2023arXiv230612514L,
       author = {{Lam}, Casey Y. and {Abrams}, Natasha and {Andrews}, Jeff and {Bachelet}, Etienne and {Bahramian}, Arash and {Bennett}, David and {Bozza}, Valerio and {Broekgaarden}, Floor and {Chakrabarti}, Sukanya and {Dawson}, William and {El-Badry}, Kareem and {Fishbach}, Maya and {Fragione}, Giacomo and {Gaudi}, Scott and {Gautam}, Abhimat and {Hirai}, Ryosuke and {Holz}, Daniel and {Hosek}, Jr., Matthew and {Huston}, Macy and {Jayasinghe}, Tharindu and {Johnson}, Samson and {Kawata}, Daisuke and {Koshimoto}, Naoki and {Lu}, Jessica R. and {Mandel}, Ilya and {Miyazaki}, Shota and {Mr{\'o}z}, Przemek and {Naoz}, Smadar and {Ranc}, Cl{\'e}ment and {Rowan}, Dominick and {Sch{\"o}del}, Rainer and {Shenar}, Tomer and {Simon}, Josh and {Street}, Rachel and {Sumi}, Takahiro and {Suzuki}, Daisuke and {Terry}, Sean},
        title = "{Roman CCS White Paper: Characterizing the Galactic population of isolated black holes}",
      journal = {arXiv e-prints},
         year = 2023,
        month = jun,
          eid = {arXiv:2306.12514},
        pages = {arXiv:2306.12514},
          doi = {10.48550/arXiv.2306.12514},
archivePrefix = {arXiv},
       eprint = {2306.12514},
 primaryClass = {astro-ph.IM},
       adsurl = {https://ui.adsabs.harvard.edu/abs/2023arXiv230612514L}
}

@ARTICLE{2023AJ....165...96S,
       author = {{Sajadian}, Sedighe and {Sahu}, Kailash C.},
        title = "{Detecting Isolated Stellar-mass Black Holes with the Roman Telescope}",
      journal = {\aj},
         year = 2023,
        month = mar,
       volume = {165},
       number = {3},
          eid = {96},
        pages = {96},
          doi = {10.3847/1538-3881/acb20f},
archivePrefix = {arXiv},
       eprint = {2301.03812},
 primaryClass = {astro-ph.GA},
       adsurl = {https://ui.adsabs.harvard.edu/abs/2023AJ....165...96S}
}

@ARTICLE{2024MNRAS.531.2433S,
       author = {{Sweeney}, David and {Tuthill}, Peter and {Krone-Martins}, Alberto and {M{\'e}rand}, Antoine and {Scalzo}, Richard and {Martinod}, Marc-Antoine},
        title = "{Observing the galactic underworld: predicting photometry and astrometry from compact remnant microlensing events}",
      journal = {\mnras},
         year = 2024,
        month = jun,
       volume = {531},
       number = {2},
        pages = {2433-2447},
          doi = {10.1093/mnras/stae1302},
archivePrefix = {arXiv},
       eprint = {2403.14612},
 primaryClass = {astro-ph.GA},
       adsurl = {https://ui.adsabs.harvard.edu/abs/2024MNRAS.531.2433S}
}

@ARTICLE{2010A&A...523A..33S,
       author = {{Sartore}, N. and {Treves}, A.},
        title = "{Probing isolated compact remnants with microlensing}",
      journal = {\aap},
         year = 2010,
        month = nov,
       volume = {523},
          eid = {A33},
        pages = {A33},
          doi = {10.1051/0004-6361/201015060},
archivePrefix = {arXiv},
       eprint = {1009.0005},
 primaryClass = {astro-ph.GA},
       adsurl = {https://ui.adsabs.harvard.edu/abs/2010A&A...523A..33S}
}

@ARTICLE{2020ApJ...889...31L,
       author = {{Lam}, Casey Y. and {Lu}, Jessica R. and {Hosek}, Jr., Matthew W. and {Dawson}, William A. and {Golovich}, Nathan R.},
        title = "{PopSyCLE: A New Population Synthesis Code for Compact Object Microlensing Events}",
      journal = {\apj},
         year = 2020,
        month = jan,
       volume = {889},
       number = {1},
          eid = {31},
        pages = {31},
          doi = {10.3847/1538-4357/ab5fd3},
archivePrefix = {arXiv},
       eprint = {1912.04510},
 primaryClass = {astro-ph.SR},
       adsurl = {https://ui.adsabs.harvard.edu/abs/2020ApJ...889...31L}
}

@ARTICLE{2021AcA....71...89M,
       author = {{Mr{\'o}z}, P. and {Wyrzykowski}, {\L}.},
        title = "{Measuring the Mass Function of Isolated Stellar Remnants with Gravitational Microlensing I. Revisiting the OGLE-III Dark Lens Candidates}",
      journal = {\actaa},
         year = 2021,
        month = jun,
       volume = {71},
       number = {2},
        pages = {89-102},
          doi = {10.32023/0001-5237/71.2.1},
archivePrefix = {arXiv},
       eprint = {2107.13701},
 primaryClass = {astro-ph.SR},
       adsurl = {https://ui.adsabs.harvard.edu/abs/2021AcA....71...89M}
}

@ARTICLE{1996ApJ...462..563N,
       author = {{Navarro}, Julio F. and {Frenk}, Carlos S. and {White}, Simon D.~M.},
        title = "{The Structure of Cold Dark Matter Halos}",
      journal = {\apj},
         year = 1996,
        month = may,
       volume = {462},
        pages = {563},
          doi = {10.1086/177173},
archivePrefix = {arXiv},
       eprint = {astro-ph/9508025},
 primaryClass = {astro-ph},
       adsurl = {https://ui.adsabs.harvard.edu/abs/1996ApJ...462..563N}
}

@ARTICLE{2017JOSS....2..388P,
       author = {{Price-Whelan}, Adrian M.},
        title = "{Gala: A Python package for galactic dynamics}",
      journal = {The Journal of Open Source Software},
         year = 2017,
        month = oct,
       volume = {2},
          eid = {388},
        pages = {388},
          doi = {10.21105/joss.00388},
       adsurl = {https://ui.adsabs.harvard.edu/abs/2017JOSS....2..388P}
}

@ARTICLE{2022AstL...48..243B,
       author = {{Bobylev}, V.~V. and {Bajkova}, A.~T. and {Karelin}, G.~M.},
        title = "{Kinematics of OB Stars with Data from the LAMOST and Gaia Catalogues}",
      journal = {Astronomy Letters},
         year = 2022,
        month = apr,
       volume = {48},
       number = {4},
        pages = {243-255},
          doi = {10.1134/S1063773722040016},
archivePrefix = {arXiv},
       eprint = {2207.01924},
 primaryClass = {astro-ph.GA},
       adsurl = {https://ui.adsabs.harvard.edu/abs/2022AstL...48..243B}
}

@ARTICLE{2014ApJS..212....6O,
       author = {{Olausen}, S.~A. and {Kaspi}, V.~M.},
        title = "{The McGill Magnetar Catalog}",
      journal = {\apjs},
         year = 2014,
        month = may,
       volume = {212},
       number = {1},
          eid = {6},
        pages = {6},
          doi = {10.1088/0067-0049/212/1/6},
archivePrefix = {arXiv},
       eprint = {1309.4167},
 primaryClass = {astro-ph.HE},
       adsurl = {https://ui.adsabs.harvard.edu/abs/2014ApJS..212....6O}
}

@ARTICLE{2003ApJ...588..400R,
       author = {{Romanova}, M.~M. and {Toropina}, O.~D. and {Toropin}, Yu. M. and {Lovelace}, R.~V.~E.},
        title = "{Magnetohydrodynamic Simulations of Accretion onto a Star in the ``Propeller'' Regime}",
      journal = {\apj},
         year = 2003,
        month = may,
       volume = {588},
       number = {1},
        pages = {400-407},
          doi = {10.1086/373990},
archivePrefix = {arXiv},
       eprint = {astro-ph/0209548},
 primaryClass = {astro-ph},
       adsurl = {https://ui.adsabs.harvard.edu/abs/2003ApJ...588..400R}
}

@ARTICLE{1985A&A...151..361W,
       author = {{Wang}, Y.-M. and {Robertson}, J.~A.},
        title = "{'Propeller' action by rotating neutron stars}",
      journal = {\aap},
         year = 1985,
        month = oct,
       volume = {151},
       number = {2},
        pages = {361-371},
       adsurl = {https://ui.adsabs.harvard.edu/abs/1985A&A...151..361W}
}

@ARTICLE{2017MNRAS.465L.119P,
       author = {{Postnov}, K. and {Oskinova}, L. and {Torrej{\'o}n}, J.~M.},
        title = "{A propelling neutron star in the enigmatic Be-star {\ensuremath{\gamma}} Cassiopeia}",
      journal = {\mnras},
         year = 2017,
        month = feb,
       volume = {465},
       number = {1},
        pages = {L119-L123},
          doi = {10.1093/mnrasl/slw223},
archivePrefix = {arXiv},
       eprint = {1610.07799},
 primaryClass = {astro-ph.HE},
       adsurl = {https://ui.adsabs.harvard.edu/abs/2017MNRAS.465L.119P}
}

@ARTICLE{2017MNRAS.469.1502S,
       author = {{Smith}, M.~A. and {Lopes de Oliveira}, R. and {Motch}, C.},
        title = "{Is there a propeller neutron star in {\ensuremath{\gamma}} Cas?}",
      journal = {\mnras},
         year = 2017,
        month = aug,
       volume = {469},
       number = {2},
        pages = {1502-1509},
          doi = {10.1093/mnras/stx926},
archivePrefix = {arXiv},
       eprint = {1704.05060},
 primaryClass = {astro-ph.HE},
       adsurl = {https://ui.adsabs.harvard.edu/abs/2017MNRAS.469.1502S}
}

@ARTICLE{2001A&A...368L...5I,
       author = {{Ikhsanov}, N.~R.},
        title = "{On the duration of the subsonic propeller state of neutron stars in wind-fed mass-exchange close binary systems}",
      journal = {\aap},
         year = 2001,
        month = mar,
       volume = {368},
        pages = {L5-L7},
          doi = {10.1051/0004-6361:20010140},
archivePrefix = {arXiv},
       eprint = {astro-ph/0111505},
 primaryClass = {astro-ph},
       adsurl = {https://ui.adsabs.harvard.edu/abs/2001A&A...368L...5I}
}

@ARTICLE{2025A&A...693A.260Z,
       author = {{Zainab}, A. and {Avakyan}, A. and {Doroshenko}, V. and {Thalhammer}, P. and {Sokolova-Lapa}, E. and {Ballhausen}, R. and {Zalot}, N. and {Stierhof}, J. and {H{\"a}mmerich}, S. and {Diez}, C.~M. and {Weber}, P. and {Dauser}, T. and {Berger}, K. and {Kretschmar}, P. and {Pottschmidt}, K. and {Pradhan}, P. and {Islam}, N. and {Maitra}, C. and {Coley}, J.~B. and {Blay}, P. and {Corbet}, R.~H.~D. and {Rothschild}, R.~E. and {Wood}, K. and {Santangelo}, A. and {Heber}, U. and {Wilms}, J.},
        title = "{Multi-wavelength study of 1eRASS J085039.9‑421151 with eROSITA, NuSTAR, and X-shooter}",
      journal = {\aap},
         year = 2025,
        month = jan,
       volume = {693},
          eid = {A260},
        pages = {A260},
          doi = {10.1051/0004-6361/202348708},
archivePrefix = {arXiv},
       eprint = {2411.02655},
 primaryClass = {astro-ph.HE},
       adsurl = {https://ui.adsabs.harvard.edu/abs/2025A&A...693A.260Z}
}

@ARTICLE{2000A&AT...19..471P,
       author = {{Popov}, S.~B. and {Colpi}, M. and {Treves}, A. and {Turolla}, R. and {Lipunov}, V.~M. and {Prokhorov}, M.~E.},
        title = "{Population synthesis of old neutron stars in the galaxy}",
      journal = {Astronomical and Astrophysical Transactions},
         year = 2000,
        month = jan,
       volume = {19},
       number = {3},
        pages = {471-478},
          doi = {10.1080/10556790008238592},
archivePrefix = {arXiv},
       eprint = {astro-ph/9910320},
 primaryClass = {astro-ph},
       adsurl = {https://ui.adsabs.harvard.edu/abs/2000A&AT...19..471P}
}

@ARTICLE{1998AN....319..109T,
       author = {{Treves}, A. and {Colpi}, M. and {Turolla}, R.},
        title = "{Can magnetic field decay explain the elusivity of old neutron stars ?}",
      journal = {Astronomische Nachrichten},
         year = 1998,
        month = jan,
       volume = {319},
       number = {1},
        pages = {109},
          doi = {10.1002/asna.2123190151},
       adsurl = {https://ui.adsabs.harvard.edu/abs/1998AN....319..109T}
}

@ARTICLE{2025ApJ...986...88S,
       author = {{Sautron}, M. and {McEwen}, A.~E. and {Younes}, G. and {P{\'e}tri}, J. and {Beniamini}, P. and {Huppenkothen}, D.},
        title = "{The Galactic Population of Magnetars: A Simulation-based Inference Study}",
      journal = {\apj},
         year = 2025,
        month = jun,
       volume = {986},
       number = {1},
          eid = {88},
        pages = {88},
          doi = {10.3847/1538-4357/add0aa},
archivePrefix = {arXiv},
       eprint = {2503.11875},
 primaryClass = {astro-ph.HE},
       adsurl = {https://ui.adsabs.harvard.edu/abs/2025ApJ...986...88S}
}

@ARTICLE{2002ApJ...566L..85B,
       author = {{Beloborodov}, Andrei M.},
        title = "{Gravitational Bending of Light Near Compact Objects}",
      journal = {\apjl},
         year = 2002,
        month = feb,
       volume = {566},
       number = {2},
        pages = {L85-L88},
          doi = {10.1086/339511},
archivePrefix = {arXiv},
       eprint = {astro-ph/0201117},
 primaryClass = {astro-ph},
       adsurl = {https://ui.adsabs.harvard.edu/abs/2002ApJ...566L..85B}
}

@ARTICLE{2001ApJ...547..355P,
       author = {{Popham}, Robert and {Sunyaev}, Rashid},
        title = "{Accretion Disk Boundary Layers around Neutron Stars: X-Ray Production in Low-Mass X-Ray Binaries}",
      journal = {\apj},
         year = 2001,
        month = jan,
       volume = {547},
       number = {1},
        pages = {355-383},
          doi = {10.1086/318336},
archivePrefix = {arXiv},
       eprint = {astro-ph/0004017},
 primaryClass = {astro-ph},
       adsurl = {https://ui.adsabs.harvard.edu/abs/2001ApJ...547..355P}
}

@ARTICLE{2026MNRAS.545f2263M,
       author = {{Meyer-Hofmeister}, Emmi and {Wang}, Yilong and {Liu}, B.~F.},
        title = "{Accretion geometry in neutron star low-mass X-ray binaries during the hard spectral state}",
      journal = {\mnras},
         year = 2026,
        month = feb,
       volume = {545},
       number = {4},
          eid = {staf2263},
        pages = {staf2263},
          doi = {10.1093/mnras/staf2263},
archivePrefix = {arXiv},
       eprint = {2512.18839},
 primaryClass = {astro-ph.HE},
       adsurl = {https://ui.adsabs.harvard.edu/abs/2026MNRAS.545f2263M}
}

@ARTICLE{2007A&ARv..15....1D,
       author = {{Done}, Chris and {Gierli{\'n}ski}, Marek and {Kubota}, Aya},
        title = "{Modelling the behaviour of accretion flows in X-ray binaries. Everything you always wanted to know about accretion but were afraid to ask}",
      journal = {\aapr},
         year = 2007,
        month = dec,
       volume = {15},
       number = {1},
        pages = {1-66},
          doi = {10.1007/s00159-007-0006-1},
archivePrefix = {arXiv},
       eprint = {0708.0148},
 primaryClass = {astro-ph},
       adsurl = {https://ui.adsabs.harvard.edu/abs/2007A&ARv..15....1D}
}

@ARTICLE{2008LRR....11...10C,
       author = {{Chamel}, Nicolas and {Haensel}, Pawel},
        title = "{Physics of Neutron Star Crusts}",
      journal = {Living Reviews in Relativity},
         year = 2008,
        month = dec,
       volume = {11},
       number = {1},
          eid = {10},
        pages = {10},
          doi = {10.12942/lrr-2008-10},
archivePrefix = {arXiv},
       eprint = {0812.3955},
 primaryClass = {astro-ph},
       adsurl = {https://ui.adsabs.harvard.edu/abs/2008LRR....11...10C}
}

@ARTICLE{2026A&A...705A.148K,
       author = {{Kurpas}, J. and {Pires}, A.~M. and {Schwope}, A.~D. and {Li}, B. and {Yin}, D. and {Haberl}, F. and {Krumpe}, M. and {Sheth}, S. and {Traulsen}, I. and {Zhang}, Z.~L.},
        title = "{X-ray, optical, and radio follow-up of five thermally emitting isolated neutron star candidates}",
      journal = {\aap},
         year = 2026,
        month = jan,
       volume = {705},
          eid = {A148},
        pages = {A148},
          doi = {10.1051/0004-6361/202556815},
archivePrefix = {arXiv},
       eprint = {2511.19591},
 primaryClass = {astro-ph.HE},
       adsurl = {https://ui.adsabs.harvard.edu/abs/2026A&A...705A.148K}
}

@ARTICLE{2020A&A...640A..24P,
       author = {{Poutanen}, Juri},
        title = "{Accurate analytic formula for light bending in Schwarzschild metric}",
      journal = {\aap},
         year = 2020,
        month = aug,
       volume = {640},
          eid = {A24},
        pages = {A24},
          doi = {10.1051/0004-6361/202037471},
archivePrefix = {arXiv},
       eprint = {1909.05732},
 primaryClass = {astro-ph.HE},
       adsurl = {https://ui.adsabs.harvard.edu/abs/2020A&A...640A..24P}
}

@ARTICLE{2004MNRAS.351..569P,
       author = {{Payne}, D.~J.~B. and {Melatos}, A.},
        title = "{Burial of the polar magnetic field of an accreting neutron star - I. Self-consistent analytic and numerical equilibria}",
      journal = {\mnras},
         year = 2004,
        month = jun,
       volume = {351},
       number = {2},
        pages = {569-584},
          doi = {10.1111/j.1365-2966.2004.07798.x},
archivePrefix = {arXiv},
       eprint = {astro-ph/0403173},
 primaryClass = {astro-ph},
       adsurl = {https://ui.adsabs.harvard.edu/abs/2004MNRAS.351..569P}
}






\end{document}